\documentclass[12pt]{article}
\RequirePackage{amsthm,amsmath,natbib}
\usepackage{amsmath, amsthm, amssymb}
\usepackage{verbatim,color,amssymb,epsfig}
\usepackage[usenames,dvipsnames]{xcolor}
\usepackage{multirow,booktabs}

\usepackage{xr}
\RequirePackage{amsthm,amsmath,natbib}
\usepackage{amsmath, amsthm, amssymb}
\usepackage[auth-sc, affil-sl]{authblk}
\usepackage{epsfig,natbib}

\renewcommand{\baselinestretch}{1.3}
\renewcommand{\baselinestretch}{1.5}
\def\boxit#1{\vbox{\hrule\hbox{\vrule\kern6pt
          \vbox{\kern6pt#1\kern6pt}\kern6pt\vrule}\hrule}}

\def\trans{^{\rm T}}
\def\th{^{th}}
\def\Normal{\hbox{Normal}}

\numberwithin{equation}{section}
\theoremstyle{plain}

\newtheorem{Theorem}{\underline{\bf Theorem}}[section]

\newtheorem{Remark}{\underline{\bf Remark}}

\newtheorem{Lemma}{\underline{\bf Lemma}}

\newtheorem{Condition}{\underline{\bf Condition}}
\newtheorem{Definition}{\underline{\bf Definition}}
\newtheorem{Algorithm}{\underline{\bf Algorithm}}[section]
\begin{document}

\thispagestyle{empty}
\baselineskip=28pt
\begin{center}
{\LARGE{\bf Sparse Fisher's discriminant analysis with thresholded linear constraints}}
\end{center}
\vskip 10mm

\baselineskip=10pt
\vskip 2mm
\begin{center}
 Ruiyan Luo\\
\vskip 2mm
Division of Epidemiology and Biostatistics, Georgia State University School of Public Health, One Park Place, Atlanta, GA 30303\\
rluo@gsu.edu\\
\hskip 5mm\\
Xin Qi\\
\vskip 2mm
Department of Mathematics and Statistics, Georgia State University, 30 Pryor Street, Atlanta, GA 30303-3083\\
xqi3@gsu.edu \\
\end{center}

\begin{abstract}
\baselineskip=15pt
Various regularized linear discriminant analysis (LDA) methods have been proposed to address the problems of the classic methods in high-dimensional settings. Asymptotic optimality has been established for some of these methods in high dimension when there are only two classes. A major difficulty in proving asymptotic optimality for multiclass classification is that the classification boundary is typically complicated and no explicit formula for classification error generally exists when the number of classes is greater than two.   For the Fisher's LDA, one additional difficulty is that  the covariance matrix is also involved in the linear constraints. The main purpose of this paper is to establish asymptotic consistency and asymptotic optimality for our sparse Fisher's LDA with thresholded linear constraints in the high-dimensional settings for arbitrary number of classes.  To address the first difficulty above, we provide asymptotic optimality and the corresponding convergence rates in high-dimensional settings for a large family of linear classification rules with arbitrary number of classes, and apply them to our method. To overcome the second difficulty, we propose a thresholding approach to avoid the estimate of the covariance matrix. We apply the  method to the classification problems for multivariate functional data through the wavelet transformations.  
\end{abstract}

 \vspace{4mm}
\par\medskip\noindent
\underline{\bf Short title}: Sparse FDA with thresholded linear constraints

\pagenumbering{arabic}
\newlength{\gnat}
\setlength{\gnat}{22pt}
\baselineskip=\gnat

\section{Introduction}
The linear discriminant analysis (LDA) has been a favored tool for classification in the settings of small $p$ and large $n$. The Fisher's  discriminant analysis is one of its important special cases. However, these classic methods face major problems for high-dimensional data. In theory, \citet{bickel2004some} and \citet{Shao-2011} showed that the usual LDA can be as bad as the random guessing when  $p>n$. In practice, the classic  methods have poor predictive performance in high-dimensional settings.  To address these problems, various regularized discriminant analysis methods have been proposed, including \citet{Friedman-1989}, \citet{Krzanowski-1995}, \citet{Dudoit-2001}, \citet{bickel2004some}, \citet{Guo-2007}, \citet{Xu-2009}, \citet{Tibshirani-2009}, \citet{Witten-2011}, \citet{Clemmensen-2011}, \citet{Shao-2011}, \citet{cai2011direct}, \citet{fan2012road}, \citet{Qi-2013} and many others. 

Asymptotic optimality has been established in some of these papers when there are two classes. \citet{Shao-2011} made   sparsity assumptions on both the difference $\boldsymbol{\delta}=\boldsymbol{\mu}_2-\boldsymbol{\mu}_1$, where $\boldsymbol{\mu}_1$ and $\boldsymbol{\mu}_2$ are the population means of the two classes, and the within-class covariance matrix ${\boldsymbol{\Sigma}}$. Then they applied thresholding procedures to both the sample estimates of $\boldsymbol{\delta}$ and  ${\boldsymbol{\Sigma}}$, and obtained the asymptotic optimality and the corresponding convergence rate for their classification rule. \citet{cai2011direct} observed that in the case of two classes, the optimal classification rule depends on ${\boldsymbol{\Sigma}}$ only through ${\boldsymbol{\Sigma}}^{-1}\boldsymbol{\delta}$. Hence, they assumed $l_1$ sparsity for ${\boldsymbol{\Sigma}}^{-1}\boldsymbol{\delta}$, proposed a sparse estimate of it
through minimizing its  $l_1$ norm with an $l_{\infty}$ constraint, and provided asymptotic optimality of their classification rule. \citet{fan2012road} imposed $l_0$ sparsity assumption on ${\boldsymbol{\Sigma}}^{-1}\boldsymbol{\delta}$, estimated it through a minimization problem with an $l_1$ constraint and derived the  asymptotic optimality.  A  difficulty preventing the derivation of asymptotic optimality of the linear classification rules for multiple classes is that for the two-class classification, the classification boundary of LDA is a hyperplane and an explicit formula for the classification error exists, however, for the multiclass classification, the classification boundary is usually complicated and no explicit formula for the classification error generally exist. The Fisher's LDA projects the original variables $\mathbf{X}$ to a low dimensional subspace to generate new predictor variables, $\mathbf{X}{\boldsymbol{\alpha}}_1$, $\mathbf{X}{\boldsymbol{\alpha}}_2, \ldots, \mathbf{X}{\boldsymbol{\alpha}}_{K-1}$, where  the coefficient vectors ${\boldsymbol{\alpha}}_1, {\boldsymbol{\alpha}}_2, \ldots, {\boldsymbol{\alpha}}_{K-1}$   satisfy the linear constraints $ \boldsymbol{\alpha}_i\trans{\boldsymbol{\Sigma}}{\boldsymbol{\alpha}}_j=0$ for any $1\le j<i <K$, and $K$ is the number of classes. These constraints imply that $\boldsymbol{\alpha}_i$ is orthogonal to the subspace spanned by $\{{\boldsymbol{\Sigma}}{\boldsymbol{\alpha}}_1, \cdots, {\boldsymbol{\Sigma}}{\boldsymbol{\alpha}}_{i-1}\}$.  

The motivation of this paper is to establish the asymptotic consistency and the asymptotic optimality of the sparse Fisher's LDA method proposed in \citet*{Qi-2013}  in the high-dimensional settings for arbitrary number of classes. However, in order to obtain the asymptotic consistency, we revise the original method because it is hard to obtain a consistent estimate for a general $\boldsymbol{\Sigma}$ in the high-dimensional settings without sparsity or other assumptions imposed on $\boldsymbol{\Sigma}$. 
Instead of aiming to estimate $\boldsymbol{\Sigma}$, we propose a soft-thresholding procedure and add it into the original method to get  a consistent estimate of the subspace $\{{\boldsymbol{\Sigma}}{\boldsymbol{\alpha}}_1, \cdots, {\boldsymbol{\Sigma}}{\boldsymbol{\alpha}}_{i-1}\}$. We establish the asymptotic consistency of the estimates of $\boldsymbol{\alpha}_i$ and the subspace $\{{\boldsymbol{\Sigma}}{\boldsymbol{\alpha}}_1, \cdots, {\boldsymbol{\Sigma}}{\boldsymbol{\alpha}}_{i-1}\}$ for the revised method. To prove the asymptotic optimality for this method, we establish the asymptotic optimality and the corresponding convergence rates in high-dimensional settings for a large family of linear classification rules with arbitrary number of classes   under the situation of multivariate normal distribution.  To assess the real performance of the revised method, we compare it with the original method and other sparse LDA methods through simulation studies. The revised method has good predictive performance as the original method and at the same time, it enjoys nice theoretical properties.  We also apply the revised method to the classification problems for multivariate functional data through the wavelet transformations.

The rest of this paper is organized as follows. In Section 2, we introduce notations and briefly review the classic Fisher's discriminant analysis. Our sparse Fisher's LDA method with thresholded linear constraints is introduced in Section 3.  In Section 4, we present the main theoretical results. Sections 5 and 6 are simulation studies and applications, respectively. The proofs of all theorems are provided in supplementary material.
 
\section{Fisher's discriminant analysis}\label{sec4}
We first introduce the notations used throughout the paper. For any vector $\mathbf{v}=(v_1,\cdots,v_p)\trans$, let $\|\mathbf{v}\|_1$, $\|\mathbf{v}\|_2$,  and $\|\mathbf{v}\|_{\infty}=\max_{1\le i\le p}|v_i|$ denote the $l_1$, $l_2$, and $l_{\infty}$ norms of $\mathbf{v}$, respectively. For any $p\times p$ symmetric matrix $\mathbf{M}$, we use $\lambda_{max}(\mathbf{M})$, $\lambda_{min}(\mathbf{M})$ and $\lambda^+_{min}(\mathbf{M})$ to denote the largest eigenvalue, the smallest eigenvalue and the smallest positive eigenvalue of $\mathbf{M}$, respectively. Now suppose that $\mathbf{M}$ is symmetric and nonnegative definite. We define two norms for $\mathbf{M}$, 
\begin{align}
\|\mathbf{M}\|=\sup_{\mathbf{v}\in\mathbb{R}^p, \|\mathbf{v}\|_2=1}\|\mathbf{M}\mathbf{v}\|_2=\lambda_{max}(\mathbf{M}), \qquad \text{ and } \qquad \|\mathbf{M}\|_{\infty}=\max_{1\le k,l\le p}|M_{kl}|\;,
\end{align}
where $M_{kl}$ is the $(k,l)$-th entry of $\mathbf{M}$.  The first norm is the usual {\it operator norm} and is also called the {\it spectral norm}. The second is  the {\it max norm}.

Throughout this paper, we assume that the number $K$ of classes is fixed and can be any positive integer.  Suppose that the population in the $i$-th class has a multivariate normal distribution $N_p(\boldsymbol{\mu}_i,  \boldsymbol{\Sigma})$, where $\boldsymbol{\mu}_i$ is the true class mean of the $i$-th class, $1\le i\le K$, and $\boldsymbol{\Sigma}$ is the true common within-class covariance matrix for all classes. We assume that the prior probabilities for all the classes are the same and equal to $1/K$. It will be seen that when we add a constant vector to all the observations, the classification results do not change for all the classification rules involved in this paper. Therefore, without loss of generality, we assume that the overall mean of the whole population is zero,  that is,
 \begin{align}
\boldsymbol{\mu}_1+\boldsymbol{\mu}_2+\cdots+\boldsymbol{\mu}_K=\mathbf{0}. \label{3}
\end{align} 
Define a $p\times K$ matrix $\mathbf{U}=[\boldsymbol{\mu}_1,\boldsymbol{\mu}_2,\cdots,\boldsymbol{\mu}_K]$, which is the collection of class means. Under the assumption $\eqref{3}$, the between-class covariance matrix is  
\begin{align}
\mathbf{B}=\sum_{i=1}^K\boldsymbol{\mu}_i\boldsymbol{\mu}_i\trans/K=\mathbf{U}\mathbf{U}\trans /K\;. \label{4}
\end{align} 
 The Fisher's discriminant analysis method (when the true class means and the true covariance matrix are  known) sequentially finds linear combinations $\mathbf{X}\boldsymbol{\alpha}_{1}, \cdots, \mathbf{X}\boldsymbol{\alpha}_{K-1}$ by solving the following generalized eigenvalue problem. Suppose that we have obtained $ \boldsymbol{\alpha}_{1}, \cdots,  \boldsymbol{\alpha}_{i-1}$, where $1\le i\le K-1$, then $\boldsymbol{\alpha}_i$ is the solution to 
\begin{align}
\max_{\boldsymbol{\alpha}\in \mathbb{R}^p}\boldsymbol{\alpha}\trans \mathbf{B}\boldsymbol{\alpha}, \quad \text{ subject to }\quad \boldsymbol{\alpha}\trans \boldsymbol{\Sigma}\boldsymbol{\alpha}=1,\quad \boldsymbol{\alpha}\trans \boldsymbol{\Sigma}\boldsymbol{\alpha}_j=0,\quad 1\le j\le i-1.\label{2220}
\end{align}
   The Fisher's classification rule is to assign a new observation $\mathbf{x}$ to class $i$ if 
\begin{align}
(\mathbf{x}-\boldsymbol{\mu}_i)\trans\mathbf{D}(\mathbf{x}-\boldsymbol{\mu}_i)<(\mathbf{x}-\boldsymbol{\mu}_j)\trans\mathbf{D}(\mathbf{x}-\boldsymbol{\mu}_j)\label{1170}
\end{align}
for all $1\le j\neq i\le K$, where $\mathbf{D}=\sum_{k=1}^{K-1}\boldsymbol{\alpha}_k\boldsymbol{\alpha}_k\trans$. 

It is well known that under our setting (that is, the population in each class has a normal distribution with the same covariance matrix and the prior probabilities for all classes are the same), the optimal classification rule is to assign a new observation $\mathbf{x}$ to class $i$ if 
\begin{align}
(\mathbf{x}-\boldsymbol{\mu}_i)\trans\boldsymbol{\Sigma}^{-1}(\mathbf{x}-\boldsymbol{\mu}_i)<(\mathbf{x}-\boldsymbol{\mu}_j)\trans\boldsymbol{\Sigma}^{-1}(\mathbf{x}-\boldsymbol{\mu}_j)
\label{optrule.eq}
\end{align}
for all $1\le j\neq i\le K$ (Theorem 6.8.1 in \citet{anderson2003introduction} or Theorem 13.2 in \citet{hardle2007applied}). Moreover, the optimal rule $\eqref{optrule.eq}$ is equivalent to the Fisher's discriminant rule $\eqref{1170}$. 

In practice, the true class means and $\boldsymbol{\Sigma}$ are unknown. Consider a training data set, $\mathbf{X}=\{\mathbf{x}_{ij}: 1\le i\le K, 1\le j\le n_i\}$, where $\mathbf{x}_{ij}$ is the $j$th observation from the $i$th class and $n_i$ is the number of the observations of the $i$th class. 
The numbers $(n_1,n_2,\cdots,n_K)$ can be either random or nonrandom. Let $n=\sum_{i=1}^K n_i$ be the total number of observations in the data. Throughout this paper, we use 
 \begin{align}
&\bar{\mathbf{x}}_i=\frac{1}{n_i}\sum_{j=1}^{n_i}\mathbf{x}_{ij}, \quad\bar{\mathbf{x}}=\frac{1}{n}\sum_{i=1}^K\sum_{j=1}^{n_i}\mathbf{x}_{ij}, \quad\widehat{\boldsymbol{\Sigma}}=\frac{1}{n-K}\sum_{i=1}^K\sum_{j=1}^{n_i}(\mathbf{x}_{ij}-\bar{\mathbf{x}}_i)(\mathbf{x}_{ij}-\bar{\mathbf{x}}_i)\trans,\notag\\
&\widehat{\mathbf{B}}=\frac{1}{n}\sum_{i=1}^Kn_i(\bar{\mathbf{x}}_i-\bar{\mathbf{x}})(\bar{\mathbf{x}}_i-\bar{\mathbf{x}})\trans,\quad 1\le i\le K,\label{1097}
\end{align}
 to denote the sample class means, the sample overall mean, the sample within-class covariance matrix and the sample between-class covariance matrix, respectively. Then the classic Fisher's discriminant analysis is to sequentially obtain the estimates $\widehat{\boldsymbol{\alpha}}_1$, $\cdots$, $\widehat{\boldsymbol{\alpha}}_{K-1}$ of $ \boldsymbol{\alpha}_{1}, \cdots, \boldsymbol{\alpha}_{K-1}$ by solving 
\begin{align}
\max_{\boldsymbol{\alpha}\in \mathbb{R}^p}\boldsymbol{\alpha}\trans \widehat{\mathbf{B}}\boldsymbol{\alpha}, \quad \text{ subject to }\quad \boldsymbol{\alpha}\trans \widehat{\boldsymbol{\Sigma}}\boldsymbol{\alpha}=1,\quad \boldsymbol{\alpha}\trans \widehat{\boldsymbol{\Sigma}}\widehat{\boldsymbol{\alpha}}_j=0,\quad 1\le j<i,\label{1240}
\end{align}
 where $1\le i\le K-1$.  The classification rule is to assign a new observation $\mathbf{x}$ to class $i$ if 
\begin{align}
(\mathbf{x}-\bar{\mathbf{x}}_i)\trans\widetilde{\mathbf{D}}(\mathbf{x}-\bar{\mathbf{x}}_i)<(\mathbf{x}-\bar{\mathbf{x}}_j)\trans\widetilde{\mathbf{D}}(\mathbf{x}-\bar{\mathbf{x}}_j),
\label{classicrule.eq}
\end{align}
 for all $1\le j\neq i\le K$, where $\widetilde{\mathbf{D}}=\sum_{k=1}^{K-1}\widehat{\boldsymbol{\alpha}}_k\widehat{\boldsymbol{\alpha}}_k\trans$. 

\section{Sparse Fisher's discriminant analysis with thresholded linear constraints}\label{sect.3}
In the high-dimensional setting, the classic Fisher's  discriminant analysis has several drawbacks. First,  $\widehat{\boldsymbol{\Sigma}}$ is not full rank, so the solution to $\eqref{1240}$ does not exist. Second, $\widehat{\mathbf{B}}$ and $\widehat{\boldsymbol{\Sigma}}$ as given in $\eqref{1097}$ are not consistent estimates in terms of the operator norm. Hence, the estimates of ${\boldsymbol{\alpha}}_k$, $1\le k\le K-1$, obtained by classic Fisher's  discriminant analysis are not consistent.  Third, suppose that we have obtained an estimate $\widetilde{\boldsymbol{\alpha}}_1$ of $\boldsymbol{\alpha}_1$, in order to estimate $ \boldsymbol{\alpha}_2$, we have to estimate the coefficient vector of the linear constraint in $\eqref{2220}$, $\boldsymbol{\Sigma}\boldsymbol{\alpha}_1$. However, even if $\widetilde{\boldsymbol{\alpha}}_1$ is  consistent, $\widehat{\boldsymbol{\Sigma}}\widetilde{\boldsymbol{\alpha}}_1$ is not a consistent estimate of $\boldsymbol{\Sigma}\boldsymbol{\alpha}_1$ due to the inconsistency of $\widehat{\boldsymbol{\Sigma}}$.  To address these drawbacks, we describe a revised method of the sparse Fisher's discriminant analysis in \citet{Qi-2013}. 
 
\subsection{The case of $K=2$}\label{sect.3.1}
When there are two classes, there is only one component $\boldsymbol{\alpha}_1$ and $\mathbf{B}=(\boldsymbol{\mu}_1\boldsymbol{\mu}_1\trans+\boldsymbol{\mu}_2\boldsymbol{\mu}_2\trans)/2=\boldsymbol{\mu}_1\boldsymbol{\mu}_1\trans$ because $\boldsymbol{\mu}_1=-\boldsymbol{\mu}_2$.  It is easily seen that $\boldsymbol{\alpha}_1=\boldsymbol{\Sigma}^{-1}\boldsymbol{\delta}/\sqrt{\boldsymbol{\delta}\trans\boldsymbol{\Sigma}^{-1}\boldsymbol{\delta}}$, where $\boldsymbol{\delta}=\boldsymbol{\mu}_2-\boldsymbol{\mu}_1$. \citet{cai2011direct} and \citet{fan2012road} imposed $l_1$ and $l_0$ sparsity assumptions on $\boldsymbol{\Sigma}^{-1}\boldsymbol{\delta}$, respectively. Equivalently, we assume that $\boldsymbol{\alpha}_1$ is sparse in terms of $l_1$ norm  as in \citet{cai2011direct}. In the case of $K=2$, it is not necessary to revise the original method in \citet{Qi-2013}. The estimate $\widehat{\boldsymbol{\alpha}}_1$ of $\boldsymbol{\alpha}_1$ is the solution to 
\begin{align}
 \max_{\boldsymbol{\alpha}\in\mathbb{R}^p}\quad\boldsymbol{\alpha}\trans \widehat{\mathbf{B}}\boldsymbol{\alpha}, \quad\text{ subject to }\quad \boldsymbol{\alpha}\trans \widehat{\boldsymbol{\Sigma}}\boldsymbol{\alpha}+\tau\|\boldsymbol{\alpha}\|^2_{\lambda}= 1, \label{1095}
\end{align}
where $\|\boldsymbol{\alpha}\|^2_\lambda=(1-\lambda)\|\boldsymbol{\alpha}\|_2^2+\lambda\|\boldsymbol{\alpha}\|_1^2$ and both $\tau\ge 0$ and $0\le \lambda \le 1$ are tuning parameters.  The introduction of $\|\boldsymbol{\alpha}\|^2_2$ overcomes the issue that $\widehat{\boldsymbol{\Sigma}}$ is not full rank in high-dimensional setting, and the term $\|\boldsymbol{\alpha}\|^2_1$ encourages the sparsity of the solution. A difference between our penalty and the usual lasso or elastic-net penalty is that we use the squared $l_1$-norm. This particular form of our penalty leads to the property that $\widehat{\boldsymbol{\alpha}}_1$ is also the solution to 
\begin{align}
 \max_{\boldsymbol{\alpha}\in \mathbb{R}^p,  \boldsymbol{\alpha}\neq \mathbf{0}}\quad\frac{\boldsymbol{\alpha}\trans \widehat{\mathbf{B}}\boldsymbol{\alpha}}{\boldsymbol{\alpha}\trans \widehat{\boldsymbol{\Sigma}}\boldsymbol{\alpha}+\tau\|\boldsymbol{\alpha}\|^2_{\lambda}}, \label{10005} 
\end{align}
where the objective function is scale-invariant. That is, for any nonzero number $t$, the vector $t\widehat{\boldsymbol{\alpha}}_1$ is also a solution to $\eqref{10005}$.  This scale-invariant property is intensively used in our theoretical development. Once we obtain $\widehat{\boldsymbol{\alpha}}_1$, our classification rule is to assign a new observation $\mathbf{x}$ to  class $i$ if $(\mathbf{x}-\bar{\mathbf{x}}_i)\trans\widehat{\mathbf{D}}(\mathbf{x}-\bar{\mathbf{x}}_i)<(\mathbf{x}-\bar{\mathbf{x}}_j)\trans\widehat{\mathbf{D}}(\mathbf{x}-\bar{\mathbf{x}}_j)$ for $1\le j\neq i\le 2$, where $\widehat{\mathbf{D}}=\widehat{\boldsymbol{\alpha}}_1\widehat{\boldsymbol{\alpha}}_1\trans$.

\subsection{The case of $K>2$}
 If $K>2$, more than one components need to be estimated.    $\boldsymbol{\alpha}_1$ is estimated in the same way as  $K=2$. Since the higher order component $\boldsymbol{\alpha}_i$, $1<i\le K-1$, satisfies the  constraints in  $\eqref{2220}$, $\boldsymbol{\alpha}_i$ is actually orthogonal to the subspace spanned by $\{\boldsymbol{\Sigma}\boldsymbol{\alpha}_1,\cdots,\boldsymbol{\Sigma}\boldsymbol{\alpha}_{i-1}\}$ in $\mathbb{R}^p$. Because $\boldsymbol{\alpha}_i$ is the eigenvector of the generalized eigenvalue problem $\eqref{2220}$, for any $1\le j< K-1$,  $\mathbf{B}\boldsymbol{\alpha}_j$ and $\boldsymbol{\Sigma}\boldsymbol{\alpha}_j$ have the same directions and only differ by a scale factor, which is the $j$-th eigenvalue. Hence,   the subspace  spanned by $\{\mathbf{B}\boldsymbol{\alpha}_1,\cdots,\mathbf{B}\boldsymbol{\alpha}_{i-1}\}$ is the same as that of $\{\boldsymbol{\Sigma}\boldsymbol{\alpha}_1,\cdots,\boldsymbol{\Sigma}\boldsymbol{\alpha}_{i-1}\}$.  

Because in the high-dimensional settings, $\widehat{\boldsymbol{\Sigma}}$ and $\widehat{\mathbf{B}}$ are not  consistent estimates of $\boldsymbol{\Sigma}$ and $\mathbf{B}$ in terms of the operator norm, respectively,  neither of the subspaces spanned by $\{\widehat{\boldsymbol{\Sigma}}\widehat{\boldsymbol{\alpha}}_1,\cdots,\widehat{\boldsymbol{\Sigma}}\widehat{\boldsymbol{\alpha}}_{i-1}\}$ and $\{\widehat{\mathbf{B}}\widehat{\boldsymbol{\alpha}}_1,\cdots,\widehat{\mathbf{B}}\widehat{\boldsymbol{\alpha}}_{i-1}\}$ is a consistent estimate of the subspace spanned by $\{\boldsymbol{\Sigma}{\boldsymbol{\alpha}}_1,\cdots,\boldsymbol{\Sigma}{\boldsymbol{\alpha}}_{i-1}\}$ (or by  $\{\mathbf{B}{\boldsymbol{\alpha}}_1,\cdots,{\mathbf{B}}{\boldsymbol{\alpha}}_{i-1}\}$), even if $\widehat{\boldsymbol{\alpha}}_j$, $1\le j\le i-1$, are consistent estimates. Therefore, in order to estimate these subspaces, in addition to the sparsity assumption on $\{ \boldsymbol{\alpha}_1,\cdots, \boldsymbol{\alpha}_{K-1}\}$, we also make sparsity assumptions on the vectors, $\boldsymbol{\Sigma}\boldsymbol{\alpha}_1, \cdots, \boldsymbol{\Sigma}\boldsymbol{\alpha}_{K-1}$,  in terms of $l_1$ norm.  Lemma \ref{lemma_8} in Section \ref{sec.4} shows that making sparsity assumptions on  $\boldsymbol{\Sigma}\boldsymbol{\alpha}_1, \cdots, \boldsymbol{\Sigma}\boldsymbol{\alpha}_{K-1}$ is equivalent to or weaker than assuming the sparsity of $\{ \boldsymbol{\mu}_{i}-\boldsymbol{\mu}_{j}, 1\le i\neq j\le K\}$ in terms of $l_1$ norm. The latter assumption has been made in  \citet{Shao-2011}.  \citet{bickel2004some}  assumes that 
$\boldsymbol{\mu}_1$ and $\boldsymbol{\mu}_2$  are sparse when $K=2$, which implies that $\boldsymbol{\mu}_1-\boldsymbol{\mu}_2$ is sparse.

 Under the above assumptions, suppose that we have obtained the estimate $\widehat{\boldsymbol{\alpha}}_j$ of $\boldsymbol{\alpha}_j$, $1\le j\le i-1$, then we 
obtain the estimate $\widehat{\boldsymbol{\xi}}_j$ of $\mathbf{B}\boldsymbol{\alpha}_j$ as the solution to
	\begin{align}
 \min_{\boldsymbol{\xi}\in\mathbb{R}^p}\left[\|\boldsymbol{\xi}-\widehat{\mathbf{B}}\widehat{\boldsymbol{\alpha}}_j\|^2_2+\kappa\|\boldsymbol{\xi}\|_1\right], \label{1122}
\end{align}	
where	  $\kappa\ge 0$   is a tuning parameter. It can be shown that the $l$-th coordinate of $\widehat{\boldsymbol{\xi}}_j$ is
	\begin{align}
 (\widehat{\boldsymbol{\xi}}_j)_l=\text{\bf sign}((\widehat{\mathbf{B}}\widehat{\boldsymbol{\alpha}}_j)_l)\left[|(\widehat{\mathbf{B}}\widehat{\boldsymbol{\alpha}}_j)_l|-\kappa/2\right]\mathbf{I}_{[|(\widehat{\mathbf{B}}\widehat{\boldsymbol{\alpha}}_j)_l|\ge \kappa/2]},\quad 1\le l\le p, \label{1098}
\end{align}	
where $\mathbf{I}_{[|(\widehat{\mathbf{B}}\widehat{\boldsymbol{\alpha}}_j)_l|\ge \kappa/2]}$ is the indicator function of $[|(\widehat{\mathbf{B}}\widehat{\boldsymbol{\alpha}}_j)_l|\ge \kappa/2]$. So we actually estimate $\mathbf{B}\boldsymbol{\alpha}_j$ by applying the soft-thresholding to $\widehat{\mathbf{B}}\widehat{\boldsymbol{\alpha}}_j$. 
We will show that the subspace spanned by $\{\widehat{\boldsymbol{\xi}}_1$, $\cdots$, $\widehat{\boldsymbol{\xi}}_{i-1}\}$ is a consistent estimate of the subspace spanned by $\{\mathbf{B}\boldsymbol{\alpha}_1,\cdots,\mathbf{B}\boldsymbol{\alpha}_{i-1}\}$ and provide the convergence rate in Section 4. An alternative way  to obtain a consistent estimate of the subspace is to apply the soft-threholding  to $\widehat{\boldsymbol{\Sigma}}\widehat{\boldsymbol{\alpha}}_1,\cdots,\widehat{\boldsymbol{\Sigma}}\widehat{\boldsymbol{\alpha}}_{i-1}$. However, it turns out that the real predictive performance of this alternative is inferior to the proposed, so we do not consider it in this paper.  Now suppose that we have obtained the estimates $\widehat{\boldsymbol{\alpha}}_1$, $\cdots$, $\widehat{\boldsymbol{\alpha}}_{i-1}$ and $\widehat{\boldsymbol{\xi}}_1$, $\cdots$, $\widehat{\boldsymbol{\xi}}_{i-1}$, then $\widehat{\boldsymbol{\alpha}}_i$ is the solution to
\begin{align}
 \max_{\boldsymbol{\alpha}\in\mathbb{R}^p}\quad\boldsymbol{\alpha}\trans \widehat{\mathbf{B}}\boldsymbol{\alpha}, \quad\text{ subject to }\quad \boldsymbol{\alpha}\trans \widehat{\boldsymbol{\Sigma}}\boldsymbol{\alpha}+\tau\|\boldsymbol{\alpha}\|^2_{\lambda}= 1,\quad \boldsymbol{\alpha}\trans  \widehat{\boldsymbol{\xi}}_j=0,\quad j<i.\label{1003}
\end{align}
  The optimization problems  $\eqref{1095}$  and $\eqref{1003}$ are both special cases of the following general problem:
\begin{align}
 \max_{\boldsymbol{\alpha}\in\mathbb{R}^p}\quad\boldsymbol{\alpha}\trans \boldsymbol{\Pi}\boldsymbol{\alpha}, \quad\text{ subject to }\quad \boldsymbol{\alpha}\trans \mathbf{C}\boldsymbol{\alpha}+\tau\|\boldsymbol{\alpha}\|^2_\lambda\le 1,\quad  \mathbf{L}\boldsymbol{\alpha}=0,\label{3100}
\end{align}
where $\boldsymbol{\Pi}$ and $\mathbf{C}$ are any two $p\times p$ nonnegative definite symmetric matrices, and $\mathbf{L}$ is either equal to zero or any matrix with $p$ columns.  $\mathbf{L}\boldsymbol{\alpha}=0$ can be viewed as linear constraints imposed on $\boldsymbol{\alpha}$. For example, $\eqref{1003}$ is the special case of $\eqref{3100}$ with $\boldsymbol{\Pi}=\widehat{\mathbf{B}}$, $\mathbf{C}=\widehat{\boldsymbol{\Sigma}}$ and  $ \mathbf{L}=(\widehat{\boldsymbol{\xi}}_1, \cdots,\widehat{\boldsymbol{\xi}}_{i-1})\trans $.  In \citet{Qi-2013}, we solve $\eqref{3100}$ by the following algorithm.

\begin{Algorithm}\label{algorithm_12} \quad
  \begin{itemize}
 \item[1.]Choose an initial vector $\boldsymbol{\alpha}^{(0)}$ with $\boldsymbol{\Pi}\boldsymbol{\alpha}^{(0)}\neq \mathbf{0}$.
 \item[2.]Iteratively compute a sequence $\boldsymbol{\alpha}^{(1)}, \boldsymbol{\alpha}^{(2)},\cdots, \boldsymbol{\alpha}^{(i)}, \cdots$ until convergence as follows: for any $i\ge 1$, compute $\boldsymbol{\alpha}^{(i)}$ by solving
\begin{align}
& \max_{\boldsymbol{\alpha}\in \mathbb{R}^p} (\boldsymbol{\Pi}\boldsymbol{\alpha}^{(i-1)})\trans \boldsymbol{\alpha}, \quad\text{ subject to }\quad \boldsymbol{\alpha}\trans \mathbf{C}\boldsymbol{\alpha}+\tau\|\boldsymbol{\alpha}\|^2_\lambda\le 1,\quad  \mathbf{L}\boldsymbol{\alpha}=\mathbf{0}.\label{2241}
\end{align}
  \end{itemize}
\end{Algorithm}

The key step $\eqref{2241}$ of Algorithm $\ref{algorithm_12}$ is a special case of the following problem with $\mathbf{c}=\boldsymbol{\Pi}\boldsymbol{\alpha}^{(i-1)}$:
\begin{align}
&\qquad \max_{\boldsymbol{\alpha}}\mathbf{c}\trans \boldsymbol{\alpha},\quad \text{ subject to }\quad \boldsymbol{\alpha}\trans \mathbf{C}\boldsymbol{\alpha}+\tau\|\boldsymbol{\alpha}\|^2_\lambda\le 1,\quad  \mathbf{L}\boldsymbol{\alpha}=0,\label{2228}
\end{align}
where $\mathbf{c}$ is any nonzero vector. The algorithm and the related theory to solve $\eqref{2228}$ have been developed and described in details in  \citet{Qi-2013}.

 Once we obtain all the estimates $\widehat{\boldsymbol{\alpha}}_1$, $\cdots$, $\widehat{\boldsymbol{\alpha}}_{K-1}$, we build the classification rule which assigns a new observation $\mathbf{x}$ to class $i$ if 
\begin{align}
(\mathbf{x}-\bar{\mathbf{x}}_i)\trans\widehat{\mathbf{D}}(\mathbf{x}-\bar{\mathbf{x}}_i)<(\mathbf{x}-\bar{\mathbf{x}}_j)\trans\widehat{\mathbf{D}}(\mathbf{x}-\bar{\mathbf{x}}_j), \label{1174}
\end{align}
for all $1\le j\neq i\le K$, where 
\begin{align}
 \widehat{\mathbf{D}}=\left(\widehat{\boldsymbol{\alpha}}_1,\cdots,\widehat{\boldsymbol{\alpha}}_{K-1}\right)  \widehat{\mathbf{K}}^{-1}\left(\widehat{\boldsymbol{\alpha}}_1,\cdots,\widehat{\boldsymbol{\alpha}}_{K-1}\right)\trans, \label{1210}
\end{align}
and $\widehat{\mathbf{K}}$ is a symmetric $(K-1)\times (K-1)$ matrix with the $(i,j)$-th entry equal to $\widehat{\boldsymbol{\alpha}}_i\trans \widehat{\boldsymbol{\Sigma}}\widehat{\boldsymbol{\alpha}}_j$. This choice of $\widehat{\mathbf{D}}$ allows us to achieve a better convergence rate than $\widetilde{\mathbf{D}}$ used in the classic Fisher’s discriminant analysis rule $\eqref{classicrule.eq}$.
 
In \citet{Qi-2013}, we proposed to estimate $\boldsymbol{\alpha}_i$ by solving
\begin{align}
 &\max_{\boldsymbol{\alpha}\in\mathbb{R}^p}\quad\boldsymbol{\alpha}\trans \widehat{\mathbf{B}}\boldsymbol{\alpha},\quad  \text{ subject to }\quad \boldsymbol{\alpha}\trans \widehat{\boldsymbol{\Sigma}}\boldsymbol{\alpha}+\tau\|\boldsymbol{\alpha}\|^2_{\lambda}= 1,\quad \boldsymbol{\alpha}\trans  \widehat{\mathbf{B}}\widehat{\boldsymbol{\alpha}}_j=0,\quad j<i,\label{sfda.eq}
\end{align}
where we used the unthresholded vector $\widehat{\mathbf{B}}\widehat{\boldsymbol{\alpha}}_j$ in the linear constraints. That method has a good empirical performance, but we cannot provide the theoretical results due to the difficulty  mentioned at the beginning of Section \ref{sect.3}.

\section{Asymptotic consistency and asymptotic optimality}\label{sec.4}
   In this section, we will provide the asymptotic results of the method described in Section \ref{sect.3}.  We first  consider two mechanisms of class label generation. The first is a random mechanism in which sample observations are randomly drawn from any of $K$ classes with equal probability $1/K$. Hence, $(n_1,n_2,\cdots,n_K)$ follows a multinomial distribution with parameters $n$ and $(1/K,\cdots,1/K)$. In this case, we have the following result.

\begin{Lemma}\label{lemma_1}
 Suppose that $(n_1,n_2,\cdots,n_K)$ follows a multinomial distribution with parameters $n$ and $(1/K,\cdots,1/K)$. Given any $(K,n,p)$ satisfying that $p\ge 2$, $K\le p+1$ and $\sqrt{K\log{p}/n}$ is bounded by some constant $d_0$,  for any $M>0$,  we have 
\begin{align}
 P\left( \max_{1\le i\le K} \left|\frac{n_i}{n}-\frac{1}{K}\right|>C\sqrt{\frac{\log{p}}{Kn}}\right) \le p^{-M}\label{1008}
\end{align}
for any $C\ge (M+3)(d_0+1)$.
\end{Lemma}

The second mechanism is nonrandom, that is, $(n_1,n_2,\cdots,n_K)$ are nonrandom numbers. In this case, we will impose the following Condition \ref{condition_1} (a) on these numbers.  

\begin{Condition}\label{condition_1}\quad
\begin{itemize}
\item[(a).] If $(n_1,n_2,\cdots,n_K)$ are nonrandom, then there exists a constant $C_0$ (independent of $n$, $p$ and $K$), such that we have $\max_{1\le i\le K} \left|n_i/n-1/K\right|\le C_0\sqrt{\log{p}/(Kn)}$ for all large enough $n$. 
\item[ (b).] There exists a constant $c_0>0$ (independent of $n$, $p$ and $K$) such that
\begin{align*}
c_0^{-1}\le  \lambda_{min}(\boldsymbol{\Sigma})\le\lambda_{max}(\boldsymbol{\Sigma})\le c_0\quad \text{ and  }\quad \max_{1\le i\le  K}\|\boldsymbol{\mu}_i\|_{\infty}\le c_0.
\end{align*}
\end{itemize}
\end{Condition}

 Lemma \ref{lemma_1} and Condition \ref{condition_1} (a) ensure that the number of observations in different classes do not differ greatly in each of the two mechanisms. The regularity condition for $\boldsymbol{\Sigma}$ in Condition \ref{condition_1} (b) has been used in \citet{Shao-2011} and \citet{cai2011direct}. The condition about $\boldsymbol{\mu}_i$ can be achieved by scaling each of the $p$ variables.   Under Condition \ref{condition_1}, we have the following two probability inequalities about $\|\widehat{\boldsymbol{\Sigma}}-\boldsymbol{\Sigma}\|_{\infty}$ and $\|\widehat{\mathbf{B}}-\mathbf{B}\|_{\infty}$, which play basic roles in our theoretical development.

\begin{Theorem}\label{theorem_11}
Suppose that Condition \ref{condition_1} holds, $p\ge 2$, $K\le p+1$ and $K\log{p}/n\to 0$ as $n\to \infty$. Then for any $M>0$, we can find $C$ large enough and independent of $n$, $p$ and $K$ such that 
\begin{align*}
&P\left(\|\widehat{\boldsymbol{\Sigma}}-\boldsymbol{\Sigma}\|_{\infty}>C\sqrt{\frac{K\log{p}}{n}}\right)\le  p^{-M},\quad P\left(\|\widehat{\mathbf{B}}-\mathbf{B}\|_{\infty}>C\sqrt{\frac{K\log{p}}{n}}\right)\le  p^{-M}
\end{align*}
for all large enough $n$.
\end{Theorem}
\begin{Remark}\label{remark_13}\quad
 Theorem \ref{theorem_11} holds even if $K\to\infty$ as $n\to \infty$. However, since we need the condition that $K$ is bounded in the following theorems, we fix $K$ in this paper. 
 \end{Remark}

 Define a $p\times p$ nonnegative definite matrix 
\begin{align}
 \boldsymbol{\Xi}=\boldsymbol{\Sigma}^{-1/2}\mathbf{B}\boldsymbol{\Sigma}^{-1/2}\;.\label{2}
\end{align}
 Solving the generalized eigenvalue problem $\eqref{2220}$ is equivalent to computing the eigenvalues and eigenvectors of $\boldsymbol{\Xi}$. In fact, because $\boldsymbol{\alpha}_k$, $1\le k\le K-1$, are the  generalized eigenvectors of the problem $\eqref{2220}$, we have 
\begin{align}
\mathbf{B}\boldsymbol{\alpha}_k= \nu_k\boldsymbol{\Sigma}\boldsymbol{\alpha}_k,\quad \text{ and hence,}\quad  \boldsymbol{\Xi}\boldsymbol{\Sigma}^{1/2}\boldsymbol{\alpha}_k= \boldsymbol{\Sigma}^{-1/2}\mathbf{B} \boldsymbol{\alpha}_k=\nu_k\boldsymbol{\Sigma}^{1/2}\boldsymbol{\alpha}_k, \label{10003}
\end{align}
for any $1\le k\le K-1$, where $\nu_k$ is the corresponding generalized eigenvalue. Therefore,  
\begin{align}
\boldsymbol{\gamma}_1=\boldsymbol{\Sigma}^{1/2}\boldsymbol{\alpha}_1,\quad   \boldsymbol{\gamma}_2=\boldsymbol{\Sigma}^{1/2}\boldsymbol{\alpha}_2,\quad\cdots,\quad \boldsymbol{\gamma}_{K-1}=\boldsymbol{\Sigma}^{1/2}\boldsymbol{\alpha}_{K-1}, \label{10023}
\end{align}
  are the eigenvectors of $\boldsymbol{\Xi}$ with corresponding eigenvalues $\nu_1,\nu_2,\ldots,\nu_{K-1}$, respectively. So they are orthogonal to each other. In the following, we will use $\lambda_k(\boldsymbol{\Xi})$, $1\le k\le K-1$, to denote the eigenvalues of $\boldsymbol{\Xi}$, which are just the above generalized eigenvalues and also equal to the maximum values of the optimization problems $\eqref{2220}$. Since  $\boldsymbol{\Xi}$ has the same rank as $\mathbf{B}$ which is not greater than  $K-1$ due to the constraint $\eqref{3}$, $\boldsymbol{\Xi}$ has at most $K-1$ positive eigenvalues. By the conditions $\boldsymbol{\alpha}_k\trans \boldsymbol{\Sigma}\boldsymbol{\alpha}_k=1$, $1\le k\le K-1$, we have $\|\boldsymbol{\gamma}_1\|_2=\|\boldsymbol{\gamma}_2\|_2=\cdots=\|\boldsymbol{\gamma}_{K-1}\|_2=1$.  Let 
	\begin{align}
\widehat{\boldsymbol{\gamma}}_1=\boldsymbol{\Sigma}^{1/2}\widehat{\boldsymbol{\alpha}}_1, \quad \widehat{\boldsymbol{\gamma}}_2=\boldsymbol{\Sigma}^{1/2}\widehat{\boldsymbol{\alpha}}_2,  \cdots   , \widehat{\boldsymbol{\gamma}}_{K-1}=\boldsymbol{\Sigma}^{1/2}\widehat{\boldsymbol{\alpha}}_{K-1}, \label{10024}
\end{align}
	 which are estimates of  $\boldsymbol{\gamma}_1$, $\cdots$, $\boldsymbol{\gamma}_{K-1}$, respectively. Since $-\widehat{\boldsymbol{\alpha}}_k$ is also the solution to the optimization problem in $\eqref{1095}$ or $\eqref{1003}$, without loss of generality, we choose the sign of $\widehat{\boldsymbol{\alpha}}_k$ such that $\widehat{\boldsymbol{\gamma}}_k\trans\boldsymbol{\gamma}_k\ge 0$, for $1\le k\le K-1$. We impose the following regularity conditions on the eigenvalues of $\boldsymbol{\Xi}$. 
	
\begin{Condition}\label{condition_2}\quad
There exist positive constants $c_1$, $c_2$ and $c_3$ which are all independent of $n$, $p$ and $K$ such that 
\begin{itemize}
\item[(a).] $\lambda_1(\boldsymbol{\Xi})\ge \lambda_{2}(\boldsymbol{\Xi})\ge\cdots\ge \lambda_{K-1}(\boldsymbol{\Xi})\ge c_1$,
\item[(b).]
$ \min\left\{\frac{\lambda_{1}(\boldsymbol{\Xi})-\lambda_{2}(\boldsymbol{\Xi})}{\lambda_{1}(\boldsymbol{\Xi})}, \quad\frac{\lambda_{2}(\boldsymbol{\Xi})-\lambda_{3}(\boldsymbol{\Xi})}{\lambda_{2}(\boldsymbol{\Xi})},\cdots, \frac{\lambda_{K-2}(\boldsymbol{\Xi})-\lambda_{K-1}(\boldsymbol{\Xi})}{\lambda_{K-2}(\boldsymbol{\Xi})} \right\}\ge c_2$,
\item[(c).] The ratio between the largest and the smallest eigenvalue satisfies $\lambda_{1}(\boldsymbol{\Xi})/\lambda_{K-1}(\boldsymbol{\Xi}) \le c_3.$
\end{itemize}
\end{Condition}
In the case of $K=2$, we will show in  Remark \ref{remark_11} (3) that $\lambda_1(\boldsymbol{\Xi})$ has the same order as $\|\boldsymbol{\mu}_2-\boldsymbol{\mu}_1\|_2^2$. Therefore, roughly speaking, Condition 2 (a) implies that the class means are not too close to each other.  Condition 2 (b) prevents the cases that the spacing between adjacent eigenvalues is too small. Condition 2 (c) excludes the situations where the effects of higher order components are dominated by those of lower order components and are negligible asymptotically.  

Now we consider the choice of the tuning parameters, $\tau$ and $\lambda$, in the penalized optimization problems $\eqref{1095}$ and $\eqref{1003}$. We will show that the choice of $\lambda$ is not essential for the asymptotic convergence rates as long as it is asymptotically bounded away from zero. In the following theorems, we will choose tuning parameters $(\tau_n,\lambda_n)$, which depend on the sample size $n$, satisfying
 \begin{align}
 0< \lambda_n< 1, \quad \liminf_{n\to\infty}\lambda_n>\lambda_0, \quad \tau_n=Cs_n,\quad \text{where} \quad s_n=\sqrt{\frac{K\log{p}}{n}}, \label{1270} 
\end{align}
$\lambda_0>0$ and $C$ are constants independent of $n$, $p$ and $K$. The constant $C$ is chosen based on Theorem  \ref{theorem_11} such that for all large enough $n$,
{\begin{align}
P\left(\|\widehat{\boldsymbol{\Sigma}}-\boldsymbol{\Sigma}\|_{\infty}>\frac{C}{C_2}s_n\right)\le  p^{-1},\quad P\left(\|\widehat{\mathbf{B}}-\mathbf{B}\|_{\infty}>\frac{C}{C_2}s_n\right)\le  p^{-1},\label{1271}
\end{align}}
 where $C_2= 2(1+c_1^{-1})/\lambda_0$ and $c_1$ is the constant in Condition \ref{condition_2} (a). Define the event  
\begin{align}
&\Omega_n=\left\{\|\widehat{\boldsymbol{\Sigma}}-\boldsymbol{\Sigma}\|_{\infty}\le \tau_n/C_2 ,\quad  \|\widehat{\mathbf{B}}-\mathbf{B}\|_{\infty}\le \tau_n/C_2\right\},\label{1033}\\
& \text{ then by $\eqref{1271}$  }  P\left(\Omega_n\right)\ge 1-2p^{-1}\;. \notag
\end{align}
We mainly consider the elements in $\Omega_n$ in proofs.

We adopt the same definition of asymptotic optimality for a linear classification rule as in  \citet{Shao-2011}, \citet{cai2011direct}, \citet{fan2012road} and other papers. Let $T_{OPT}$ denote the optimal linear classification rule $\eqref{1170}$ or $\eqref{optrule.eq}$ and $R_{OPT}$ represent its misclassification error rate. Let $T$ be any linear classification rule  based on $\mathbf{X}$. The conditional misclassification rate of $T$ given $\mathbf{X}$ is defined as
\begin{align*}
&R_{T}(\mathbf{X})= \sum_{i=1}^KP \left(\left\{\text{$\mathbf{x}_{\rm new}$ belongs to the $i$-th class but $T(\mathbf{x}_{\rm new})\neq i$}\right\}\bigg\vert\mathbf{X} \right)
\end{align*}  
where $\mathbf{x}_{\rm new}$ is a new observation independent of $\mathbf{X}$. Therefore,  $R_{T}(\mathbf{X})$ is  a function of $\mathbf{X}$.

 \begin{Definition}\label{definition_1}
\renewcommand{\baselinestretch}{1.5}
 Let $T$ be a linear classification rule with conditional misclassification rate $R_T(\mathbf{X})$. Then $T$ is asymptotically optimal if 
\begin{align}
&\frac{R_{T}(\mathbf{X})}{R_{{OPT}}}-1= o_p(1).\label{112}
\end{align}
\end{Definition}

Since $0\le R_{{OPT}}\le R_{T}(\mathbf{X})\le 1$ for any $\mathbf{X}$, $\eqref{112}$ implies that $ 0\le R_{T}(\mathbf{X})-R_{{OPT}}= o_p(1)$. Hence we have $R_{T}(\mathbf{X})\to R_{{OPT}}$ in probability and $E[R_{T}(\mathbf{X})]\to R_{{OPT}}$, which have been used to define the consistency of a classification rule by  \citet{devroye1996probabilistic} and others. If $R_{{OPT}}$ is bounded away from 0, then $  R_{T}(\mathbf{X})-R_{{OPT}}= o_p(1)$ also implies $\eqref{112}$. However, if $R_{{OPT}}\to 0$, $\eqref{112}$ is stronger than  $  R_{T}(\mathbf{X})-R_{{OPT}}=o_p(1)$.  

In the following, we will consider the asymptotic properties of our method and  assume that $K$ is fixed, $p\to \infty$ and $s_n=\sqrt{K\log{p}/n}\to 0$ as $n\to \infty$. The following theorem provides an upper bound for the $l_1$ sparsity and the consistency of the estimator $\widehat{\boldsymbol{\alpha}}_1$ obtained from $\eqref{1095}$. 

\begin{Theorem}\label{theorem_12}
Suppose that   Conditions \ref{condition_1} and \ref{condition_2} hold. If  $\|\boldsymbol{\alpha}_1\|^2_1s_n\to 0$ as $n, p\to\infty$, then for all large enough $n$, we have, in $\Omega_n$, 
\begin{align}
  &\|\widehat{\boldsymbol{\alpha}}_1\|^2_1\le 6\|\boldsymbol{\alpha}_1\|^2_1/\lambda_0, \quad \|\widehat{\boldsymbol{\gamma}}_1-\boldsymbol{\gamma}_1\|_2^2 \le  C_5\|\boldsymbol{\alpha}_1\|^2_1s_n, \label{1036}\\
	& \|\widehat{\boldsymbol{\alpha}}_1-\boldsymbol{\alpha}_1\|_2^2 \le  c_0C_5\|\boldsymbol{\alpha}_1\|^2_1s_n, \notag
\end{align}
where $C_5$ is a constant independent of $n$ and $p$, $\lambda_0$ is the constant in $\eqref{1270}$, and $c_0$ is the constant in Condition 1 (b). Therefore, $\widehat{\boldsymbol{\alpha}}_1$ is a consistent estimate of $\boldsymbol{\alpha}_1$.
\end{Theorem}
 By Theorem \ref{theorem_12}, the estimate $\widehat{\boldsymbol{\alpha}}_1$ has the same order of $l_1$ sparsity as $\boldsymbol{\alpha}_1$ and  in order that $\widehat{\boldsymbol{\alpha}}_1$ is consistent, we  need $\|\boldsymbol{\alpha}_1\|^2_1$ is $o(\sqrt{n/\log{p}})$. In the following, we will consider the cases of $K=2$ and $K>2$, separately.

\subsection{The case of $K=2$}
When $K=2$, there exists only one component $\boldsymbol{\alpha}_1$ and $\boldsymbol{\Xi}$ has one postive eigenvalue $\lambda_1(\boldsymbol{\Xi})$. Therefore,  Conditions \ref{condition_2} (b)-(c) are not necessary. We provide explicit formulas for the misclassification errors of the optimal rule with $\mathbf{D}=\boldsymbol{\alpha}_1\boldsymbol{\alpha}_1\trans$ and our rule with $\widehat{\mathbf{D}}=\widehat{\boldsymbol{\alpha}}_1\widehat{\boldsymbol{\alpha}}_1\trans$, and prove the asymptotic optimality of our method in the following theorem.

\begin{Theorem}\label{theorem_14}
Suppose that $K=2$ and Conditions \ref{condition_1} and \ref{condition_2} (a) hold. Then the misclassification rate of the optimal rule $\eqref{1170}$ and the conditional misclassification rate of our sparse LDA rule  in Section \ref{sect.3.1} are given by
\begin{align}
  &R_{OPT}=\Phi\left(-\frac{\boldsymbol{\delta}\trans\mathbf{D}\boldsymbol{\delta}}{2\|\boldsymbol{\delta}\trans\mathbf{D}\boldsymbol{\Sigma}^{1/2}\|_2}\right),\label{1176}\\
	&R(\mathbf{X})=\frac{1}{2}\Phi\left(-\frac{\widehat{\boldsymbol{\delta}}\trans\widehat{\mathbf{D}}(2 \boldsymbol{\mu}_2-\bar{\mathbf{x}}_1-\bar{\mathbf{x}}_2)}{2\|\widehat{\boldsymbol{\delta}}\trans\widehat{\mathbf{D}}\boldsymbol{\Sigma}^{1/2}\|_2}\right)+\frac{1}{2}\Phi\left(-\frac{\widehat{\boldsymbol{\delta}}\trans\widehat{\mathbf{D}}(\bar{\mathbf{x}}_1+\bar{\mathbf{x}}_2-2 \boldsymbol{\mu}_1)}{2\|\widehat{\boldsymbol{\delta}}\trans\widehat{\mathbf{D}}\boldsymbol{\Sigma}^{1/2}\|_2}\right),\notag
\end{align}
respectively, where $\Phi$ is the cumulative distribution function of the standard normal distribution, $\boldsymbol{\delta}=\boldsymbol{\mu}_2-\boldsymbol{\mu}_1$ and $\widehat{\boldsymbol{\delta}}=\bar{\mathbf{x}}_2-\bar{\mathbf{x}}_1$. Moreover, if  $\lambda_1(\boldsymbol{\Xi})\|\boldsymbol{\alpha}_1\|^2_1s_n\to 0$ as $n, p\to\infty$, our method is asymptotically optimal and we have
\begin{align}
  &\frac{R(\mathbf{X})}{R_{OPT}}-1= O_p\left(\lambda_1(\boldsymbol{\Xi})\|\boldsymbol{\alpha}_1\|^2_1s_n\right).\label{1193}
\end{align}
\end{Theorem}

 \begin{Remark}\label{remark_11}\quad  
 \begin{itemize}
\item[(1).] The misclassification rate of the optimal rule is expressed as $R_{OPT}=\Phi\left(-\sqrt{\boldsymbol{\delta}\trans\boldsymbol{\Sigma}^{-1}\boldsymbol{\delta}}/2\right)$ in Equation (1) in \citet{Shao-2011} and Equation (5) in \citet{cai2011direct}. Since by Lemma \ref{lemma_3} (Supplementary Material), $\boldsymbol{\Sigma}^{-1}\boldsymbol{\delta}=\mathbf{D}\boldsymbol{\delta}$,  the $R_{OPT}$ in $\eqref{1176}$ is the same as in those papers. 
\item[(2).] Under the $l_1$ sparsity on $\boldsymbol{\Sigma}^{-1}\boldsymbol{\delta}$,
\citet{cai2011direct} obtained the convergence rate
\begin{align}
  &\frac{R(\mathbf{X})}{R_{OPT}}-1= O_p\left\{\left(\|\boldsymbol{\Sigma}^{-1}\boldsymbol{\delta}\|_1\sqrt{\Delta_p}+\|\boldsymbol{\Sigma}^{-1}\boldsymbol{\delta}\|_1^2\right)\sqrt{\frac{\log{p}}{n}}\right\},\label{1194}
\end{align}
 in their Theorem 3, where $\Delta_p=\boldsymbol{\delta}\trans\boldsymbol{\Sigma}^{-1}\boldsymbol{\delta}$. When $K=2$,  $\boldsymbol{\alpha}_1=\boldsymbol{\Sigma}^{-1}\boldsymbol{\delta}/\sqrt{\boldsymbol{\delta}\trans\boldsymbol{\Sigma}^{-1}\boldsymbol{\delta}}$. By $\eqref{1187}$ (Supplementary Material) in the proof of Theorem \ref{theorem_14}, we have $\boldsymbol{\delta}\trans\mathbf{D}\boldsymbol{\delta}=\boldsymbol{\delta}\trans\boldsymbol{\Sigma}^{-1}\boldsymbol{\delta}=4\lambda_1(\boldsymbol{\Xi})$. Hence, our convergence rate on the right hand side of $\eqref{1193}$ is
\begin{align*}
  O_p\left(\lambda_1(\boldsymbol{\Xi})\|\boldsymbol{\alpha}_1\|^2_1s_n\right)&=O_p\left\{(\boldsymbol{\delta}\trans\boldsymbol{\Sigma}^{-1}\boldsymbol{\delta})\left\|\frac{\boldsymbol{\Sigma}^{-1}\boldsymbol{\delta}}{\sqrt{\boldsymbol{\delta}\trans\boldsymbol{\Sigma}^{-1}\boldsymbol{\delta}}}\right\|_1^2\sqrt{\frac{K\log{p}}{n}}\right\}\\&=O_p\left( \|\boldsymbol{\Sigma}^{-1}\boldsymbol{\delta}\|_1^2 \sqrt{\frac{\log{p}}{n}}\right).
\end{align*}
Compared to the convergence rate in $\eqref{1194}$, our convergence rate does not have the first term in  $\eqref{1194}$.
\end{itemize}
\end{Remark}

\subsection{The case of $K>2$}
 We first illustrates the relationship between sparsity assumptions on  $\boldsymbol{\Sigma}\boldsymbol{\alpha}_1, \cdots, \boldsymbol{\Sigma}\boldsymbol{\alpha}_{K-1}$ and  $\{ \boldsymbol{\mu}_{i}-\boldsymbol{\mu}_{j}, 1\le i\neq j\le K\}$ in the following lemma. 
\begin{Lemma}\label{lemma_8}
Suppose that Conditions \ref{condition_1}-\ref{condition_2} hold. Then we have
 \begin{align*}
\frac{1}{(K-1)c_0\sqrt{2K\lambda_1(\boldsymbol{\Xi})}}\left(\max_{1\le i\neq j\le K}\|\boldsymbol{\mu}_i-\boldsymbol{\mu}_j\|_1\right)
\le \max_{1\le i\le K-1}\|\boldsymbol{\Sigma}\boldsymbol{\alpha}_i\|_1 \\
\le \frac{
\sqrt{c_3}}{\sqrt{\lambda_1(\boldsymbol{\Xi})}} \left(\max_{1\le i\neq j\le K}\|\boldsymbol{\mu}_i-\boldsymbol{\mu}_j\|_1\right).
\end{align*}
\end{Lemma}
 Since $\lambda_1(\boldsymbol{\Xi})\ge c_1$ by Condition \ref{condition_2} (a), Lemma \ref{lemma_8} implies that if $\lambda_1(\boldsymbol{\Xi})$ is bounded from the above, then $\max_{1\le i\le K-1}\|\boldsymbol{\Sigma}\boldsymbol{\alpha}_i\|_1$ has the same order as  $\max_{1\le i\neq j\le K}\|\boldsymbol{\mu}_i-\boldsymbol{\mu}_j\|_1$. If $\lambda_1(\boldsymbol{\Xi})\to \infty$, we have $\max_{1\le i\le K-1}\|\boldsymbol{\Sigma}\boldsymbol{\alpha}_i\|_1/\max_{1\le i\neq j\le K}\|\boldsymbol{\mu}_i-\boldsymbol{\mu}_j\|_1\to 0$. Therefore, making sparsity assumptions on  $\boldsymbol{\Sigma}\boldsymbol{\alpha}_1, \cdots, \boldsymbol{\Sigma}\boldsymbol{\alpha}_{K-1}$ is equivalent to or weaker than assuming the sparsity of $\{ \boldsymbol{\mu}_{i}-\boldsymbol{\mu}_{j}, 1\le i\neq j\le K\}$ in $l_1$ norm. 

 We define the following measurement of sparsity on $\boldsymbol{\alpha}_{i}$ and  $\boldsymbol{\Sigma}\boldsymbol{\alpha}_i$, $1\le i\le K-1$: 
 \begin{align}
\Lambda_p=\max_{ 1\le i\le K-1} \{\|\boldsymbol{\alpha}_{i}\|_1, \|\boldsymbol{\Sigma}\boldsymbol{\alpha}_i\|_1 \}. \label{1002}
\end{align}
  In the following theorem, we  show that for each $1\le i\le K-1$, the $l_1$ sparsity of the estimate $\widehat{\boldsymbol{\alpha}}_i$ is bounded by $\Lambda_p$ multiplied by a constant which does not depend on $n$ and $p$, and $\widehat{\boldsymbol{\alpha}}_i$ is a consistent estimate. Moreover, we show that the subspace spanned by $\{\widehat{\boldsymbol{\xi}}_1,\cdots,\widehat{\boldsymbol{\xi}}_i\}$ is a consistent estimate of the subspace spanned by $\{\mathbf{B}\boldsymbol{\alpha}_1,\cdots,\mathbf{B}\boldsymbol{\alpha}_{i}\}$ (or equivalently the subspace spanned by $\{\boldsymbol{\Sigma}\boldsymbol{\alpha}_1,\cdots,\boldsymbol{\Sigma}\boldsymbol{\alpha}_{i}\}$) and provide the convergence rates, where $\widehat{\boldsymbol{\xi}}_j$ is the solution to $\eqref{1122}$. In this paper, to measure whether two subspaces  with the same dimensions in $\mathbb{R}^p$ are close to each other, we use the operator norm of the difference between the orthogonal projection matrices onto the two subspaces. 

\begin{Theorem}\label{theorem_10}
Suppose that Conditions \ref{condition_1}-\ref{condition_2} hold. We choose the tuning parameter in the optimization problem $\eqref{1122}$ as $\kappa_n=\widetilde{C}\lambda_1(\boldsymbol{\Xi})\Lambda_ps_n$, where $\widetilde{C}$ is a constant large enough and independent of $n$ and $p$.  For any $1\le i\le K-1$, let $\mathbf{Q}_i$ and $\widehat{\mathbf{Q}}_i$ be the orthogonal projection matrices onto  the following subspaces of $\mathbb{R}^{p}$, respectively,  
\begin{align}
	\mathbf{W}_i={\rm span}\{\boldsymbol{\xi}_1,\boldsymbol{\xi}_2, \cdots, \boldsymbol{\xi}_i\},\quad \widehat{\mathbf{W}}_i={\rm span}\{\widehat{\boldsymbol{\xi}}_1,\widehat{\boldsymbol{\xi}}_2, \cdots, \widehat{\boldsymbol{\xi}}_i\},\label{10026}
\end{align}
where $\boldsymbol{\xi}_i=\mathbf{B}\boldsymbol{\alpha}_i=\lambda_i(\boldsymbol{\Xi})\boldsymbol{\Sigma}\boldsymbol{\alpha}_i$. If  $\Lambda_p^2s_n\to 0$ as $n, p\to\infty$,  then for each $1\le i\le K-1$, there exist constants $D_{i,1}$, $D_{i,2}$  and $D_{i,3}$ independent of $n$ and $p$ such that in $\Omega_n$,
 \begin{align}
  & \|\widehat{\boldsymbol{\alpha}}_i\|_1\le D_{i,1}\Lambda_p ,\quad \|\widehat{\boldsymbol{\alpha}}_i-\boldsymbol{\alpha}_i\|^2_2\le D_{i,2}\Lambda_p^2s_n, \quad \|\mathbf{Q}_i-\widehat{\mathbf{Q}}_i\|^2\le D_{i,3}\Lambda_p^2s_n.\label{1195}
\end{align} 
Hence, for each $1\le i\le K-1$, $\widehat{\boldsymbol{\alpha}}_i$ is a consistent estimate of $\boldsymbol{\alpha}_i$, and the projection matrix $\widehat{\mathbf{Q}}_i$ is a consistent estimate of $\mathbf{Q}_i$.
\end{Theorem}

Based on Theorem \ref{theorem_10}, we will prove the asymptotic optimality of our classification rule and provide the corresponding convergence rate. When $K>2$,  the classification boundary of a linear classification rule is typically complicated and no explicit formula for the error generally exist. In the following,  we first prove a theorem which provides the conditions for asymptotic optimality and the corresponding convergence rates for a large family of linear classification rules. Then by applying the general result to our method, we obtain the asymptotic optimality results.  

We consider a family of linear classification rules motivated by the following observation. The optimal classification rule $T_{OPT}$ can be rewritten in the following way. Let
\begin{align}
\mathbf{a}_{ji}=\boldsymbol{\Sigma}^{-1/2}(\boldsymbol{\mu}_j-\boldsymbol{\mu}_i), \quad \mathbf{b}_{ji}=\frac{1}{2}(\boldsymbol{\mu}_j+\boldsymbol{\mu}_i),\label{22}
\end{align}  
where $1\le i, j\le K$. Then $T_{OPT}$ assigns a new observation $\mathbf{x}$ to the $i$th class if $\mathbf{a}_{ji}\trans\boldsymbol{\Sigma}^{-1/2}(\mathbf{x}-\mathbf{b}_{ji})<0$ for all $j\neq i$. Based on this observation, we consider a family of classification rules having the form, 
\begin{align}
T \text{: to assign a new  $\mathbf{x}$ to the $i$th class if  } \widehat{\mathbf{a}}_{ji}\trans\boldsymbol{\Sigma}^{-1/2}(\mathbf{x}- \widehat{\mathbf{b}}_{ji})<0, \text{ for all } j\neq i, \label{23} 
\end{align}  
where $\widehat{\mathbf{a}}_{ji}$ and $\widehat{\mathbf{b}}_{ji}$ are  $p$-dimensional vectors which may depend on the sample $\mathbf{X}$, and  satisfy
\begin{align}
\widehat{\mathbf{a}}_{ji}=-\widehat{\mathbf{a}}_{ij},\quad \widehat{\mathbf{b}}_{ji}=\widehat{\mathbf{b}}_{ij},\label{24}
\end{align}
for all $1\le i\neq j\le K$.  Typically, $\widehat{\mathbf{a}}_{ji}$ and $\widehat{\mathbf{b}}_{ji}$  are estimates of $\mathbf{a}_{ji}$ and $\mathbf{b}_{ji}$, respectively. In addition to the optimal rule, many linear classification rules in practice belong to this family. For example,  the classic Fisher's rule $\eqref{classicrule.eq}$ is of the form $\eqref{23}$ with $\widehat{\mathbf{a}}_{ji}=\boldsymbol{\Sigma}^{1/2}\widetilde{\mathbf{D}}(\bar{\mathbf{x}}_j-\bar{\mathbf{x}}_i)$ and $\widehat{\mathbf{b}}_{ji}=\frac{1}{2}(\bar{\mathbf{x}}_j+\bar{\mathbf{x}}_i)$. The rule of our sparse Fisher's discriminant analysis method is also a special case of $\eqref{23}$ with
\begin{align}
\widehat{\mathbf{a}}_{ji}=\boldsymbol{\Sigma}^{1/2}\widehat{\mathbf{D}}(\bar{\mathbf{x}}_j-\bar{\mathbf{x}}_i),\quad \widehat{\mathbf{b}}_{ji}=\frac{1}{2}(\bar{\mathbf{x}}_j+\bar{\mathbf{x}}_i),\label{1196}
\end{align}  
where $\widehat{\mathbf{D}}$ is defined in $\eqref{1210}$.  Now we study the asymptotic optimality of a classification rule $T$ in this family. It is relatively easy to calculate the convergence rates of $\widehat{\mathbf{a}}_{ji}$ and $\widehat{\mathbf{b}}_{ji}$ in a given $T$.  We will establish the asymptotic optimality of $T$ and the convergence rate for $R_T(\mathbf{X})/R_{OPT}-1$ based on the convergence rates of $\widehat{\mathbf{a}}_{ji}$ and $\widehat{\mathbf{b}}_{ji}$, where $R_T(\mathbf{X})$ is the conditional misclassification rate of  $T$ given the training sample $\mathbf{X}$. 

\begin{Theorem}\label{theorem_7}
Suppose that Conditions \ref{condition_1} and \ref{condition_2} hold and the general classification rule $T$ in $\eqref{23}$ satisfies: $\widehat{\mathbf{a}}_{ji}=-\widehat{\mathbf{a}}_{ij}$ and $\widehat{\mathbf{b}}_{ji}=\widehat{\mathbf{b}}_{ij}$. Let $\{\delta_n: n\ge 1\}$ be a sequence of nonrandom positive numbers with $\delta_n\to 0$ and $\lambda_1(\boldsymbol{\Xi})\delta_n\to 0$ as $n\to \infty$. For any $1\le j\neq i\le K$, let 
\begin{align}
\mathbf{a}_{ji}=t_{ji}\widehat{\mathbf{a}}_{ji}+(\mathbf{a}_{ji})_\perp\label{100400}
\end{align}
be an orthogonal decomposition of $\mathbf{a}_{ji}$, where $t_{ji}\widehat{\mathbf{a}}_{ji}$ is the orthogonal projection of $\mathbf{a}_{ji}$ along the direction of $\widehat{\mathbf{a}}_{ji}$, $t_{ji}$ is a real number, and $(\mathbf{a}_{ji})_\perp$ is orthogonal to $t_{ji}\widehat{\mathbf{a}}_{ji}$.  Let
\begin{align}
&  \widehat{d}_{ji}=\widehat{\mathbf{a}}_{ji}\trans\boldsymbol{\Sigma}^{-1/2}(\widehat{\mathbf{b}}_{ji}-\boldsymbol{\mu}_i)\;, \quad d_{ji}=\mathbf{a}_{ji}\trans\boldsymbol{\Sigma}^{-1/2}(\mathbf{b}_{ji}-\boldsymbol{\mu}_i)=\frac{1}{2}\|\mathbf{a}_{ji}\|_2^2.\label{10040}
\end{align}  
If the following conditions are satisfied,
 \begin{align}
&\|\mathbf{a}_{ji}\|^2_2-\|\widehat{\mathbf{a}}_{ji}\|_2^2=\|\mathbf{a}_{ji}\|_2^2O_p(\delta_n),\quad t_{ji}=1+O_p(\delta_n), \label{104}\\
&  d_{ji}-\widehat{d}_{ji}
=\|\widehat{\mathbf{a}}_{ji}\|_2^2O_p(\delta_n),\notag
\end{align} 
then the classification rule $T$ is asymptotically optimal and we have 
\begin{align}
&\frac{R_{T}(\mathbf{X})}{R_{{OPT}}}-1 = O_p\left(\sqrt{\lambda_1(\boldsymbol{\Xi})\delta_n\log{\left[\{\lambda_1(\boldsymbol{\Xi})\delta_n\}^{-1}\right]}} \right).\label{105}
\end{align}
\end{Theorem}

To apply  Theorem \ref{theorem_7} to a specific linear classification rule with the form $\eqref{23}$, we need to determine the sequence $\delta_n$ and verify
 the conditions $\eqref{104}$.  For our classification rule $\eqref{1174}$, which is a special case of   $\eqref{23}$ with $\widehat{\mathbf{a}}_{ji}$ and $\widehat{\mathbf{b}}_{ji}$ as given in $\eqref{1196}$, it turns out that we can choose $\delta_n=\Lambda_p^2s_n$ which is the convergence rate in Theorem \ref{theorem_10}.

\begin{Theorem}\label{theorem_8}
Suppose that Conditions \ref{condition_1} and \ref{condition_2} hold,    and $\lambda_1(\boldsymbol{\Xi})\Lambda_p^2s_n \to 0$ as $n,p \to\infty$. Then our classification rule $\eqref{1174}$  is asymptotically optimal.  Moreover, we have 
\begin{align}
&\frac{R_{T}(\mathbf{X})}{R_{{OPT}}}-1= O_p\left(\sqrt{\lambda_1(\boldsymbol{\Xi})\Lambda_p^2s_n\log{\left[\{\lambda_1(\boldsymbol{\Xi})\Lambda_p^2s_n\}^{-1}\right]}} \right).\label{2336}
\end{align}
\end{Theorem}
 
Comparing Theorem \ref{theorem_8} with Theorem \ref{theorem_14}, we find that the convergence rate in $\eqref{2336}$ is slower than  that for $K=2$. This may be due to the complicated classification boundary when $K>2$. It is a future direction to investigate whether the convergence rates  in Theorems \ref{theorem_7} and \ref{theorem_8} can be improved.

\section{Simulation studies}\label{sec.5}
In the previous section, we have shown that the revised sparse Fisher's discriminant analysis method with soft thresholding (SFDA-threshold) has good theoretical properties. In this  and the following section, we will show that SFDA-threshold  also has good predictive performance as the original method  (SFDA) in \citet{Qi-2013} by comparing  them with regularized discriminant analysis (RDA) (\citet{Guo-2007}, R package ``{\sf rda}'') and penalized discriminant analysis (PDA) (\citet{Witten-2011}, R package ``{\sf penalizedLDA}'') through simulation studies and applications to real data sets.  

Three simulation models are considered. In each simulation, 50 independent data sets are simulated each of which has 1500 observations and three classes. In each dataset, for each observation, we randomly select a class label and then generate the value of $\mathbf{x}$ based on the distribution of that class.   Then the 1500 observations in each dataset are randomly split into the training set with 150 observations and the test set with 1350 observations. There are 500 features ($p=500$) in these datasets. For our methods, SFDA-threshold and SFDA, we use the usual cross-validation procedure to select tuning parameters $\tau$ from $\{0.5,1,5,10\}$, and  $\lambda$ from $\{0.01, 0.05, 0.1, 0.2, 0.3, 0.4\}$. For SFDA-threshold, we choose $\kappa$ in $\eqref{1122}$ from the three values which are equal to $\|\hat{\mathbf{B}}\hat{\alpha}_j\|_1$ multiplied by 0, 0.001 and 0.01, respectively.  For RDA and PDA, the default cross-validation procedure in the corresponding packages are used.  The details of the three simulation studies are provided below.
\begin{itemize}
\item[(a).]{\it Simulation 1:} There is no overlap between the features for different classes. There are correlations among some feature variables. Specifically, let $x_{ij}$ be the $i\th$  observation on the $j\th$  variable, $1\le j\le 500$ and $1\le i\le 1500$. If the $i\th$  observation is in class $k(=1,2,3)$, then $x_{ij}=\mu_{kj}+Z_{i}+\epsilon_{ij}$ if $1\le j\le 30$, and $x_{ij}=\mu_{kj}+\epsilon_{ij}$ if $ j\ge 31$, where $Z_{i}\sim \Normal(0,1)$ and $\epsilon_{ij}\sim \Normal(0, \sigma^2)$ are independent. Here $\mu_{1j}\sim \Normal(1,0.8^2)$ if $1\le j\le 20$, $\mu_{2j}\sim \Normal(4,0.8^2)$ if $21\le j\le 30$, $\mu_{3j}\sim \Normal(1,0.8^2)$ if $31\le j\le 50$ and $\mu_{kj}=0$ otherwise. We consider the cases that $\sigma^2=1$, $1.5^2$ and $4$, respectively.
\item[(b).]{\it Simulation 2:} There are overlaps between the features for different classes and the variables are correlated. The $i\th$ observation, $\mathbf{x}_i=(x_{i1},x_{i2},\cdots, x_{i,500})\sim \Normal(\boldsymbol{\mu}_k,\boldsymbol{\Sigma})$, where $\boldsymbol{\mu}_k=(\mu_{k,1},\mu_{k,2},\cdots,\mu_{k,500})$, if it is in class $k$, $1\le k\le 3$. The covariance matrix $\boldsymbol{\Sigma}$ is block diagonal, with five blocks each of dimension $100\times 100$. The five blocks are the same  and have $(j,j^\prime)$ element $0.6^{|j-j^\prime|} \times \sigma^2$. Also, $\mu_{1j}\sim \Normal(1,1)$, $\mu_{2j}\sim \Normal(2,1)$ and $\mu_{3j}\sim \Normal(3,1)$ if $1\le j\le 10$ or $101\le j\le 110$ and $\mu_{kj}=0$ otherwise. We consider $\sigma^2=1$, $2$ and $3$.
\item[(c).]{\it Simulation 3:}  Observations from different classes have different distributions about the class means. If the $i\th$  observation is in class $k$, $\mathbf{x}_i\sim \Normal(\boldsymbol{\mu}_k,\boldsymbol{\Sigma}_k)$. We take $\mu_{1j}=3$ if $1\le j\le 10$, $\mu_{2j}=2$ if $1\le j\le 20$, $\mu_{3j}=1$ if $1\le j\le 30$, and $\mu_{kj}=0$ otherwise. The covariance matrix $\boldsymbol{\Sigma}_1$ is diagonal with the diagonal elements generated from the uniform distribution in $(0.5,2) \times \sigma^2 $. $\boldsymbol{\Sigma}_2$ is block diagonal, with five blocks each of dimension $100\times 100$. The blocks have $(j,j^\prime)$ element $0.9^{|j-j^\prime|}\times \sigma^2$.  And $\boldsymbol{\Sigma}_3$ is block diagonal, with five blocks each of dimension $100\times 100$. The blocks have $(j,j^\prime)$ element $0.6 \times \sigma^2$ if $j\neq j^\prime$ and $\sigma^2$ otherwise. We consider $\sigma^2=1$, $2$ and $3$.
\end{itemize}
The mean misclassification rates (percentages) of 50 data sets for each simulation are shown in Table $\ref{table_8}$, with standard deviations in parentheses. The PDA has the highest misclassification rate in all simulations. SFDA-threshold performs similarly with SFDA and both methods have good prediction accuracies in all the simulations.  
 
\begin{table}[h]
\caption{\baselineskip=10pt The averages and standard deviations (in  parentheses) of the misclassification rates (\%) for the simulations in Section \ref{sec.5}. }
\vspace{5mm}
\label{table_8}\centering
\addtolength{\tabcolsep}{-1pt}
\begin{tabular}{|c|c|c|c|c|c|}
\hline
 & $\sigma^2$  &SFDA-threshold & SFDA  &RDA & PDA\\\hline
Simulation 1 & 1 & 0.21(0.26) & 0.24(0.26) & 0.32(0.39) & 2.37(1.46)\\ 
& $1.5^2$ & 1.52(0.77) & 1.54(0.71) & 1.75(0.96) & 5.40(2.07)\\
& 4 & 8.78(4.06) & 8.60(3.71) & 10.20(4.41)&  12.73(4.32)\\\hline
Simulation 2 & 1 & 0.48(0.43) & 0.48(0.47) & 0.79(0.73)& 0.86(0.57) \\ 
& 2 & 3.15(2.40) & 3.29(2.38) & 3.61(2.15) & 4.84(2.45)\\
& 3 & 5.05(2.57) & 5.10(2.43) & 6.05(2.99) & 8.55(3.52)\\\hline
Simulation 3 & 1 & 4.86(1.12) & 4.85(1.12) & 7.71(2.03)  & 9.51(4.20)\\ 
& 2 & 13.02(2.73) & 12.84(2.79) & 18.74(2.84) & 20.42(5.72)\\
& 3 & 21.49(3.45) & 21.48(3.35) & 26.56(3.58) &29.74(7.61)\\
\hline
\end{tabular}
\end{table}

\section{Application to multivariate functional data}\label{sec.6}
With the advance of techniques, multiple curves can be extracted and recorded simultaneously for one subject in a single experiment. In this section, we consider two real datasets where observations are classified into multiple categories and for each subject, multiple curves were measured. We first apply the wavelet transformation to those curves, and then apply our method to the obtained wavelet coefficients. The setting for the tuning parameters is the same as that in the simulation studies.

\subsection{Daily and sports activities data}\label{sec.6.1}
This  motion sensor data set, available in UCI Machine Learning Repository \citep{Bache+Lichman:2013}, recorded several daily and sports activities each performed by 8 subjects \citep{altun2010comparative, barshan2013recognizing, altun2010human} in 60 time segments. Nine sensors (x, y, z accelerometers, x, y, z gyroscopes, x, y, z magnetometers) were placed on each of five body parts (torso, right arm, left arm, right leg, left leg) and calibrated to acquire data at 25 Hz sampling frequency. Therefore, for each activity, there are 480 observations. In each observation, 45 curves are recorded and each of them has $125$ discrete time points.   The purpose of the study is to build a classification rule to identify the corresponding activity based on the observed curves.
 
 We first apply the Fast Fourier Transformation to each of 45 curves to convert it from time domain to the frequency domain and get its spectrum curve. After filtering out the higher frequency, we use the first 64 frequency points for each of 45 frequency curves. Then we apply wavelet transformation with 64 wavelet basis functions to  each of 45 spectrum curves and obtain 64 wavelet coefficients. In this way,
 for each observation, a vector with $64\times 45=2880$ wavelet coefficients is obtained as the features to make classifications. 

We consider nine activities which can be divided into three groups. Group 1 includes three activities: walking in a parking lot, ascending and descending stairs; Group 2 has three activities: running on a treadmill with a speed of 8 km/h, exercising on a stepper and exercising on a cross trainer;  Group 3 includes rowing, jumping and playing basketball. We will consider seven classification problems. In each of the first three problems, we consider the classification of the three activities in each of the three groups. In each of the next three problems, we combine any two of the three groups and consider the classification of the six activities in the combined groups. The last problem is the classification of all nine activities.  In each problem, for each class, we randomly select 30 observations as the training sample and all the other 450 observations as the test sample. The procedure is repeated 50 times for each of the seven problems and the averages and standard deviations of misclassification rates are reported in Table \ref{table6_1}.  
 SFDA-threshold performs similarly with SFDA and both methods have higher prediction accuracies than RDA and PDA in all cases.

\begin{table}[h]
\caption{\baselineskip=10pt The averages and standard deviations (in parentheses) of the misclassification rates (\%) for the daily and sports activities data.  }
\label{table6_1}
\vspace{5mm}
\centering
 \small\addtolength{\tabcolsep}{-3pt}
\begin{tabular}{c|c|c|c|c}
\hline
 Classes included & SFDA-threshold  & SFDA   & RDA& PDA   \\
\hline
 Group 1& 0.23(0.23) &0.23(0.23) &1.94(1.91)& 1.96(2.10)   \\
Group 2 &	0.14(0.43) &0.14(0.44) &0.58(0.66) &0.21(0.58) 		\\
Group 3 &	0.12(0.07)& 0.12(0.08) &0.58(1.08) &0.23(0.36) 		\\
 \hline		
Group 1+2 &	0.45(0.44)& 0.46(0.43)& 1.13(0.79)& 2.39(1.52) 		\\
Group 1+3 &	1.50(0.84) &1.54(0.96)& 1.92(0.99)& 4.79(2.33)  		\\
Group 2+3 &0.53(0.26)& 0.54(0.24) &1.06(0.72)& 0.80(0.37)		\\
 \hline	
Group 1+2+3 &1.63(0.60)&1.53(0.63)&1.78(0.65)&4.20(2.01)		\\
 \hline	
\end{tabular}
\end{table}

\subsection{Australian sign language data}\label{sec.6.2}
The data is available in UCI Machine Learning Repository and the details of the experiments can be founded in \citet{Kadous-2002}. This data set consists of samples of Auslan (Australian Sign Language) signs. Twenty seven examples of each sign were captured from a native signer using high-quality position trackers and instrumented gloves. This was a two-hand system. For each hand, 
11 time series curves were recorded simultaneously, including the measurements of x, y, z positions, the direction of palm and five finger bends. The frequency curve of each of the 22 curves were extracted by the Fast Fourier Transformation and then were transformed by 16 wavelet basis functions. Hence, for each sign, we obtained 352 features. We choose nine signs and divide them into three groups: Group 1 contains the three signs with meanings ``innocent'', ``responsible'' and ``not-my-problem'', respectively; Group 2 contains ``read'', ``write'' and ``draw''; Group 3 contains ``hear'', ``answer'' and ``think''. As in the previous example, we consider seven classification problems. For each class, we randomly choose 20 observations as the training sample and the other 7 as the test sample. The procedure is repeated 50 times and the averages and standard deviations of misclassification rates are reported in Table \ref{table6_2}. As in previous studies, SFDA-threshold performs similarly with SFDA and both methods have higher prediction accuracies than RDA and PDA in all cases.

\begin{table}[h]
 \caption{\baselineskip=10pt The averages and standard deviations (in  parentheses) of the misclassification rates (\%) for the Australian sign language data.  }
\label{table6_2} \centering
\vspace{5mm}
 \small\addtolength{\tabcolsep}{-3pt}
\begin{tabular}{c|c|c|c|c}
\hline
 Classes included & SFDA-threshold  & SFDA & RDA& PDA   \\
\hline
 Group 1& 0(0)& 0(0) &1.24(2.32) &0.19(0.94)   \\
Group 2 &	0(0) & 0(0) & 1.43(2.76) &4.57(5.61) 		\\
Group 3 &	1.24(2.11) & 1.14(2.05) & 3.05(3.94) & 3.9(6.5) 		\\
 \hline		
Group 1+2 &	0.19(0.65) &0.62(1.26) &0.76(1.31) &3.81(2.93) 		\\
Group 1+3 &	0.81(1.71)& 0.62(1.16)& 1.29(1.94)& 2.24(2.38)  	\\
Group 2+3 &0.93(1.45)&1.06(1.68)&1.72(2.13)& 5.16(4.34)		\\
 \hline	
Group 1+2+3 &0.73(1.02)&0.57(0.95)&1.14(1.11)&6.0(2.78)  		\\
 \hline	
\end{tabular}
\end{table}

\section*{Acknowledgments}
 Xin Qi is supported by NSF DMS 1208786.


\end{document}


\begin{center}
{\LARGE{\bf  Supplementary materials for ``Sparse Fisher's discriminant analysis with thresholded linear constraints''}}
\end{center}
\vskip 2mm

\baselineskip=14pt
\vskip 2mm
\begin{center}
 Ruiyan Luo\\
\vskip 2mm
Divison of Epidemiology and Biostatistics, Georgia State University School of Public Health, One Park Place, Atlanta, GA 30303\\
rluo@gsu.edu\\
\hskip 5mm\\
Xin Qi\\
\vskip 2mm
Department of Mathematics and Statistics, Georgia State University, 30 Pryor Street, Atlanta, GA 30303-3083\\
xqi3@gsu.edu \\
\end{center}
  
\vskip 20mm
 \renewcommand{\proof}{\noindent\underline{\bf Proof of Theorem}}

In this supplementary material, we provide the proofs of all the theorems in Appendix A and the proofs of all technical lemmas in Appendix B.

\section*{Appendix A: Proofs of theorems}
 \renewcommand{\thesubsection}{A.\arabic{subsection}}
  
\baselineskip=18pt

 We first provide several lemmas whose proofs can be found in Appendix B.
\begin{Lemma}\label{lemma_3}
For any $1\le i\neq j\le K-1$, we have
$\boldsymbol{\Sigma}^{-1}\boldsymbol{\delta}_{ij}=\mathbf{D}\boldsymbol{\delta}_{ij}$, where $\boldsymbol{\delta}_{ij}=\boldsymbol{\mu}_j-\boldsymbol{\mu}_i$.\\
\end{Lemma}

\begin{Lemma}\label{lemma_6}
Define  a $K\times K$ matrix,
\begin{align*}
\boldsymbol{\Delta}=\mathbf{U}\trans\boldsymbol{\Sigma}^{-1}\mathbf{U}.
\end{align*}
Under Condition \ref{condition_2}, $\boldsymbol{\Delta}$ has $K-1$ positive eigenvalues denoted by $\lambda^+_{min}(\boldsymbol{\Delta})=\lambda_{K-1}(\boldsymbol{\Delta})\le \cdots\le \lambda_{2}(\boldsymbol{\Delta})\le \lambda_{1}(\boldsymbol{\Delta})=\lambda_{max}(\boldsymbol{\Delta})$. Then we have  $\lambda_1(\boldsymbol{\Delta})=K\lambda_1(\boldsymbol{\Xi})$, $\cdots$, $\lambda_{K-1}(\boldsymbol{\Delta})=K\lambda_{K-1}(\boldsymbol{\Xi})$, and 
 \begin{align*}
&2Kc_1\le2\lambda^+_{min}(\boldsymbol{\Delta})\le \min_{i\neq j}(\boldsymbol{\mu}_i-\boldsymbol{\mu}_j)\trans\boldsymbol{\Sigma}^{-1}(\boldsymbol{\mu}_i-\boldsymbol{\mu}_j)\notag\\
\le &\max_{i\neq j}(\boldsymbol{\mu}_i-\boldsymbol{\mu}_j)\trans\boldsymbol{\Sigma}^{-1}(\boldsymbol{\mu}_i-\boldsymbol{\mu}_j)\le 2\lambda_{max}(\boldsymbol{\Delta}).
\end{align*}
\end{Lemma}

\begin{Lemma}\label{lemma_2}
Suppose that Condition \ref{condition_1} holds, $p\ge 2$, $K\le p+1$ and $K\log{p}/n\to 0$ as $n\to \infty$.  Then for any $M>0$, we can find a constant $C$ large enough and independent of $n$, $p$ and $K$, such that
 \begin{align}
& P\left(\max_{1\le j\le K}\|\bar{\mathbf{x}}_j-\boldsymbol{\mu}_j\|_{\infty}> C\sqrt{\frac{K\log{p}}{n}}\right)\le p^{-M},\label{134}
\end{align}
for all $n$ large enough.
\end{Lemma}

Now we prove the main theorems.

\subsection{Proof of Theorem  $\ref{theorem_11}$}
 
We only consider the case that $(n_1,n_2,\cdots,n_K)$ follows a multinomial distribution. For the nonrandom case, a similar argument can prove the theorem.  Let $\widehat{\sigma}_{kl}$ and $\sigma_{kl}$ be the $(k,l)$ element of $\widehat{\boldsymbol{\Sigma}}$ and $\boldsymbol{\Sigma}$, respectively, $1\le k,l\le p$. By the definition of $\widehat{\boldsymbol{\Sigma}}$ in $\eqref{1097}$, 
\begin{align}
&\left(\frac{n-K}{n}\right)\left|\widehat{\sigma}_{kl}-\sigma_{kl}\right|=\left|\frac{1}{n}\sum_{i=1}^K\sum_{j=1}^{n_i}(\mathbf{x}^{k}_{ij}-\bar{\mathbf{x}}^{k}_i)(\mathbf{x}^{l}_{ij}-\bar{\mathbf{x}}^{k}_i)-(1-\frac{K}{n})\sigma_{kl}\right|\label{28}\\
=&\left|\frac{1}{n}\sum_{i=1}^K\sum_{j=1}^{n_i}(\mathbf{x}^{k}_{ij}-\boldsymbol{\mu}^{k}_i)(\mathbf{x}^{l}_{ij}-\boldsymbol{\mu}^{l}_i)-\frac{1}{n}\sum_{i=1}^K n_i(\bar{\mathbf{x}}^{k}_i-\boldsymbol{\mu}^{k}_i)(\bar{\mathbf{x}}^{l}_i-\boldsymbol{\mu}^{l}_i)-(1-\frac{K}{n})\sigma_{kl}\right|\notag\\
\le & \frac{1}{n}\left|\sum_{i=1}^K\sum_{j=1}^{n_i}\left[(\mathbf{x}^{k}_{ij}-\boldsymbol{\mu}^{k}_i)(\mathbf{x}^{l}_{ij}-\boldsymbol{\mu}^{l}_i)  -\sigma_{kl}\right]\right|+ \frac{1}{n}\left| \sum_{i=1}^K\left[n_i(\bar{\mathbf{x}}^{k}_i-\boldsymbol{\mu}^{k}_i)(\bar{\mathbf{x}}^{l}_i-\boldsymbol{\mu}^{l}_i)-\sigma_{kl}\right] \right|,\notag
\end{align}
where $\mathbf{x}^{k}_{ij}$ and $\boldsymbol{\mu}^{k}_i$ denotes the $k$-th coordinate of $\mathbf{x}_{ij}$ and $\boldsymbol{\mu}_i$, respectively. Note that both $\mathbf{x}_{ij}-\boldsymbol{\mu}_i$ and $\sqrt{n_i}(\bar{\mathbf{x}}_i-\boldsymbol{\mu}_i)$ have the distributions $N(\mathbf{0}, \boldsymbol{\Sigma})$. By Lemma A.3. in \citet{bickel2008regularized}, we have 
\begin{align}
&P\left(\left|\sum_{i=1}^K\sum_{j=1}^{n_i}\left[(\mathbf{x}^{k}_{ij}-\boldsymbol{\mu}^{k}_i)(\mathbf{x}^{l}_{ij}-\boldsymbol{\mu}^{l}_i)  -\sigma_{kl}\right]\right|>n\nu_1\right)\le C_1\exp(-C_2n\nu_1^2),\text{ and hence}\notag\\
&P\left(\max_{k,l}\left|\sum_{i=1}^K\sum_{j=1}^{n_i}\left[(\mathbf{x}^{k}_{ij}-\boldsymbol{\mu}^{k}_i)(\mathbf{x}^{l}_{ij}-\boldsymbol{\mu}^{l}_i)  -\sigma_{kl}\right]\right|>n\nu_1\right)\le C_1p^2\exp(-C_2n\nu_1^2),\label{55}
\end{align}
for any $\nu_1$ less than a constant $\delta$, where $C_1$, $C_2$ and $\delta$ are constants only depending on the upper bound $c_0$ of the eigenvalues of $\boldsymbol{\Sigma}$. For any $C>0$, taking $\nu_1=C\sqrt{\frac{\log{p}}{n}}$, we can obtain 
\begin{align}
P\left(\max_{k,l}\frac{1}{n}\left|\sum_{i=1}^K\sum_{j=1}^{n_i}\left[(\mathbf{x}^{k}_{ij}-\boldsymbol{\mu}^{k}_i)(\mathbf{x}^{l}_{ij}-\boldsymbol{\mu}^{l}_i)  -\sigma_{kl}\right]\right|>C\sqrt{\frac{\log{p}}{n}}\right)\le C_1p^2p^{-C^2C_2}.\label{31}
\end{align}
For any $M>0$, we can find $C$ large enough such that the right hand side of $\eqref{31}$ is less than $p^{-M}$ for all $p>1$. For the second term in in the last line of $\eqref{28}$, define $Z_{ik}=\sqrt{n_i}(\bar{\mathbf{x}}^{k}_i-\boldsymbol{\mu}^{k}_i)$, for any $1\le i\le K$ and $1\le k\le p$. Then
\begin{align}
  \sum_{i=1}^K\left[n_i(\bar{\mathbf{x}}^{k}_i-\boldsymbol{\mu}^{k}_i)(\bar{\mathbf{x}}^{l}_i-\boldsymbol{\mu}^{l}_i)-\sigma_{kl}\right]=\sum_{i=1}^K\left[Z_{ik}Z_{il}-\sigma_{kl}\right]\label{29}\\
= \frac{1}{4}\sum_{i=1}^K[(Z_{ik}+Z_{il})^2-(\sigma_{kk}+\sigma_{ll}+2\sigma_{kl})]\notag\\
-\frac{1}{4}\sum_{i=1}^K[(Z_{ik}-Z_{il})^2-(\sigma_{kk}+\sigma_{ll}-2\sigma_{kl})].\notag
\end{align}
We will derive the upper bound for the first sum in the last line of $\eqref{29}$.  Let 
$$Y_i=(Z_{ik}+Z_{il})^2/(\sigma_{kk}+\sigma_{ll}+2\sigma_{kl})-1.$$
 Then $Y_1,\cdots,Y_K$, are i.i.d. random variables with the distribution $\chi^2_1-1$. We will apply the Bernstein's inequality (see Lemma 2.2.11 in \citet{van1996weak} or page 855 of \citet{shorack2009empirical}) for unbounded random variables to $Y_1+\cdots+Y_K$. We first verify the moment condition required by the Bernstein's inequality. For any positive integer $m\ge 3$, noting that $Var(Y_i)=2$, we have 
\begin{align*}
&E[|Y_i|^m]=E[|\chi^2_1-1|^m]\le 2^{m-1}\left(E[|\chi^2_1|^m]+1\right)\le 2^{m}E[|\chi^2_1|^m]\notag\\
=&2^m[1\cdot 3\cdot 5\cdots (2m-1)]\le 2^m[2\cdot 4\cdot 6\cdots (2m)]= 2^m[2^mm!]\notag\\
=&4^mm!Var(Y_i)/2\le D^{m-2}m!Var(Y_i)/2,
\end{align*}
where $D=64$. When $m=2$,  $E[|Y_i|^m]=Var(Y_i)=D^{m-2}m!Var(Y_i)/2$. Hence,  the moment condition for the Bernstein's inequality holds for $Y_i$. Now by the Bernstein's inequality, for any $\nu_2>0$, let $x=n\nu_2/(\sigma_{kk}+\sigma_{ll}+2\sigma_{kl})$, then we have
\begin{align*}
&P\left( \left|\sum_{i=1}^K[(Z_{ik}+Z_{il})^2-(\sigma_{kk}+\sigma_{ll}+2\sigma_{kl})]\right|>n\nu_2\right)\notag\\
=&P\left( \left|Y_1+\cdots+Y_K\right|>\frac{n\nu_2}{\sigma_{kk}+\sigma_{ll}+2\sigma_{kl}}\right) =P\left( \left|Y_1+\cdots+Y_K\right|>x\right)\\
\le& 2\exp{(-\frac{1}{2}\frac{x^2}{KVar(Y_1)+Dx})}=2\exp{(-\frac{1}{2}\frac{x^2}{2K+Dx})}.
 \end{align*}
  Note that $-x^2/(2K+Dx)$ is a decreasing function for $x>0$, and that
	\begin{align}
  \sigma_{kk}+\sigma_{ll}+2\sigma_{kl}=\mathbf{v}_{kl}\trans \boldsymbol{\Sigma} \mathbf{v}_{kl}\le \lambda_{\max}(\boldsymbol{\Sigma} )\|\mathbf{v}_{kl}\|_2^2= 2\lambda_{\max}(\boldsymbol{\Sigma} )\le 2c_0, \label{10011}
 \end{align}
	where $\mathbf{v}_{kl}$ is the p-dimensional vector with all coordinates equal to 0 except the $k$-th and $l$-th coordinates which are equal to 1 and  the last inequality is due to Condition \ref{condition_1} (b). Then we have $x\ge n\nu_2/(2c_0)$ and 
\begin{align*}
  \exp{(-\frac{1}{2}\frac{x^2}{2K+Dx})}\le \exp{(-\frac{1}{2}\frac{(n\nu_2)^2}{8c_0^2K+2c_0Dn\nu_2})}\;.
 \end{align*}
 Hence,
\begin{align*}
&P\left( \max_{k,l}\left|\sum_{i=1}^K[(Z_{ik}+Z_{il})^2-(\sigma_{kk}+\sigma_{ll}+2\sigma_{kl})]\right|>n\nu_2\right)\notag\\
\le& 2p^2\exp{\left(-\frac{1}{2}\frac{(n\nu_2)^2}{8c_0^2K+2c_0Dn\nu_2}\right)}.
 \end{align*}
 For any $C>0$, let $\nu_2=C\sqrt{\frac{\log{p}}{n}}$. Since $\log{p}/n\to 0$, when $n$ is large enough, we have $\log{p}\le n$, and hence
\begin{align}
 2p^2\exp{\left(-\frac{1}{2}\frac{(n\nu_2)^2}{8c_0^2K+2c_0Dn\nu_2}\right)}=2p^2\exp{\left(-\frac{1}{2}\frac{C^2n\log{p}}{8c_0^2K+2c_0DC\sqrt{n\log(p)}}\right)}\notag\\
\le 2p^2\exp{\left(-\frac{1}{2}\frac{C^2n\log{p}}{8c_0^2n+2c_0DC\sqrt{nn}}\right)}=2p^2p^{ -\frac{1}{2}\frac{C^2}{8c_0^2+2c_0DC}},\label{30}
 \end{align}
where we use  $K\le n$ as $n$ is large enough due to the condition $K\log{p}/n\to 0$. Now for any $M>0$, we can find $C$ large enough such that the right hand side of $\eqref{30}$ is less than $p^{-M}$ for all $p\ge 2$. We can obtain the similar result for the second sum in the last line of $\eqref{29}$. Hence, for any $M>0$, we can find $C$ large enough such that 
\begin{align}
P\left(\max_{k,l}\frac{1}{n}\left|\sum_{i=1}^K\left[n_i(\bar{\mathbf{x}}^{k}_i-\boldsymbol{\mu}^{k}_i)(\bar{\mathbf{x}}^{l}_i-\boldsymbol{\mu}^{l}_i)-\sigma_{kl}\right]\right|>C\sqrt{\frac{\log{p}}{n}}\right)\le  p^{-M}.\label{1031}
\end{align}
It follows from $\eqref{28}$,  $\eqref{31}$ and $\eqref{1031}$,
for any $M>0$, we can find $C$ large enough and independent of $n$, $p$ and $K$, such that 
\begin{align*}
P\left(\max_{k,l} \left|\widehat{\sigma}_{kl}-\sigma_{kl}\right|>C\sqrt{\frac{K\log{p}}{n}}\right)\le P\left(\max_{k,l} \left|\widehat{\sigma}_{kl}-\sigma_{kl}\right|>C\sqrt{\frac{\log{p}}{n}}\right)\le  p^{-M},
\end{align*}
 for all $n$ large enough, where we use $K/n\to 0$ due to the condition $K\log{p}/n\to 0$. \\

In order to estimate $\|\widehat{\mathbf{B}}-\mathbf{B}\|_{\infty}$, we first calculate the $(k,l)$ element of $\widehat{\mathbf{B}}-\mathbf{B}$. 
\begin{align}
&\left|\frac{1}{n}\sum_{i=1}^Kn_i(\bar{\mathbf{x}}^k_i-\bar{\mathbf{x}}^k)(\bar{\mathbf{x}}^l_i-\bar{\mathbf{x}}^l)-\frac{1}{K}\sum_{i=1}^K\boldsymbol{\mu}^{k}_i\boldsymbol{\mu}^{l}_i\right|\notag\\
=&\left|\frac{1}{n}\sum_{i=1}^Kn_i\bar{\mathbf{x}}^k_i\bar{\mathbf{x}}^l_i-\bar{\mathbf{x}}^k\bar{\mathbf{x}}^l-\frac{1}{K}\sum_{i=1}^K\boldsymbol{\mu}^{k}_i\boldsymbol{\mu}^{l}_i\right|\notag\\
=&\left|\frac{1}{n}\sum_{i=1}^Kn_i (\bar{\mathbf{x}}^{k}_i-\boldsymbol{\mu}^{k}_i)(\bar{\mathbf{x}}^{l}_i-\boldsymbol{\mu}^{l}_i)+\frac{1}{n}\sum_{i=1}^Kn_i(\bar{\mathbf{x}}^k_i-\boldsymbol{\mu}^{k}_i)\boldsymbol{\mu}^{l}_i+\frac{1}{n}\sum_{i=1}^Kn_i\boldsymbol{\mu}^{k}_i(\bar{\mathbf{x}}^l_i-\boldsymbol{\mu}^{l}_i)\right\notag\\
&\left+\frac{1}{n}\sum_{i=1}^Kn_i\boldsymbol{\mu}^{k}_i\boldsymbol{\mu}^{l}_i-\bar{\mathbf{x}}^k\bar{\mathbf{x}}^l-\frac{1}{K}\sum_{i=1}^K\boldsymbol{\mu}^{k}_i\boldsymbol{\mu}^{l}_i\right|\notag\\
\le &\frac{K}{n}|\sigma_{kl}|+\left|\frac{1}{n}\sum_{i=1}^K[n_i(\bar{\mathbf{x}}^{k}_i-\boldsymbol{\mu}^{k}_i)(\bar{\mathbf{x}}^{l}_i-\boldsymbol{\mu}^{l}_i)-\sigma_{kl}]\right|+\left|\frac{1}{n}\sum_{i=1}^Kn_i(\bar{\mathbf{x}}^k_i-\boldsymbol{\mu}^{k}_i)\boldsymbol{\mu}^{l}_i\right|\notag\\
&\qquad\qquad +\left|\frac{1}{n}\sum_{i=1}^Kn_i\boldsymbol{\mu}^{k}_i(\bar{\mathbf{x}}^l_i-\boldsymbol{\mu}^{l}_i)\right|+\left|\sum_{i=1}^K\left(\frac{n_i}{n}-\frac{1}{K}\right)\boldsymbol{\mu}^{k}_i\boldsymbol{\mu}^{l}_i\right|+|\bar{\mathbf{x}}^k\bar{\mathbf{x}}^l|\notag\\
=&\frac{K}{n}|\sigma_{kl}|+I+II+III+IV+V\le \frac{K}{n}c_0+I+II+III+IV+V.\label{1264}
\end{align}
Note that the term $I$ is just that in $\eqref{1031}$. By Condition \ref{condition_1} (b), 
\begin{align*}
&\max_{k,l}II\le \frac{1}{n}\sum_{i=1}^Kn_i\max_{1\le j\le K}\|\bar{\mathbf{x}}_j-\boldsymbol{\mu}_j\|_{\infty}\max_{1\le j\le K}\|\boldsymbol{\mu}_j\|_{\infty}\le c_0\max_{1\le j\le K}\|\bar{\mathbf{x}}_j-\boldsymbol{\mu}_j\|_{\infty}, 
 \end{align*}
which combined with Lemma \ref{lemma_2} gives that for any $M>0$, we can find a constant $C$ large enough and independent of $n$, $p$ and $K$, such that
\begin{align*}
& P\left(\max_{k,l}II> C\sqrt{\frac{K\log{p}}{n}}\right)\le p^{-M}, \label{1011}
\end{align*}
for all $n$ large enough. The same bound can be obtained for the term $III$. As to the term $IV$, by Lemma \ref{lemma_1} and Condition \ref{condition_1} (b),  for any $M>0$, we can find a constant $C$ such that
  \begin{align}
  P\left( \max_{1\le k,l\le p} \left|\sum_{i=1}^K\left(\frac{n_i}{n}-\frac{1}{K}\right)\boldsymbol{\mu}^{k}_i\boldsymbol{\mu}^{l}_i\right|>C\sqrt{\frac{K\log{p}}{n}}\right) \label{1012}\\
 \le P\left( Kc_0^2\max_{1\le i\le K} \left|\frac{n_i}{n}-\frac{1}{K}\right|>C\sqrt{\frac{K\log{p}}{n}}\right)\notag\\
 =P\left( \max_{1\le i\le K} \left|\frac{n_i}{n}-\frac{1}{K}\right|>\frac{C}{c_0^2}\sqrt{\frac{\log{p}}{Kn}}\right) \le p^{-M}\notag
\end{align} 
for all $n$ large enough. For the term $V$, because 
 \begin{align}
 &\bar{\mathbf{x}}=\frac{1}{n}\sum_{i=1}^K\sum_{j=1}^{n_i}\mathbf{x}_{ij}=\bar{\mathbf{y}}+\frac{1}{n}\sum_{i=1}^Kn_i\boldsymbol{\mu}_i=\bar{\mathbf{y}}+\sum_{i=1}^K\left(\frac{n_i}{n}-\frac{1}{K}\right)\boldsymbol{\mu}_i,\label{10009}
\end{align} 
where $\bar{\mathbf{y}}=\frac{1}{n}\sum_{i=1}^K\sum_{j=1}^{n_i}(\mathbf{x}_{ij}-\boldsymbol{\mu}_i)$ and the last equality is due to $\eqref{3}$. Then we have
 \begin{align}
 &\bar{\mathbf{x}}^k\bar{\mathbf{x}}^l=\bar{\mathbf{y}}^k\bar{\mathbf{y}}^l +\sum_{i=1}^K\left(\frac{n_i}{n}-\frac{1}{K}\right)\boldsymbol{\mu}^l_i\bar{\mathbf{y}}^k+\sum_{i=1}^K\left(\frac{n_i}{n}-\frac{1}{K}\right)\boldsymbol{\mu}^k_i\bar{\mathbf{y}}^l\label{10010}\\
&+\left[\sum_{i=1}^K\left(\frac{n_i}{n}-\frac{1}{K}\right)\boldsymbol{\mu}^l_i\right]\left[\sum_{i=1}^K\left(\frac{n_i}{n}-\frac{1}{K}\right)\boldsymbol{\mu}^k_i\right].\notag
\end{align} 
We will consider the four terms on the right hand side of $\eqref{10010}$, respectively. Note that $\bar{\mathbf{y}}$ has the normal distribution with mean zero and covariance matrix $\boldsymbol{\Sigma}/n$. $\bar{\mathbf{y}}^k\bar{\mathbf{y}}^l=(\bar{\mathbf{y}}^k+\bar{\mathbf{y}}^l)^2/4-(\bar{\mathbf{y}}^k-\bar{\mathbf{y}}^l)^2/4$. Note that $\bar{\mathbf{y}}^k+\bar{\mathbf{y}}^l$ has a normal distribution with mean zero and variance $(\sigma_{kk}+\sigma_{ll}+2\sigma_{kl})/n\le 2c_0/n$ by $\eqref{10011}$. By the same arguments as in the proof of Lemma \ref{lemma_2}, we can show that for any $M>0$, we can find $C$ large enough such that 
\begin{align*}
P\left(\max_{k,l} \frac{1}{\sqrt{Kc_0^2}}\left|\frac{\bar{\mathbf{y}}^k+\bar{\mathbf{y}}^l}{2}\right|>C\sqrt{\frac{\log{p}}{n}}\right)\le  p^{-M}, 
\end{align*}
for all $n$ large enough. Since when $n$ is large enough, we have $Cc_0\sqrt{K\log{p}/n}\le 1$, then,
 \begin{align}
 P\left(\max_{k,l} \left|\frac{(\bar{\mathbf{y}}^l+\bar{\mathbf{y}}^l)^2}{4}\right|>Cc_0\sqrt{\frac{K\log{p}}{n}}\right)\notag\\
\le P\left(\max_{k,l} \left|\frac{(\bar{\mathbf{y}}^l+\bar{\mathbf{y}}^l)^2}{4}\right|>\left[Cc_0\sqrt{\frac{K\log{p}}{n}}\right]^2\right)\notag\\
= P\left(\max_{k,l} \frac{1}{\sqrt{Kc_0^2}}\left|\frac{\bar{\mathbf{y}}^k+\bar{\mathbf{y}}^l}{2}\right|>C\sqrt{\frac{\log{p}}{n}}\right)\le  p^{-M}.\label{1010}
\end{align} 
 and the same inequality for $(\bar{\mathbf{y}}^k-\bar{\mathbf{y}}^l)^2/4$. Therefore, we have that 
for any $M>0$, we can find $C$ large enough and independent of $n$, $p$ and $K$, such that 
\begin{align}
P\left(\max_{k,l}  \left|\bar{\mathbf{y}}^k\bar{\mathbf{y}}^l\right|>C\sqrt{\frac{K\log{p}}{n}}\right)\le  p^{-M}.\label{10012}
\end{align}
Using the same arguments as in $\eqref{1012}$, we can obtain the same probability bounds for the last three terms on the right hand side of $\eqref{10010}$. Then by combining $\eqref{1264}$-$\eqref{10012}$ and using Lemmas \ref{lemma_1} and  \ref{lemma_2}, for any $M>0$, we can find $C$ large enough such that 
 \begin{align}
 P\left(\|\widehat{\mathbf{B}}-\mathbf{B}\|_{\infty}>C\sqrt{\frac{K\log{p}}{n}}\right)&\label{1268} \\
= P\left(\max_{k,l} \left|\frac{1}{n}\sum_{i=1}^Kn_i(\bar{\mathbf{x}}^k_i-\bar{\mathbf{x}}^k)(\bar{\mathbf{x}}^l_i-\bar{\mathbf{x}}^l)-\frac{1}{K}\sum_{i=1}^K\boldsymbol{\mu}^{k}_i\boldsymbol{\mu}^{l}_i\right|>C\sqrt{\frac{K\log{p}}{n}}\right)&\le  p^{-M},\notag
\end{align} 
for all $n$ large enough.  \\

\subsection{Proof of Theorem  $\ref{theorem_12}$}
 
In this proof, we only consider  elements in the event $\Omega_n$. First, note that $\boldsymbol{\alpha}_1$ and $\widehat{\boldsymbol{\alpha}}_1$ are the solutions to
\begin{align}
\max_{\boldsymbol{\alpha}\in \mathbb{R}^p,  \boldsymbol{\alpha}\neq \mathbf{0}}\frac{\boldsymbol{\alpha}\trans \mathbf{B}\boldsymbol{\alpha}}{\boldsymbol{\alpha}\trans \boldsymbol{\Sigma}\boldsymbol{\alpha}},\quad \text{ and }\quad \max_{\boldsymbol{\alpha}\in \mathbb{R}^p,  \boldsymbol{\alpha}\neq \mathbf{0}}\frac{\boldsymbol{\alpha}\trans \widehat{\mathbf{B}}\boldsymbol{\alpha}}{\boldsymbol{\alpha}\trans \widehat{\boldsymbol{\Sigma}}\boldsymbol{\alpha}+\tau_n\|\boldsymbol{\alpha}\|^2_{\lambda_n}},\label{1014}
\end{align}
respectively, with 
\begin{align}
\boldsymbol{\alpha}_1\trans \boldsymbol{\Sigma}\boldsymbol{\alpha}_1=1,\qquad \widehat{\boldsymbol{\alpha}}_1\trans \widehat{\boldsymbol{\Sigma}}\widehat{\boldsymbol{\alpha}}_1 +\tau_n\|\widehat{\boldsymbol{\alpha}}_1 \|_{\lambda_n}=1,\label{10014}
\end{align}
 Hence, we have
 \begin{align}
 &\frac{\widehat{\boldsymbol{\alpha}}_1\trans  {\mathbf{B}}\widehat{\boldsymbol{\alpha}}_1}{\widehat{\boldsymbol{\alpha}}_1\trans  {\boldsymbol{\Sigma}}\widehat{\boldsymbol{\alpha}}_1 }\le \frac{\boldsymbol{\alpha}_1\trans  {\mathbf{B}}\boldsymbol{\alpha}_1}{\boldsymbol{\alpha}_1\trans  {\boldsymbol{\Sigma}}\boldsymbol{\alpha}_1 }=\boldsymbol{\alpha}_1\trans  {\mathbf{B}}\boldsymbol{\alpha}_1,\notag\\
  &  \widehat{\boldsymbol{\alpha}}_1\trans \widehat{\mathbf{B}}\widehat{\boldsymbol{\alpha}}_1=\frac{\widehat{\boldsymbol{\alpha}}_1\trans \widehat{\mathbf{B}}\widehat{\boldsymbol{\alpha}}_1}{\widehat{\boldsymbol{\alpha}}_1\trans \widehat{\boldsymbol{\Sigma}}\widehat{\boldsymbol{\alpha}}_1+\tau_n\|\widehat{\boldsymbol{\alpha}}_1\|^2_{\lambda_n}}\ge \frac{\boldsymbol{\alpha}_1\trans \widehat{\mathbf{B}}\boldsymbol{\alpha}_1}{\boldsymbol{\alpha}_1\trans \widehat{\boldsymbol{\Sigma}}\boldsymbol{\alpha}_1+\tau_n\|\boldsymbol{\alpha}_1\|^2_{\lambda_n}}.\label{1054}
\end{align}
 The first inequality in $\eqref{1054}$ leads to
\begin{align}
   \widehat{\boldsymbol{\alpha}}_1\trans  {\mathbf{B}}\widehat{\boldsymbol{\alpha}}_1\le  (\boldsymbol{\alpha}_1\trans  {\mathbf{B}}\boldsymbol{\alpha}_1 )(\widehat{\boldsymbol{\alpha}}_1\trans  {\boldsymbol{\Sigma}}\widehat{\boldsymbol{\alpha}}_1).\label{1020}
\end{align}
 By the definition of $\Omega_n$ in $\eqref{1033}$,
 \begin{align}
  &|\widehat{\boldsymbol{\alpha}}_1\trans \widehat{\mathbf{B}}\widehat{\boldsymbol{\alpha}}_1- \widehat{\boldsymbol{\alpha}}_1\trans  \mathbf{B}\widehat{\boldsymbol{\alpha}}_1|\le \|\widehat{\mathbf{B}}-\mathbf{B}\|_{\infty}\|\widehat{\boldsymbol{\alpha}}_1\|^2_1=\frac{1}{C_2}\tau_n\|\widehat{\boldsymbol{\alpha}}_1\|^2_1,\quad\text{ and similarly,}\notag\\
	&|\widehat{\boldsymbol{\alpha}}_1\trans \widehat{\boldsymbol{\Sigma}}\widehat{\boldsymbol{\alpha}}_1-\widehat{\boldsymbol{\alpha}}_1\trans  \boldsymbol{\Sigma}\widehat{\boldsymbol{\alpha}}_1|\le\frac{1}{C_2}\tau_n \|\widehat{\boldsymbol{\alpha}}_1\|^2_1,\notag\\
	&|\boldsymbol{\alpha}_1\trans \widehat{\mathbf{B}}\boldsymbol{\alpha}_1-\boldsymbol{\alpha}_1\trans  \mathbf{B}\boldsymbol{\alpha}_1|\le \frac{1}{C_2}\tau_n \|\boldsymbol{\alpha}_1\|^2_1, \quad |\boldsymbol{\alpha}_1\trans \widehat{\boldsymbol{\Sigma}}\boldsymbol{\alpha}_1-\boldsymbol{\alpha}_1\trans  \boldsymbol{\Sigma}\boldsymbol{\alpha}_1|\le \frac{1}{C_2}\tau_n \|\boldsymbol{\alpha}_1\|^2_1.\label{1017}
\end{align} 
By Condition \ref{condition_2} (a),
\begin{align}
  \boldsymbol{\alpha}_1\trans  {\mathbf{B}}\boldsymbol{\alpha}_1=\lambda_1(\boldsymbol{\Xi})\ge c_1. \label{1000017}
\end{align}
   Moreover, we have 
 \begin{align}
  \lambda_n\|\widehat{\boldsymbol{\alpha}}\|^2_1\le \|\widehat{\boldsymbol{\alpha}}_1\|^2_{\lambda_n}=(1-\lambda_n) \|\widehat{\boldsymbol{\alpha}}\|^2_2+\lambda_n\|\widehat{\boldsymbol{\alpha}}\|^2_1\le \|\widehat{\boldsymbol{\alpha}}\|^2_1. \label{10017}
\end{align}
Then by $\eqref{10014}$ , $\eqref{1020}$, $\eqref{1017}$, $\eqref{1000017}$ and  $\eqref{10017}$,   
\begin{align}
  & \widehat{\boldsymbol{\alpha}}_1\trans \widehat{\mathbf{B}}\widehat{\boldsymbol{\alpha}}_1\le \widehat{\boldsymbol{\alpha}}_1\trans  {\mathbf{B}}\widehat{\boldsymbol{\alpha}}_1+\frac{1}{C_2}\tau_n\|\widehat{\boldsymbol{\alpha}}_1\|^2_1\le (\boldsymbol{\alpha}_1\trans  {\mathbf{B}}\boldsymbol{\alpha}_1 )(\widehat{\boldsymbol{\alpha}}_1\trans  {\boldsymbol{\Sigma}}\widehat{\boldsymbol{\alpha}}_1)+\frac{1}{C_2}\tau_n\|\widehat{\boldsymbol{\alpha}}_1\|^2_1\label{1064}\\
	\le &(\boldsymbol{\alpha}_1\trans  {\mathbf{B}}\boldsymbol{\alpha}_1 )\left(\widehat{\boldsymbol{\alpha}}_1\trans  \widehat{\boldsymbol{\Sigma}}\widehat{\boldsymbol{\alpha}}_1+\frac{1}{C_2}\tau_n\|\widehat{\boldsymbol{\alpha}}_1\|^2_1\right)+\frac{\boldsymbol{\alpha}_1\trans  {\mathbf{B}}\boldsymbol{\alpha}_1}{c_1}\frac{1}{C_2}\tau_n\|\widehat{\boldsymbol{\alpha}}_1\|^2_1\notag\\
=&(\boldsymbol{\alpha}_1\trans  {\mathbf{B}}\boldsymbol{\alpha}_1 )\left(1-\tau_n\|\widehat{\boldsymbol{\alpha}}_1\|^2_{\lambda_n} +\frac{1+c_1^{-1}}{C_2}\tau_n\|\widehat{\boldsymbol{\alpha}}_1\|^2_1\right)\notag \\
\le&(\boldsymbol{\alpha}_1\trans  {\mathbf{B}}\boldsymbol{\alpha}_1 )\left(1-\tau_n\lambda_n\|\widehat{\boldsymbol{\alpha}}_1\|^2_1 +\frac{1+c_1^{-1}}{C_2}\tau_n\|\widehat{\boldsymbol{\alpha}}_1\|^2_1\right)\notag\\
=&(\boldsymbol{\alpha}_1\trans  {\mathbf{B}}\boldsymbol{\alpha}_1 )\left[1-\tau_n(\lambda_n-\lambda_0/2)\|\widehat{\boldsymbol{\alpha}}_1\|^2_1  \right],\notag
\end{align}
where the last equality is due to the definition  of $C_2$ in $\eqref{1271}$. By $\eqref{1054}$ and $\eqref{1017}$,  
\begin{align}
  & \widehat{\boldsymbol{\alpha}}_1\trans \widehat{\mathbf{B}}\widehat{\boldsymbol{\alpha}}_1\ge \frac{\boldsymbol{\alpha}_1\trans \widehat{\mathbf{B}}\boldsymbol{\alpha}_1}{\boldsymbol{\alpha}_1\trans \widehat{\boldsymbol{\Sigma}}\boldsymbol{\alpha}_1+\tau_n\|\boldsymbol{\alpha}_1\|^2_{\lambda_n}}\ge \frac{\boldsymbol{\alpha}_1\trans  {\mathbf{B}}\boldsymbol{\alpha}_1-\frac{1}{C_2}\tau_n \|\boldsymbol{\alpha}_1\|^2_1}{\boldsymbol{\alpha}_1\trans  {\boldsymbol{\Sigma}}\boldsymbol{\alpha}_1+\frac{1}{C_2}\tau_n \|\boldsymbol{\alpha}_1\|^2_1+\tau_n\|\boldsymbol{\alpha}_1\|^2_1}\notag\\
	&\ge \frac{\boldsymbol{\alpha}_1\trans  {\mathbf{B}}\boldsymbol{\alpha}_1-\frac{\boldsymbol{\alpha}_1\trans  {\mathbf{B}}\boldsymbol{\alpha}_1}{c_1}\frac{1}{C_2}\tau_n \|\boldsymbol{\alpha}_1\|^2_1}{\boldsymbol{\alpha}_1\trans  {\boldsymbol{\Sigma}}\boldsymbol{\alpha}_1+\frac{1}{C_2}\tau_n \|\boldsymbol{\alpha}_1\|^2_1+\tau_n\|\boldsymbol{\alpha}_1\|^2_1}= \frac{\boldsymbol{\alpha}_1\trans  {\mathbf{B}}\boldsymbol{\alpha}_1\left[1-\frac{c_1^{-1}}{C_2}\tau_n \|\boldsymbol{\alpha}_1\|^2_1\right]}{1+\frac{1}{C_2}\tau_n \|\boldsymbol{\alpha}_1\|^2_1+\tau_n\|\boldsymbol{\alpha}_1\|^2_1},\label{1080}
\end{align}
which together with  $\eqref{1064}$ leads to
 \begin{align}
  &   \tau_n(\lambda_n-\lambda_0/2)\|\widehat{\boldsymbol{\alpha}}_1\|^2_1 \le \frac{\left(1+\frac{1+c_1^{-1}}{C_2}\right)\tau_n\| {\boldsymbol{\alpha}}_1\|^2_1}{1+\left(1+\frac{1}{C_2}\right)\tau_n\| {\boldsymbol{\alpha}}_1\|^2_1}=\frac{  (1+\frac{\lambda_0}{2}) \tau_n\| {\boldsymbol{\alpha}}_1\|^2_1}{1+\left(1+\frac{1}{C_2}\right)\tau_n\| {\boldsymbol{\alpha}}_1\|^2_1}.\label{10018}
\end{align}
By $\eqref{1270}$ and the conditions in the theorem,  when $n$ is large enough, we have $\lambda_0< \lambda_n<1$ and $ \tau_n\|\boldsymbol{\alpha}_1\|^2_1=C\|\boldsymbol{\alpha}_1\|^2_1s_n\to 0$. Therefore, for all $n$ large enough, by $\eqref{10018}$, we have  
  \begin{align}
  \|\widehat{\boldsymbol{\alpha}}_1\|^2_1\le 6\|\boldsymbol{\alpha}_1\|^2_1/\lambda_0,\label{1081}
\end{align}
which together with $\eqref{1064}$ give
\begin{align}
  & \frac{\widehat{\boldsymbol{\alpha}}_1\trans \widehat{\mathbf{B}}\widehat{\boldsymbol{\alpha}}_1}{\boldsymbol{\alpha}_1\trans  {\mathbf{B}}\boldsymbol{\alpha}_1 }\le 1-\tau_n(\lambda_n-\lambda_0/2)\|\widehat{\boldsymbol{\alpha}}_1\|^2_1\le 1-\frac{1}{2}\lambda_0 \tau_n \|\widehat{\boldsymbol{\alpha}}_1\|^2_1\le 1.\label{1082}
\end{align}
  On the other hand,  $\eqref{1080}$  implies
\begin{align*}
  & \frac{\widehat{\boldsymbol{\alpha}}_1\trans \widehat{\mathbf{B}}\widehat{\boldsymbol{\alpha}}_1}{\boldsymbol{\alpha}_1\trans  {\mathbf{B}}\boldsymbol{\alpha}_1 }\ge \frac{1-\frac{c_1^{-1}}{C_2}\tau_n \|\boldsymbol{\alpha}_1\|^2_1}{1+(\frac{1}{C_2}+1)\|\boldsymbol{\alpha}_1\|^2_1}\ge \left(1-\frac{c_1^{-1}}{C_2}\tau_n \|\boldsymbol{\alpha}_1\|^2_1\right)\left(1-(1+\frac{1}{C_2})\tau_n \|\boldsymbol{\alpha}_1\|^2_1 \right)\\
	&\ge 1-\left(1+\frac{1+c_1^{-1}}{C_2}\right)\tau_n\| {\boldsymbol{\alpha}}_1\|^2_1=1-\left(1+\lambda_0/2\right)\tau_n\| {\boldsymbol{\alpha}}_1\|^2_1\ge 1-3\tau_n\| {\boldsymbol{\alpha}}_1\|^2_1/2,
\end{align*}
which together with $\eqref{1082}$ leads to 
\begin{align}
  & \left|\frac{\widehat{\boldsymbol{\alpha}}_1\trans \widehat{\mathbf{B}}\widehat{\boldsymbol{\alpha}}_1}{\boldsymbol{\alpha}_1\trans  {\mathbf{B}}\boldsymbol{\alpha}_1 }- 1\right|\le 3\tau_n \| \boldsymbol{\alpha}_1\|^2_1/2=3C \| \boldsymbol{\alpha}_1\|^2_1s_n/2.\label{1083}
\end{align}
It follows from $\eqref{1017}$ and $\eqref{1081}$,
\begin{align*}
  &|\widehat{\boldsymbol{\alpha}}_1\trans \widehat{\mathbf{B}}\widehat{\boldsymbol{\alpha}}_1- \widehat{\boldsymbol{\alpha}}_1\trans  \mathbf{B}\widehat{\boldsymbol{\alpha}}_1|\le  \frac{1}{C_2}\tau_n\|\widehat{\boldsymbol{\alpha}}_1\|^2_1\le \frac{(\boldsymbol{\alpha}_1\trans \mathbf{B}\boldsymbol{\alpha}_1)}{c_1} \frac{1}{C_2}\tau_n\|\widehat{\boldsymbol{\alpha}}_1\|^2_1\notag\\
	\le&   (\boldsymbol{\alpha}_1\trans \mathbf{B}\boldsymbol{\alpha}_1) \frac{6Cc_1^{-1}}{\lambda_0 C_2}\|{\boldsymbol{\alpha}}_1\|^2_1s_n, 
\end{align*}
which together with $\eqref{1083}$ imply
\begin{align}
    |\boldsymbol{\alpha}_1\trans \mathbf{B}\boldsymbol{\alpha}_1-\widehat{\boldsymbol{\alpha}}_1\trans  {\mathbf{B}}\widehat{\boldsymbol{\alpha}}_1|\le (\boldsymbol{\alpha}_1\trans \mathbf{B}\boldsymbol{\alpha}_1)C_3\|\boldsymbol{\alpha}_1\|^2_1s_n,\label{1044}
\end{align}
  where $C_3=3C/2+6Cc_1^{-1}/(\lambda_0C_2)$.  Recall that  $\widehat{\boldsymbol{\gamma}}_k=\boldsymbol{\Sigma}^{1/2}\widehat{\boldsymbol{\alpha}}_k$, $1\le k\le K-1$, defined in Section \ref{sec.4}}. Let $\widehat{\boldsymbol{\gamma}}_1=d_1\boldsymbol{\gamma}_1+d_2\boldsymbol{\gamma}_2+\cdots+d_{K-1}\boldsymbol{\gamma}_{K-1}+\widehat{c}\widehat{\boldsymbol{\beta}}$ be the orthogonal expansion of $\widehat{\boldsymbol{\gamma}}_1$, where $\widehat{\boldsymbol{\beta}}$ is an vector orthogonal to each of $\boldsymbol{\gamma}_1$, $\cdots$, $\boldsymbol{\gamma}_{K-1}$, with $\|\widehat{\boldsymbol{\beta}}\|_2=1$. Because $\boldsymbol{\Xi}$ has only $K-1$ nonzero eigenvalues with the corresponding eigenvectors,   $\boldsymbol{\gamma}_1$, $\cdots$, $\boldsymbol{\gamma}_{K-1}$, we have $\boldsymbol{\Xi}\widehat{\boldsymbol{\beta}}=\mathbf{0}$. Then $\widehat{\boldsymbol{\alpha}}_1\trans  {\mathbf{B}}\widehat{\boldsymbol{\alpha}}_1= \widehat{\boldsymbol{\gamma}}_1\trans\boldsymbol{\Xi}\widehat{\boldsymbol{\gamma}}_1=d_1^2\lambda_1(\boldsymbol{\Xi})+d_2^2\lambda_2(\boldsymbol{\Xi})+\cdots+d_{K-1}^2\lambda_{K-1}(\boldsymbol{\Xi})$. By $\eqref{1044}$ and $\eqref{1000017}$,
 \begin{align}
    &\lambda_1(\boldsymbol{\Xi})C_3\|\boldsymbol{\alpha}_1\|_1^2s_n=(\boldsymbol{\alpha}_1\trans \mathbf{B}\boldsymbol{\alpha}_1)C_3\|\boldsymbol{\alpha}_1\|_1^2s_n\ge |\widehat{\boldsymbol{\alpha}}_1\trans  {\mathbf{B}}\widehat{\boldsymbol{\alpha}}_1-\boldsymbol{\alpha}_1\trans \mathbf{B}\boldsymbol{\alpha}_1|\notag\\
		=&\biggr\rvert d_1^2\lambda_1(\boldsymbol{\Xi})+d_2^2\lambda_2(\boldsymbol{\Xi})+\cdots+d_{K-1}^2\lambda_{K-1}(\boldsymbol{\Xi})-\lambda_1(\boldsymbol{\Xi})\biggr\rvert\ge\biggr\rvert d_1^2-1\biggr\rvert\lambda_1(\boldsymbol{\Xi})-\lambda_2(\boldsymbol{\Xi})\sum_{i=2}^{K-1}d_i^2 \notag\\
		\ge&\biggr\rvert d_1^2-1\biggr\rvert\lambda_1(\boldsymbol{\Xi})-\biggr\rvert d_1^2-1\biggr\rvert \lambda_2(\boldsymbol{\Xi})- (d_1^2-1)\lambda_2(\boldsymbol{\Xi})-\lambda_2(\boldsymbol{\Xi})\sum_{i=2}^{K-1}d_i^2 \notag\\
	=& \biggr\rvert d_1^2-1\biggr\rvert[\lambda_1(\boldsymbol{\Xi})-\lambda_2(\boldsymbol{\Xi})]-\lambda_2(\boldsymbol{\Xi})\left[\sum_{i=1}^{K-1}d_i^2-1\right]\notag\\
	\ge&\biggr\rvert d_1^2-1\biggr\rvert[\lambda_1(\boldsymbol{\Xi})-\lambda_2(\boldsymbol{\Xi})]-\lambda_2(\boldsymbol{\Xi})\left(\|\widehat{\boldsymbol{\gamma}}_1\|_2^2-1\right)\notag\\
	\ge& \biggr\rvert d_1^2-1\biggr\rvert[\lambda_1(\boldsymbol{\Xi})-\lambda_2(\boldsymbol{\Xi})]-\lambda_1(\boldsymbol{\Xi})\left|\|\widehat{\boldsymbol{\gamma}}_1\|_2^2-1\right|.\label{1050}
\end{align} 
 By  $\eqref{1081}$ and  $\eqref{1017}$,
\begin{align}
  &\biggr\rvert\|\widehat{\boldsymbol{\gamma}}_1\|_2^2-1\biggr\rvert=\biggr\rvert \widehat{\boldsymbol{\gamma}}_1\trans\widehat{\boldsymbol{\gamma}}_1-1\biggr\rvert=\biggr\rvert\widehat{\boldsymbol{\alpha}}_1\trans\boldsymbol{\Sigma}\widehat{\boldsymbol{\alpha}}_1-1\biggr\rvert\le\biggr\rvert\widehat{\boldsymbol{\alpha}}_1\trans\widehat{\boldsymbol{\Sigma}}\widehat{\boldsymbol{\alpha}}_1-1\biggr\rvert+ \|\widehat{\boldsymbol{\Sigma}}-\boldsymbol{\Sigma}\|_{\infty}\|\widehat{\boldsymbol{\alpha}}_1\|^2_1\notag\\
	\le&\tau_n\|\widehat{\boldsymbol{\alpha}}_1\|^2_{\lambda_n}+\frac{1}{C_2}\tau_n\|\widehat{\boldsymbol{\alpha}}_1\|^2_1\le (1+\frac{1}{C_2})\tau_n\|\widehat{\boldsymbol{\alpha}}_1\|^2_1 =6(1+C_2^{-1})C\|\boldsymbol{\alpha}_1\|_1^2s_n/\lambda_0.\label{1085}
\end{align}
Then by $\eqref{1050}$, $\eqref{1085}$ and Condition \ref{condition_2} (b),
\begin{align}
   \biggr\rvert d_1^2-1\biggr\rvert\le \left(\frac{\lambda_1(\boldsymbol{\Xi})-\lambda_2(\boldsymbol{\Xi})}{\lambda_1(\boldsymbol{\Xi})}\right)^{-1}[C_3+6(1+C_2^{-1})C/\lambda_0]\|\boldsymbol{\alpha}_1\|_1^2s_n\le C_4\|\boldsymbol{\alpha}_1\|_1^2s_n,\label{1051}
\end{align}
where $C_4=c_2^{-1}[C_3+6(1+C_2^{-1})C/\lambda_0]$.  Since $\widehat{\boldsymbol{\gamma}}_1\trans\boldsymbol{\gamma}_1=d_1>0$, by $\eqref{1051}$,
 \begin{align}
   \biggr\rvert \widehat{\boldsymbol{\gamma}}_1\trans\boldsymbol{\gamma}_1-\| \boldsymbol{\gamma}_1\|_2^2\biggr\rvert=\biggr\rvert d_1-1\biggr\rvert\le  \biggr\rvert d_1-1\biggr\rvert(d_1+1)=\biggr\rvert d_1^2-1\biggr\rvert\le C_4\|\boldsymbol{\alpha}_1\|_1^2s_n.\label{1052}
\end{align}
Now by combining $\eqref{1085}$ and $\eqref{1052}$, we obtain
 \begin{align*}
   \|\widehat{\boldsymbol{\gamma}}_1-\boldsymbol{\gamma}_1\|_2^2=\biggr\rvert \| \widehat{\boldsymbol{\gamma}}_1\|_2^2-2\widehat{\boldsymbol{\gamma}}_1\trans\boldsymbol{\gamma}_1+\| \boldsymbol{\gamma}_1\|_2^2\biggr\rvert\le \biggr\rvert \| \widehat{\boldsymbol{\gamma}}_1\|_2^2- \| \boldsymbol{\gamma}_1\|_2^2\biggr\rvert+\biggr\rvert    -2\widehat{\boldsymbol{\gamma}}_1\trans\boldsymbol{\gamma}_1+2\| \boldsymbol{\gamma}_1\|_2^2\biggr\rvert\notag\\
	=\biggr\rvert\|\widehat{\boldsymbol{\gamma}}_1\|_2^2-1\biggr\rvert+2\biggr\rvert \widehat{\boldsymbol{\gamma}}_1\trans\boldsymbol{\gamma}_1-\| \boldsymbol{\gamma}_1\|_2^2\biggr\rvert\le  C_5\|\boldsymbol{\alpha}_1\|_1^2s_n, 
\end{align*}
where $C_5=6(1+C_2^{-1})C/\lambda_0+2C_4$. Moreover, by Condition \ref{condition_1}, it follows from the above inequality
 \begin{align*}
   &\|\widehat{\boldsymbol{\alpha}}_1-\boldsymbol{\alpha}_1\|_2^2=(\widehat{\boldsymbol{\alpha}}_1-\boldsymbol{\alpha}_1)\trans(\widehat{\boldsymbol{\alpha}}_1-\boldsymbol{\alpha}_1)=(\widehat{\boldsymbol{\gamma}}_1-\boldsymbol{\gamma}_1)\trans\boldsymbol{\Sigma}^{-1}(\widehat{\boldsymbol{\gamma}}_1-\boldsymbol{\gamma}_1)\\
	\le& \|\boldsymbol{\Sigma}^{-1}\|\|\widehat{\boldsymbol{\gamma}}_1-\boldsymbol{\gamma}_1\|_2^2\le c_0C_5\|\boldsymbol{\alpha}_1\|_1^2s_n.
\end{align*} 
We have proved the theorem.\\

\subsection{Proof of Theorem  $\ref{theorem_14}$}
 
When $K=2$, a new observation $\mathbf{x}$ is assigned to Class 1 by the optimal rule $\eqref{1170}$ if and only if $(\boldsymbol{\mu}_2-\boldsymbol{\mu}_1)\trans\mathbf{D}[\mathbf{x}-(\boldsymbol{\mu}_2+\boldsymbol{\mu}_1)/2]<0$. Hence,
\begin{align}
&P\left(\mathbf{x} \text{ is assigned to Class 1}|\mathbf{x}\in \text{ Class 2}\right)\label{1175}\\
=&P\left((\boldsymbol{\mu}_2-\boldsymbol{\mu}_1)\trans\mathbf{D}[\mathbf{x}-(\boldsymbol{\mu}_2+\boldsymbol{\mu}_1)/2]<0|\mathbf{x}\in \text{ Class 2}\right)\notag\\
=&P\left( \boldsymbol{\delta}\trans\mathbf{D}\boldsymbol{\Sigma}^{1/2}\boldsymbol{\Sigma}^{-1/2}[\mathbf{x}-\boldsymbol{\mu}_2]<-\frac{\boldsymbol{\delta}\trans\mathbf{D}\boldsymbol{\delta}}{2}|\mathbf{x}\in \text{ Class 2}\right)\notag\\
=&P\left( \boldsymbol{\delta}\trans\mathbf{D}\boldsymbol{\Sigma}^{1/2}\mathbf{Z}<-\frac{\boldsymbol{\delta}\trans\mathbf{D}\boldsymbol{\delta}}{2} \right),\notag
\end{align}
where $\mathbf{Z}=\boldsymbol{\Sigma}^{-1/2}[\mathbf{x}-\boldsymbol{\mu}_2]\sim N(\mathbf{0},\mathbf{I})$ given $\mathbf{x}$ in Class 2. Hence, the above probability is equal to $\Phi(-\frac{\boldsymbol{\delta}\trans\mathbf{D}\boldsymbol{\delta}}{2\|\boldsymbol{\delta}\trans\mathbf{D}\boldsymbol{\Sigma}^{1/2}\|_2})$. The same result is true for 
$$P\left(\mathbf{x} \text{ is assigned to Class 2}|\mathbf{x}\in \text{ Class 1}\right).$$ 
Hence,
\begin{align*}
  &R_{OPT}=\frac{1}{2}P\left(\mathbf{x} \text{ is assigned to Class 1}|\mathbf{x}\in \text{ Class 2}\right)\\
	&+\frac{1}{2}P\left(\mathbf{x} \text{ is assigned to Class 2}|\mathbf{x}\in \text{ Class 1}\right)=\Phi\left(-\frac{\boldsymbol{\delta}\trans\mathbf{D}\boldsymbol{\delta}}{2\|\boldsymbol{\delta}\trans\mathbf{D}\boldsymbol{\Sigma}^{1/2}\|_2}\right).
\end{align*}
On the other hand, a new observation $\mathbf{x}$ is assigned to Class 1 by sparse LDA rule $\eqref{1174}$ if and only if $(\bar{\mathbf{x}}_2-\bar{\mathbf{x}}_1)\trans\widehat{\mathbf{D}}[\mathbf{x}-(\bar{\mathbf{x}}_1+\bar{\mathbf{x}}_2)/2]<0$. A similar argument as in $\eqref{1175}$ leads to 
\begin{align*}
&P_{\cdot|\mathbf{X}}\left(\mathbf{x} \text{ is assigned to Class 1}|\mathbf{x}\in \text{ Class 2}\right)=\Phi\left(-\frac{\widehat{\boldsymbol{\delta}}\trans\widehat{\mathbf{D}}[2 \boldsymbol{\mu}_2-\bar{\mathbf{x}}_1-\bar{\mathbf{x}}_2]}{2\|\widehat{\boldsymbol{\delta}}\trans\widehat{\mathbf{D}}\boldsymbol{\Sigma}^{1/2}\|_2}\right),
\end{align*}
where $P_{\cdot|\mathbf{X}}$ means the probability given the training sample $\mathbf{X}$. Similarly,
\begin{align*}
&P_{\cdot|\mathbf{X}}\left(\mathbf{x} \text{ is assigned to Class 2}|\mathbf{x}\in \text{ Class 1}\right)=\Phi\left(-\frac{\widehat{\boldsymbol{\delta}}\trans\widehat{\mathbf{D}}[\bar{\mathbf{x}}_1+\bar{\mathbf{x}}_2-2 \boldsymbol{\mu}_1]}{2\|\widehat{\boldsymbol{\delta}}\trans\widehat{\mathbf{D}}\boldsymbol{\Sigma}^{1/2}\|_2}\right).
\end{align*}
Then $\eqref{1176}$ follows. We will use the following inequality (see page 850 of \citet{shorack2009empirical}):
\begin{align}
(1-\frac{1}{x^2})\phi(x)\le x[1-\Phi(x)]=x\Phi(-x)\le \phi(x),\quad \forall x>0,\label{10020}
\end{align}  
where $\phi$ is the density function of the standard normal distribution. Therefore, if $x>\sqrt{2}$,
\begin{align}
\Phi(-x)\ge (1-\frac{1}{x^2})\phi(x)x^{-1}\ge \frac{1}{2}\phi(x)x^{-1},\label{10021}
\end{align}
 if $0<x\le \sqrt{2}$, $\Phi(-x)\ge \Phi(-\sqrt{2})\ge \Phi(-\sqrt{2})\phi(0)^{-1}\phi(x)$. Hence, we have for any $x>0$,
\begin{align}
\Phi(-x)\ge \frac{C_1}{1+2C_1x}\phi(x), \quad \text{where }\quad C_1= \Phi(-\sqrt{2})\phi(0)^{-1} \label{10022}
\end{align}  
By $\eqref{10022}$, for any $x>0$ and $\epsilon$ with $x+\epsilon>0$ ($\epsilon$ can be negative or positive),
\begin{align}
 \left|\frac{\Phi(-(x+\epsilon))}{\Phi(-x)}-1\right|=\frac{\left|\Phi(-(x+\epsilon))-\Phi(-x)\right|}{\Phi(-x)}=\frac{\left|\int_{-x}^{-(x+\epsilon)}\phi(y)dy\right|}{\Phi(-x)}\notag\\
= \frac{\left|-\epsilon\phi(-(x+\widetilde{\epsilon}))\right|}{\Phi(-x)}\le \frac{(1+2C_1x)|\epsilon|\phi(-(x+\widetilde{\epsilon}))}{ C_1\phi(x)}= \frac{ (1+2C_1x)|\epsilon|  }{C_1 }e^{-\frac{(x+\widetilde{\epsilon})^2-x^2}{2}}\notag\\
= \frac{ (1+2C_1x)|\epsilon|  }{C_1 }e^{-\frac{2x\widetilde{\epsilon}+\widetilde{\epsilon}^2}{2}}\le \frac{(1+2C_1x)|\epsilon|}{C_1 }e^{x|\epsilon|},\label{1177}
\end{align}
where $\widetilde{\epsilon}$  is a number between $0$ and $\epsilon$. We will apply $\eqref{1177}$ to 
 \begin{align}
  &x=\frac{\boldsymbol{\delta}\trans\mathbf{D}\boldsymbol{\delta}}{2\|\boldsymbol{\delta}\trans\mathbf{D}\boldsymbol{\Sigma}^{1/2}\|_2}, \quad \epsilon=\frac{\widehat{\boldsymbol{\delta}}\trans\widehat{\mathbf{D}}[2 \boldsymbol{\mu}_2-\bar{\mathbf{x}}_1-\bar{\mathbf{x}}_2]}{2\|\widehat{\boldsymbol{\delta}}\trans\widehat{\mathbf{D}}\boldsymbol{\Sigma}^{1/2}\|_2}-\frac{\boldsymbol{\delta}\trans\mathbf{D}\boldsymbol{\delta}}{2\|\boldsymbol{\delta}\trans\mathbf{D}\boldsymbol{\Sigma}^{1/2}\|_2}.\label{1191}
\end{align} 
By Lemma \ref{lemma_2}, we can choose a constant $\widetilde{C}$ such that
\begin{align*}
& P\left(\max_{1\le j\le K}\|\bar{\mathbf{x}}_j-\boldsymbol{\mu}_j\|_{\infty}> \widetilde{C}\sqrt{\frac{K\log{p}}{n}}\right)\le p^{-1}, 
\end{align*}
 Define
\begin{align}
\widetilde{\Omega}_n=\left\{\max_{1\le j\le K}\|\bar{\mathbf{x}}_j-\boldsymbol{\mu}_j\|_{\infty}\le  \widetilde{C}\sqrt{\frac{K\log{p}}{n}}=\widetilde{C}s_n\right\}, \text{ then } P\left(\widetilde{\Omega}_n\right)\ge 1-p^{-1}.\label{1184}
\end{align}
In the rest of the proof, we only consider the elements in ${\Omega}_n\bigcap \widetilde{\Omega}_n$ which has a probability greater than $1-3p^{-1}$ by $\eqref{1033}$ and $\eqref{1184}$.  Note that by $\eqref{3}$, we have $\boldsymbol{\mu}_1=-\boldsymbol{\mu}_2$. Because
 \begin{align*} 
  & \widehat{\boldsymbol{\delta}}\trans\widehat{\mathbf{D}}[2 \boldsymbol{\mu}_2-\bar{\mathbf{x}}_1-\bar{\mathbf{x}}_2]=[(\bar{\mathbf{x}}_2-\boldsymbol{\mu}_2)-(\bar{\mathbf{x}}_1-\boldsymbol{\mu}_1)+2\boldsymbol{\mu}_2]\trans\widehat{\mathbf{D}}[2 \boldsymbol{\mu}_2-(\bar{\mathbf{x}}_1-\boldsymbol{\mu}_1)-(\bar{\mathbf{x}}_2-\boldsymbol{\mu}_2)]\\
	=&4\boldsymbol{\mu}_2\trans\widehat{\mathbf{D}} \boldsymbol{\mu}_2-(\bar{\mathbf{x}}_2-\boldsymbol{\mu}_2)\trans\widehat{\mathbf{D}} (\bar{\mathbf{x}}_2-\boldsymbol{\mu}_2)+(\bar{\mathbf{x}}_1-\boldsymbol{\mu}_1)\trans\widehat{\mathbf{D}} (\bar{\mathbf{x}}_1-\boldsymbol{\mu}_1)-4(\bar{\mathbf{x}}_1-\boldsymbol{\mu}_1)\trans\widehat{\mathbf{D}}\boldsymbol{\mu}_2,
\end{align*} 
we have
 \begin{align} 
&\left|\widehat{\boldsymbol{\delta}}\trans\widehat{\mathbf{D}}[2 \boldsymbol{\mu}_2-\bar{\mathbf{x}}_1-\bar{\mathbf{x}}_2]-\boldsymbol{\delta}\trans\mathbf{D}\boldsymbol{\delta}\right|\le \left|4\boldsymbol{\mu}_2\trans\widehat{\mathbf{D}} \boldsymbol{\mu}_2-\boldsymbol{\delta}\trans\mathbf{D}\boldsymbol{\delta}\right|+(\bar{\mathbf{x}}_2-\boldsymbol{\mu}_2)\trans\widehat{\mathbf{D}} (\bar{\mathbf{x}}_2-\boldsymbol{\mu}_2)\notag\\
&+(\bar{\mathbf{x}}_1-\boldsymbol{\mu}_1)\trans\widehat{\mathbf{D}} (\bar{\mathbf{x}}_1-\boldsymbol{\mu}_1)+4\left| (\bar{\mathbf{x}}_1-\boldsymbol{\mu}_1) \trans\widehat{\mathbf{D}}\boldsymbol{\mu}_2\right|=I+II+III+IV.\label{1280}
\end{align} 
We estimate each of the four terms. Because $\boldsymbol{\delta}=\boldsymbol{\mu}_2-\boldsymbol{\mu}_1=2\boldsymbol{\mu}_2$, by $\eqref{1044}$, the first term
 \begin{align} 
& I=\left|4\boldsymbol{\mu}_2\trans\widehat{\mathbf{D}} \boldsymbol{\mu}_2-\boldsymbol{\delta}\trans\mathbf{D}\boldsymbol{\delta}\right|=\left|4\boldsymbol{\mu}_2\trans\widehat{\boldsymbol{\alpha}}_1\widehat{\boldsymbol{\alpha}}_1\trans\boldsymbol{\mu}_2-4\boldsymbol{\mu}_2\trans\boldsymbol{\alpha}_1\boldsymbol{\alpha}_1\trans\boldsymbol{\mu}_2\right|\notag\\
=& \left|4\widehat{\boldsymbol{\alpha}}_1\trans\boldsymbol{\mu}_2\boldsymbol{\mu}_2\trans\widehat{\boldsymbol{\alpha}}_1-4\boldsymbol{\alpha}_1\trans\boldsymbol{\mu}_2\boldsymbol{\mu}_2\trans\boldsymbol{\alpha}_1\right|=4\left|\widehat{\boldsymbol{\alpha}}_1\trans\mathbf{B} \widehat{\boldsymbol{\alpha}}_1-\boldsymbol{\alpha}_1\trans\mathbf{B} \boldsymbol{\alpha}_1\right|\notag\\
\le& 4C_3\boldsymbol{\alpha}_1\trans\mathbf{B} \boldsymbol{\alpha}_1\|\boldsymbol{\alpha}_1\|^2_1s_n= 4C_3\lambda_1(\boldsymbol{\Xi})\|\boldsymbol{\alpha}_1\|^2_1s_n,\label{1181}
\end{align} 
and  we have
 \begin{align} 
&  \boldsymbol{\delta}\trans\mathbf{D}\boldsymbol{\delta}=4\boldsymbol{\alpha}_1\trans\mathbf{B} \boldsymbol{\alpha}_1=4\lambda_1(\boldsymbol{\Xi}).\label{1187}
\end{align} 
For the second term, by the definition of $\widetilde{\Omega}_n$ in $\eqref{1184}$ and Theorem \ref{theorem_12}
 \begin{align} 
& II=(\bar{\mathbf{x}}_2-\boldsymbol{\mu}_2)\trans\widehat{\mathbf{D}} (\bar{\mathbf{x}}_2-\boldsymbol{\mu}_2)=\left|(\bar{\mathbf{x}}_2-\boldsymbol{\mu}_2)\trans\widehat{\boldsymbol{\alpha}}_1\right|^2\le\|\bar{\mathbf{x}}_2-\boldsymbol{\mu}_2\|_{\infty}^2\|\widehat{\boldsymbol{\alpha}}_1\|^2_1\notag\\
\le& \widetilde{C}^2s_n^26\|\boldsymbol{\alpha}_1\|^2_1/\lambda_0 .\label{1185}
\end{align} 
The same bound for the third term. For the last one, by  Condition \ref{condition_1} (b),
 \begin{align} 
& IV =4\left|  (\bar{\mathbf{x}}_1-\boldsymbol{\mu}_1) \trans\widehat{\boldsymbol{\alpha}}_1\right|\left| \widehat{\boldsymbol{\alpha}}_1\trans\boldsymbol{\mu}_2\right|\le 4\left[\max_{1\le i\le K}\|\bar{\mathbf{x}}_i-\boldsymbol{\mu}_i\|_{\infty}\right]\|\widehat{\boldsymbol{\alpha}}_1\|_1\|\widehat{\boldsymbol{\alpha}}_1\|_1\|\boldsymbol{\mu}_2\|_{\infty}\notag\\
\le& 24c_0\widetilde{C}s_n\|\boldsymbol{\alpha}_1\|^2_1/\lambda_0 .\label{1186}
\end{align} 
By $\eqref{1280}$-$\eqref{1186}$, $s_n\to 0$ and $\lambda_1(\boldsymbol{\Xi})\ge c_1$ (see Condition \ref{condition_2}), we have
 \begin{align} 
&\left|\widehat{\boldsymbol{\delta}}\trans\widehat{\mathbf{D}}[2 \boldsymbol{\mu}_2-\bar{\mathbf{x}}_1-\bar{\mathbf{x}}_2]-\boldsymbol{\delta}\trans\mathbf{D}\boldsymbol{\delta}\right|\le C_4\lambda_1(\boldsymbol{\Xi})\|\boldsymbol{\alpha}_1\|^2_1s_n,\label{1180}
\end{align} 
where $C_4$ is a constant independent of $n$ and $p$.

Next, by $\eqref{1085}$ and $\boldsymbol{\alpha}_1\trans\boldsymbol{\Sigma}\boldsymbol{\alpha}_1=1$,
 \begin{align}
  & \left|\|\widehat{\boldsymbol{\delta}}\trans\widehat{\mathbf{D}}\boldsymbol{\Sigma}^{1/2}\|_2^2- \|\boldsymbol{\delta}\trans\mathbf{D}\boldsymbol{\Sigma}^{1/2}\|_2^2\right|=\left|\widehat{\boldsymbol{\delta}}\trans\widehat{\mathbf{D}}\boldsymbol{\Sigma}\widehat{\mathbf{D}}\widehat{\boldsymbol{\delta}}-\boldsymbol{\delta}\trans\mathbf{D}\boldsymbol{\Sigma}\mathbf{D}\boldsymbol{\delta}\right|\label{1190}\\
	=&\left|\widehat{\boldsymbol{\delta}}\trans\widehat{\boldsymbol{\alpha}}_1\widehat{\boldsymbol{\alpha}}_1\trans\boldsymbol{\Sigma}\widehat{\boldsymbol{\alpha}}_1\widehat{\boldsymbol{\alpha}}_1\trans\widehat{\boldsymbol{\delta}}-\boldsymbol{\delta}\trans\boldsymbol{\alpha}_1\boldsymbol{\alpha}_1\trans\boldsymbol{\Sigma}\boldsymbol{\alpha}_1\boldsymbol{\alpha}_1\trans\boldsymbol{\delta}\right|=|\widehat{\boldsymbol{\alpha}}_1\trans\boldsymbol{\Sigma}\widehat{\boldsymbol{\alpha}}_1-1|\widehat{\boldsymbol{\delta}}\trans\widehat{\mathbf{D}}\widehat{\boldsymbol{\delta}}+|\widehat{\boldsymbol{\delta}}\trans\widehat{\mathbf{D}}\widehat{\boldsymbol{\delta}}- \boldsymbol{\delta}\trans\mathbf{D}\boldsymbol{\delta}|\notag\\
	=&|\widehat{\boldsymbol{\gamma}}_1\trans\widehat{\boldsymbol{\gamma}}_1-1|\widehat{\boldsymbol{\delta}}\trans\widehat{\mathbf{D}}\widehat{\boldsymbol{\delta}}+|\widehat{\boldsymbol{\delta}}\trans\widehat{\mathbf{D}}\widehat{\boldsymbol{\delta}}- \boldsymbol{\delta}\trans\mathbf{D}\boldsymbol{\delta}|\notag\\
	\le& 6(1+C_2^{-1})C\lambda_0^{-1}\|\boldsymbol{\alpha}_1\|^2_1s_n \widehat{\boldsymbol{\delta}}\trans\widehat{\mathbf{D}}\widehat{\boldsymbol{\delta}}+|\widehat{\boldsymbol{\delta}}\trans\widehat{\mathbf{D}}\widehat{\boldsymbol{\delta}}- \boldsymbol{\delta}\trans\mathbf{D}\boldsymbol{\delta}|.\notag
\end{align} 
By a similar argument as those for $\eqref{1180}$, we can show that 
 \begin{align}
  &   |\widehat{\boldsymbol{\delta}}\trans\widehat{\mathbf{D}}\widehat{\boldsymbol{\delta}}- \boldsymbol{\delta}\trans\mathbf{D}\boldsymbol{\delta}|\le C_5\lambda_1(\boldsymbol{\Xi})\|\boldsymbol{\alpha}_1\|^2_1s_n, \notag\\
	&\widehat{\boldsymbol{\delta}}\trans\widehat{\mathbf{D}}\widehat{\boldsymbol{\delta}}=(1+o(1))\boldsymbol{\delta}\trans\mathbf{D}\boldsymbol{\delta}=(1+o(1))4\lambda_1(\boldsymbol{\Xi}), \label{1189}
\end{align} 
 where the last equality is due to $\eqref{1187}$ and $C_5$ is a constant independent of $n$ and $p$. By Lemma \ref{lemma_3} and $\eqref{1187}$, we have  $\|\boldsymbol{\delta}\trans\mathbf{D}\boldsymbol{\Sigma}^{1/2}\|_2=\sqrt{\boldsymbol{\delta}\trans\mathbf{D}\boldsymbol{\delta}}=\sqrt{4\lambda_1(\boldsymbol{\Xi})}$ which together with  $\eqref{1190}$, $\eqref{1189}$ give 
 \begin{align*}
&\left|\|\widehat{\boldsymbol{\delta}}\trans\widehat{\mathbf{D}}\boldsymbol{\Sigma}^{1/2}\|_2^2- \|\boldsymbol{\delta}\trans\mathbf{D}\boldsymbol{\Sigma}^{1/2}\|_2^2\right|\le \widetilde{C_5}\lambda_1(\boldsymbol{\Xi})\|\boldsymbol{\alpha}_1\|^2_1s_n,\quad \|\widehat{\boldsymbol{\delta}}\trans\widehat{\mathbf{D}}\boldsymbol{\Sigma}^{1/2}\|_2=\sqrt{4\lambda_1(\boldsymbol{\Xi})}+o(1),\notag\\
   &\text{ and }\quad \left|\frac{\boldsymbol{\delta}\trans\mathbf{D}\boldsymbol{\delta}}{2\|\widehat{\boldsymbol{\delta}}\trans\widehat{\mathbf{D}}\boldsymbol{\Sigma}^{1/2}\|_2}-\frac{\boldsymbol{\delta}\trans\mathbf{D}\boldsymbol{\delta}}{2\|\boldsymbol{\delta}\trans\mathbf{D}\boldsymbol{\Sigma}^{1/2}\|_2}\right|\notag\\
	=&|\boldsymbol{\delta}\trans\mathbf{D}\boldsymbol{\delta}|\frac{\left|\|\widehat{\boldsymbol{\delta}}\trans\widehat{\mathbf{D}}\boldsymbol{\Sigma}^{1/2}\|_2^2- \|\boldsymbol{\delta}\trans\mathbf{D}\boldsymbol{\Sigma}^{1/2}\|_2^2\right|}{2\|\widehat{\boldsymbol{\delta}}\trans\widehat{\mathbf{D}}\boldsymbol{\Sigma}^{1/2}\|_2\|\boldsymbol{\delta}\trans\mathbf{D}\boldsymbol{\Sigma}^{1/2}\|_2\left(\|\widehat{\boldsymbol{\delta}}\trans\widehat{\mathbf{D}}\boldsymbol{\Sigma}^{1/2}\|_2+\|\boldsymbol{\delta}\trans\mathbf{D}\boldsymbol{\Sigma}^{1/2}\|_2\right)}\le C_6\sqrt{\lambda_1(\boldsymbol{\Xi})}\|\boldsymbol{\alpha}_1\|^2_1s_n,
\end{align*} 
which together with $\eqref{1180}$ imply
 \begin{align}
  &|\epsilon|=\left|\frac{\widehat{\boldsymbol{\delta}}\trans\widehat{\mathbf{D}}[\widehat{\boldsymbol{\delta}}-2(\bar{\mathbf{x}}_2-\boldsymbol{\mu}_2)]}{2\|\widehat{\boldsymbol{\delta}}\trans\widehat{\mathbf{D}}\boldsymbol{\Sigma}^{1/2}\|_2}-\frac{\boldsymbol{\delta}\trans\mathbf{D}\boldsymbol{\delta}}{2\|\boldsymbol{\delta}\trans\mathbf{D}\boldsymbol{\Sigma}^{1/2}\|_2}\right|\le \frac{|\widehat{\boldsymbol{\delta}}\trans\widehat{\mathbf{D}}[\widehat{\boldsymbol{\delta}}-2(\bar{\mathbf{x}}_2-\boldsymbol{\mu}_2)]-\boldsymbol{\delta}\trans\mathbf{D}\boldsymbol{\delta}|}{2\|\widehat{\boldsymbol{\delta}}\trans\widehat{\mathbf{D}}\boldsymbol{\Sigma}^{1/2}\|_2}\notag\\
	+&\left|\frac{\boldsymbol{\delta}\trans\mathbf{D}\boldsymbol{\delta}}{2\|\widehat{\boldsymbol{\delta}}\trans\widehat{\mathbf{D}}\boldsymbol{\Sigma}^{1/2}\|_2}-\frac{\boldsymbol{\delta}\trans\mathbf{D}\boldsymbol{\delta}}{2\|\boldsymbol{\delta}\trans\mathbf{D}\boldsymbol{\Sigma}^{1/2}\|_2}\right|\le C_7\sqrt{\lambda_1(\boldsymbol{\Xi})}\|\boldsymbol{\alpha}_1\|^2_1s_n,\label{300000}
\end{align} 
 where $\widetilde{C_5}$, $C_6$ and $C_7$ are constants independent of $n$ and $p$. By $\eqref{1177}$, $\eqref{1191}$ and $\eqref{300000}$ and noting that $x=\sqrt{\lambda_1(\boldsymbol{\Xi})}\ge \sqrt{c_1}$, we have $|x\epsilon|\le C_7\lambda_1(\boldsymbol{\Xi})\|\boldsymbol{\alpha}_1\|^2_1s_n=o(1)$ and hence 
 \begin{align*}
  &\Phi\left(-\frac{\widehat{\boldsymbol{\delta}}\trans\widehat{\mathbf{D}}[\widehat{\boldsymbol{\delta}}-2(\bar{\mathbf{x}}_2-\boldsymbol{\mu}_2)]}{2\|\widehat{\boldsymbol{\delta}}\trans\widehat{\mathbf{D}}\boldsymbol{\Sigma}^{1/2}\|_2}\right)/\Phi\left(-\frac{\boldsymbol{\delta}\trans\mathbf{D}\boldsymbol{\delta}}{2\|\boldsymbol{\delta}\trans\mathbf{D}\boldsymbol{\Sigma}^{1/2}\|_2}\right)-1=\frac{\Phi(-(x+\epsilon))}{\Phi(-x)}-1\notag\\
	\le &\frac{(1+2C_1x)|\epsilon|}{C_1 }e^{x|\epsilon|}\le C_8\lambda_1(\boldsymbol{\Xi})\|\boldsymbol{\alpha}_1\|^2_1s_n 
\end{align*} 
where $C_8$ is a constant independent of $n$ and $p$. Similarly, we have 
 \begin{align*}
  &\Phi\left(-\frac{\widehat{\boldsymbol{\delta}}\trans\widehat{\mathbf{D}}[\widehat{\boldsymbol{\delta}}+2(\bar{\mathbf{x}}_1-\boldsymbol{\mu}_1)]}{2\|\widehat{\boldsymbol{\delta}}\trans\widehat{\mathbf{D}}\boldsymbol{\Sigma}^{1/2}\|_2}\right)/\Phi\left(-\frac{\boldsymbol{\delta}\trans\mathbf{D}\boldsymbol{\delta}}{2\|\boldsymbol{\delta}\trans\mathbf{D}\boldsymbol{\Sigma}^{1/2}\|_2}\right)-1  \le C_9\lambda_1(\boldsymbol{\Xi})\|\boldsymbol{\alpha}_1\|^2_1s_n.
\end{align*} 
where $C_9$ is a constant independent of $n$ and $p$. Therefore, the above two inequalities together with $\eqref{1176}$ give $\eqref{1193}$.\\

\subsection{Proof of Theorem  $\ref{theorem_10}$}
 
Due to the constraints of the optimization problems $\eqref{1095}$ and $\eqref{1003}$ and the definitions of $\mathbf{W}_i$ and $\widehat{\mathbf{W}}_i$ in $\eqref{10026}$, for any $1\le i\le K-1$ and $j<i$, we have  
\begin{align}
  &\boldsymbol{\alpha}_i\trans \boldsymbol{\Sigma}\boldsymbol{\alpha}_i=1,\quad    \widehat{\boldsymbol{\alpha}}_i\trans \widehat{\boldsymbol{\Sigma}}\widehat{\boldsymbol{\alpha}}_i +\tau_n\|\widehat{\boldsymbol{\alpha}}_i \|_{\lambda_n}=1,\notag\\
&\boldsymbol{\alpha}_i\trans \boldsymbol{\Sigma}\boldsymbol{\alpha}_j=0,\quad \boldsymbol{\alpha}_i\trans \boldsymbol{\xi}_j=0,\quad \widehat{\boldsymbol{\alpha}}_i\trans  \widehat{\boldsymbol{\xi}}_j=0.\label{1037}
\end{align}
 Then by the definitions of $\boldsymbol{\gamma}_k$ and $\widehat{\boldsymbol{\gamma}}_k$ in $\eqref{10023}$ and $\eqref{10024}$, for any $1\le i\le K-1$ and $j<i$, we have  
\begin{align}
  &\boldsymbol{\gamma}_i=\boldsymbol{\Sigma}^{1/2}\boldsymbol{\alpha}_i,\quad \|\boldsymbol{\gamma}_i\|_2=1 ,\quad \boldsymbol{\gamma}_i\trans \boldsymbol{\gamma}_j=0,\quad \widehat{\boldsymbol{\gamma}}_i=\boldsymbol{\Sigma}^{1/2}\widehat{\boldsymbol{\alpha}}_i,\quad      \widehat{\boldsymbol{\gamma}}_i\trans \boldsymbol{\Sigma}^{-1/2} \widehat{\boldsymbol{\xi}}_j=0.\label{10025}
\end{align}
 For any $1\le i\le K-1$, in addition to $\mathbf{W}_i$ and $\widehat{\mathbf{W}}_i$, we define the following two subspaces of $\mathbb{R}^{p}$,  
\begin{align*}
   \mathbf{V}_i=\text{span}\{\boldsymbol{\gamma}_1,\boldsymbol{\gamma}_2, \cdots, \boldsymbol{\gamma}_i\},\quad \widehat{\mathbf{V}}_i=\text{span}\{\widehat{\boldsymbol{\zeta}}_1,\widehat{\boldsymbol{\zeta}}_2, \cdots, \widehat{\boldsymbol{\zeta}}_i\}, 
\end{align*}
where  $\widehat{\boldsymbol{\zeta}}_i=\lambda_i(\boldsymbol{\Xi})^{-1}\boldsymbol{\Sigma}^{-1/2}\widehat{\boldsymbol{\xi}}_i$. Let $\mathbf{P}_i$ and $\widehat{\mathbf{P}}_i$ be the orthogonal projection matrices onto $\mathbf{V}_i$ and $\widehat{\mathbf{V}}_i$, respectively. By $\eqref{1037}$ and $\eqref{10025}$ and the definitions of $\mathbf{W}_i$ and $\widehat{\mathbf{W}}_i$ in $\eqref{10026}$,  
\begin{align}
  \boldsymbol{\gamma}_k\in \mathbf{V}_{i}^\perp,  \quad \boldsymbol{\alpha}_k\in \mathbf{W}_{i}^\perp,\quad \widehat{\boldsymbol{\alpha}}_k\in \widehat{\mathbf{W}}_{i}^\perp, \quad \widehat{\boldsymbol{\gamma}}_k\in \widehat{\mathbf{V}}_{i}^\perp,\quad \text{for any } k> i,\label{1154}
\end{align}
where $\mathbf{V}_i^\perp$, $\widehat{\mathbf{V}}_i^\perp$, $\mathbf{W}_i^\perp$ and $\widehat{\mathbf{W}}_i^\perp$ are  orthogonal complementary subspaces of $\mathbf{V}_i $, $\widehat{\mathbf{V}}_i $, $\mathbf{W}_i $ and $\widehat{\mathbf{W}}_i $, respectively. We will prove that in the event $\Omega_n$ (defined in $\eqref{1033}$), for each $1\le i\le K-1$, there exist constants, $C_{i,1}$, $C_{i,2}$, $C_{i,3}$, $C_{i,4}$, $C_{i,5}$ and $C_{i,6}$ independent of $n$ and $p$ such that
\begin{align}
   &\|\widehat{\boldsymbol{\alpha}}_i\|_1\le C_{i,1}\Lambda_p ,\quad \|\widehat{\boldsymbol{\gamma}}_i-\boldsymbol{\gamma}_i\|^2_2\le C_{i,2}\Lambda_p^2s_n, \quad\|\mathbf{P}_i-\widehat{\mathbf{P}}_i\|^2\le C_{i,3}\Lambda_p^2s_n,\notag\\
	&\|\mathbf{Q}_i-\widehat{\mathbf{Q}}_i\|^2\le C_{i,4}\Lambda_p^2s_n,\quad \|\widehat{\boldsymbol{\xi}}_i\|_1\le C_{i,5}\lambda_1(\boldsymbol{\Xi})\Lambda_p,\quad \|\widehat{\boldsymbol{\xi}}_i- \boldsymbol{\xi}_i\|^2_2 \le C_{i,6}\lambda_1(\boldsymbol{\Xi})^2\Lambda_p^2s_n,\label{1053}
\end{align}
as $n$ is large enough. We proceed by induction. When $i=1$, the first two inequalities  in $\eqref{1053}$ follow from $\eqref{1036}$ in Theorem \ref{theorem_12} by setting $C_{1,1}=6/\lambda_0$ and $C_{1,2}= C_5$, and the last two inequalities  follow from the following lemma by setting $C_{1,5}= C_7$ and $C_{1,6}= C_6^2$ in $\eqref{1126}$.

\begin{Lemma}\label{lemma_10}
Under the conditions of the theorem, we have, in $\Omega_n$, 
\begin{align}
 & \|\widehat{\boldsymbol{\xi}}_1- \boldsymbol{\xi}_1\| _2=\|\widehat{\boldsymbol{\xi}}_1- \mathbf{B}\boldsymbol{\alpha}_1\|_2\le C_6\lambda_1(\boldsymbol{\Xi})\sqrt{\Lambda_p^2s_n}, \quad  \|\widehat{\boldsymbol{\xi}}_1\|_1\le C_7\lambda_1(\boldsymbol{\Xi})\Lambda_p. \label{1126}
\end{align}
where $C_6$ and $C_7$ are constants independent of $p$ and $n$.
\end{Lemma}

 On the other hand, since $\|\boldsymbol{\gamma}_1\|_2=1$, we have $\mathbf{P}_1=\boldsymbol{\gamma}_1\boldsymbol{\gamma}_1\trans$ and $\widehat{\mathbf{P}}_1=\widehat{\boldsymbol{\zeta}}_1\widehat{\boldsymbol{\zeta}}_1\trans/\|\widehat{\boldsymbol{\zeta}}_1\|_2^2$. Then
 \begin{align}
    &\|\mathbf{P}_1-\widehat{\mathbf{P}}_1\|=\left\|\boldsymbol{\gamma}_1\boldsymbol{\gamma}_1\trans-\frac{1}{\|\widehat{\boldsymbol{\zeta}}_1\|_2^2}\widehat{\boldsymbol{\zeta}}_1\widehat{\boldsymbol{\zeta}}_1\trans\right\|\le  \|(\boldsymbol{\gamma}_1-\widehat{\boldsymbol{\zeta}}_1)\boldsymbol{\gamma}_1\trans\|+ \|\widehat{\boldsymbol{\zeta}}_1(\boldsymbol{\gamma}_1-\widehat{\boldsymbol{\zeta}}_1)\trans\|\notag\\
		&+\left\|\left(1-\frac{1}{\|\widehat{\boldsymbol{\zeta}}_1\|_2^2}\right)\widehat{\boldsymbol{\zeta}}_1\widehat{\boldsymbol{\zeta}}_1\trans\right\|\le \|\boldsymbol{\gamma}_1-\widehat{\boldsymbol{\zeta}}_1\|_2(\|\boldsymbol{\gamma}_1\|_2+\|\widehat{\boldsymbol{\zeta}}_1\|_2)+\left|1-\|\widehat{\boldsymbol{\zeta}}_1\|_2^2\right|.\label{1125}
\end{align} 
Note that by Condition \ref{condition_1} (b), $\|\boldsymbol{\Sigma}^{-1/2}\|=\lambda_{max}(\boldsymbol{\Sigma}^{-1/2})=\lambda_{min}(\boldsymbol{\Sigma})^{-1/2} \le c_0^{1/2}$. Then by $\eqref{1126}$,
\begin{align}
    & \|\boldsymbol{\gamma}_1-\widehat{\boldsymbol{\zeta}}_1\|_2 =\|\boldsymbol{\Sigma}^{1/2}\boldsymbol{\alpha}_1-\lambda_1(\boldsymbol{\Xi})^{-1}\boldsymbol{\Sigma}^{-1/2}\widehat{\boldsymbol{\xi}}_1\|_2\le \lambda_1(\boldsymbol{\Xi})^{-1}\|\boldsymbol{\Sigma}^{-1/2}\| \|\lambda_i(\boldsymbol{\Xi})\boldsymbol{\Sigma}\boldsymbol{\alpha}_1- \widehat{\boldsymbol{\xi}}_1\|_2\notag\\
		=&\lambda_1(\boldsymbol{\Xi})^{-1}\|\boldsymbol{\Sigma}^{-1/2}\| \|\mathbf{B}\boldsymbol{\alpha}_1- \widehat{\boldsymbol{\xi}}_1\|_2\le c_0^{1/2} C_6\sqrt{\Lambda_p^2s_n}.\label{1127}
\end{align}
  Therefore, $\|\widehat{\boldsymbol{\zeta}}_1\|_2\le 1+c_0^{1/2} C_6\sqrt{\Lambda_p^2s_n}$ and 
\begin{align}
    & \left|1-\|\widehat{\boldsymbol{\zeta}}_1\|_2^2\right|=\left|1-\|\widehat{\boldsymbol{\zeta}}_1\|_2\right|\left(1+\|\widehat{\boldsymbol{\zeta}}_1\|_2\right)=\left|\|\boldsymbol{\gamma}_1\|_2-\|\widehat{\boldsymbol{\zeta}}_1\|_2\right|\left(1+\|\widehat{\boldsymbol{\zeta}}_1\|_2\right)\notag\\
		\le &\|\boldsymbol{\gamma}_1-\widehat{\boldsymbol{\zeta}}_1\|_2(2+c_0^{1/2} C_6\sqrt{\Lambda_p^2s_n})\le c_0^{1/2} C_6\sqrt{\Lambda_p^2s_n}(2+c_0^{1/2} C_6\sqrt{\Lambda_p^2s_n}).\label{1128}
\end{align}
Since $\Lambda_p^2s_n}\to 0$, as $n$ is large enough, by $\eqref{1125}$-$\eqref{1128}$, we can find $C_{1,3}$ large enough and independent of $n$ and $p$ such that $\|\mathbf{P}_1-\widehat{\mathbf{P}}_1\|^2\le C_{1,3}\Lambda_p^2s_n$. Note that
 \begin{align}
    &\| \boldsymbol{\xi}_1\|_2^2=\| \mathbf{B}\boldsymbol{\alpha}_1\|_2^2=\| \lambda_1(\boldsymbol{\Xi})\boldsymbol{\Sigma}\boldsymbol{\alpha}_1\|_2^2=\lambda_1(\boldsymbol{\Xi})^2\boldsymbol{\gamma}_1\trans\boldsymbol{\Sigma}\boldsymbol{\gamma}_1, \label{1145}\\
	&\|\mathbf{Q}_1-\widehat{\mathbf{Q}}_1\|=\left\|\frac{1}{\| \boldsymbol{\xi}_1\|_2^2}\boldsymbol{\xi}_1\boldsymbol{\xi}_1\trans-\frac{1}{\|\widehat{\boldsymbol{\xi}}_1\|_2^2}\widehat{\boldsymbol{\xi}}_1\widehat{\boldsymbol{\xi}}_1\trans\right\|,\notag
\end{align} 
Therefore, we have $c_0^{-1}\lambda_1(\boldsymbol{\Xi})^2\le \| \boldsymbol{\xi}_1\|_2^2\le c_0\lambda_1(\boldsymbol{\Xi})^2$ and by $\eqref{1126}$, $\eqref{1145}$ and the same argument as in $\eqref{1125}$-$\eqref{1128}$, we can find a constant $C_{1,4}$  independent of $n$ and $p$ such that $\|\mathbf{Q}_1-\widehat{\mathbf{Q}}_1\|^2\le C_{1,4}\Lambda_p^2s_n$. Hence, $\eqref{1053}$ is true for $i=1$. Now let $1<k\le K-1$. We will show that  under the assumption that all the inequalities $\eqref{1053}$ are true for all $1\le i\le k-1$ and all large enough $n$, they are also true for $k$ and all large enough $n$. Because the proof is long and technical, we summarize the results in the following Lemma and  provide the proof in Appendix B.
 \begin{Lemma}\label{lemma_7}
In $\Omega_n$, suppose that $\eqref{1053}$ is true for all $1\le i\le k-1$ and all large enough $n$. Then $\eqref{1053}$ is also true for $k$ and all large enough $n$.
\end{Lemma}
Hence, it follows from  \ref{lemma_7} that the inequalities in $\eqref{1053}$ are true for all $1\le k\le K-1$. Based on $\eqref{1053}$, in order to prove the theorem, we only need to show 
 \begin{align*}
   \|\widehat{\boldsymbol{\alpha}}_i -\boldsymbol{\alpha}_i\|^2_2\le \|\boldsymbol{\Sigma}^{-1/2}\widehat{\boldsymbol{\gamma}}_i -\boldsymbol{\Sigma}^{-1/2}\boldsymbol{\gamma}_i\|^2_2\le\|\boldsymbol{\Sigma}^{-1}\| \|\widehat{\boldsymbol{\gamma}}_i -\boldsymbol{\gamma}_i\|^2_2\le c_0C_{i,2}\Lambda_p^2s_n.  
\end{align*} 
Then we can obtain $\eqref{1195}$ by setting $D_{i,1}=C_{i,1}$, $D_{i,2}=c_0C_{i,2}$ and $D_{i,3}=C_{i,4}$. \\

\subsection{Proof of Theorem  $\ref{theorem_7}$}
 
Given a new observation $\mathbf{x}$, $T_{OPT}(\mathbf{x})$ and $T(\mathbf{x})$ denote the classes to which $\mathbf{x}$ is assigned by the rules $T_{OPT}$ and $T$, respectively. We use $P_{\cdot|\mathbf{X}}$ to denote the conditional probability given the training sample $\mathbf{X}$.

{\small\begin{align}
&R_{T}(\mathbf{X})-R_{{OPT}}=(1-R_{{OPT}})-(1-R_{T}(\mathbf{X}))\label{150}\\
=&\sum_{i=1}^KP\left(T_{OPT}(\mathbf{x})=i|\mathbf{x}\in \text{the $i$th class}\right)P\left(\mathbf{x}\in \text{the $i$th class}\right)\notag\\
&-\sum_{i=1}^KP_{\cdot|\mathbf{X}}\left(T(\mathbf{x})=i|\mathbf{x}\in \text{the $i$th class}\right)P\left(\mathbf{x}\in \text{the $i$th class}\right)\notag\\
=&\frac{1}{K}\sum_{i=1}^K\left[P \left(T_{OPT}(\mathbf{x})=i|\mathbf{x}\in \text{the $i$th class}\right)-P_{\cdot|\mathbf{X}}\left(T(\mathbf{x})=i|\mathbf{x}\in \text{the $i$th class}\right)\right]\notag\\
=&\frac{1}{K}\sum_{i=1}^K\left[\sum_{j\neq i}P_{\cdot|\mathbf{X}}\left(T_{OPT}(\mathbf{x})=i, T(\mathbf{x})=j|\mathbf{x}\in \text{the $i$th class}\right)\notag\\
& \left +P_{\cdot|\mathbf{X}}\left(T_{OPT}(\mathbf{x})=i, T(\mathbf{x})=i|\mathbf{x}\in \text{the $i$th class}\right)-P_{\cdot|\mathbf{X}}\left(T(\mathbf{x})=i|\mathbf{x}\in \text{the $i$th class}\right)\right]\notag\\
\le &\frac{1}{K}\sum_{i=1}^K\sum_{j\neq i}P_{\cdot|\mathbf{X}}\left(T_{OPT}(\mathbf{x})=i, T(\mathbf{x})=j|\mathbf{x}\in \text{the $i$th class}\right)\notag\\
= &\frac{1}{K}\sum_{i=1}^K\sum_{j\neq i}P_{\cdot|\mathbf{X}}\left(T(\mathbf{x})=j, T_{OPT}(\mathbf{x})=i|\mathbf{x}\in \text{the $i$th class}\right).\notag
\end{align} } 
We use $P_i(\cdot)$ to denote the conditional probability $P_{\cdot|\mathbf{X}}\left(\cdot|\mathbf{x}\in \text{the $i$th class}\right)$. Then
\begin{align}
&P_{\cdot|\mathbf{X}}\left(T(\mathbf{x})=j, T_{OPT}(\mathbf{x})=i|\mathbf{x}\in \text{the $i$th class}\right)=P_i\left(T(\mathbf{x})=j, T_{OPT}(\mathbf{x})=i\right)\label{1335}\\
=& P_i\left(\widehat{\mathbf{a}}_{kj}\trans\boldsymbol{\Sigma}^{-1/2}(\mathbf{x}- \widehat{\mathbf{b}}_{kj})<0, \forall k\neq j, \text{ and } \mathbf{a}_{li}\trans\boldsymbol{\Sigma}^{-1/2}(\mathbf{x}-\mathbf{b}_{li})<0, \forall l\neq i \right)\notag\\
\le & P_i\left(\widehat{\mathbf{a}}_{ij}\trans\boldsymbol{\Sigma}^{-1/2}(\mathbf{x}- \widehat{\mathbf{b}}_{ij})<0, \text{ and } \mathbf{a}_{ji}\trans\boldsymbol{\Sigma}^{-1/2}(\mathbf{x}-\mathbf{b}_{ji})<0 \right)\notag\\
=& P_i\left(\widehat{\mathbf{a}}_{ji}\trans\boldsymbol{\Sigma}^{-1/2}(\mathbf{x}- \widehat{\mathbf{b}}_{ji})>0, \text{ and } \mathbf{a}_{ji}\trans\boldsymbol{\Sigma}^{-1/2}(\mathbf{x}-\mathbf{b}_{ji})<0\right)\notag\\
=& P_{\mathbf{Z}}\left(\widehat{\mathbf{a}}_{ji}\trans\mathbf{Z}>\widehat{\mathbf{a}}_{ji}\trans\boldsymbol{\Sigma}^{-1/2}(\widehat{\mathbf{b}}_{ji}-\boldsymbol{\mu}_i), \text{ and } \mathbf{a}_{ji}\trans\mathbf{Z}<\mathbf{a}_{ji}\trans\boldsymbol{\Sigma}^{-1/2}(\mathbf{b}_{ji}-\boldsymbol{\mu}_i)\right),\notag
\end{align}  
where $\mathbf{Z}=\boldsymbol{\Sigma}^{-1/2}(\mathbf{x}-\boldsymbol{\mu}_i)\sim N(\mathbf{0},\mathbf{I}_p)$ and independent of the training sample $\mathbf{X}$, $P_{\mathbf{Z}}$ is the probability measure with respect to $\mathbf{Z}$ given $\mathbf{X}$, and in the fourth line, we use $\widehat{\mathbf{a}}_{ij}=-\widehat{\mathbf{a}}_{ji}$. 

To calculate the probability ${\small P_{\mathbf{Z}}\left(\widehat{\mathbf{a}}_{ji}\trans\mathbf{Z}>\widehat{\mathbf{a}}_{ji}\trans\boldsymbol{\Sigma}^{-1/2}(\widehat{\mathbf{b}}_{ji}-\boldsymbol{\mu}_i), \quad \mathbf{a}_{ji}\trans\mathbf{Z}<\mathbf{a}_{ji}\trans\boldsymbol{\Sigma}^{-1/2}(\mathbf{b}_{ji}-\boldsymbol{\mu}_i)\right)}$, we note that it is equal to
$P_{\mathbf{Z}}\left(\widehat{\mathbf{a}}_{ji}\trans\mathbf{Z}>\widehat{d}_{ji}, \mathbf{a}_{ji}\trans\mathbf{Z}<d_{ji}\right)$ by the definitions of $d_{ji}$ and $\widehat{d}_{ji}$. First, we note that by the definition $\eqref{22}$ of $\mathbf{a}_{ji}$ and Lemma \ref{lemma_3},
\begin{align*}
&\|\mathbf{a}_{ji}\|^2_2=\|\boldsymbol{\Sigma}^{1/2}\mathbf{D}(\boldsymbol{\mu}_j-\boldsymbol{\mu}_i)\|^2_2=\|\boldsymbol{\Sigma}^{1/2}\boldsymbol{\Sigma}^{-1}(\boldsymbol{\mu}_j-\boldsymbol{\mu}_i)\|^2_2=\|\boldsymbol{\Sigma}^{-1/2} (\boldsymbol{\mu}_j-\boldsymbol{\mu}_i)\|^2_2\notag\\
=& (\boldsymbol{\mu}_j-\boldsymbol{\mu}_i)\trans\boldsymbol{\Sigma}^{-1}(\boldsymbol{\mu}_j-\boldsymbol{\mu}_i) 
\end{align*} 
which together with Lemma \ref{lemma_6} give
\begin{align}
& 2Kc_1\le \|\mathbf{a}_{ji}\|^2_2 \le 2\lambda_{max}(\boldsymbol{\Delta})\label{1304}
\end{align} 
Moreover, we have 
\begin{align}
&   d_{ji}=\mathbf{a}_{ji}\trans\boldsymbol{\Sigma}^{-1/2}(\mathbf{b}_{ji}-\boldsymbol{\mu}_i)=(\boldsymbol{\mu}_j-\boldsymbol{\mu}_i)\trans\boldsymbol{\Sigma}^{-1/2}\boldsymbol{\Sigma}^{-1/2}(\boldsymbol{\mu}_j-\boldsymbol{\mu}_i)/2=\frac{1}{2}\|\mathbf{a}_{ji}\|_2^2.\label{1004000}
\end{align}  
Since the subscript $ij$ is fixed during the calculation, for simplicity, we omit it in the following. We also omit the subscript $\mathbf{Z}$ in $P_{\mathbf{Z}}$. Note the orthogonal decomposition $\mathbf{a}=t\widehat{\mathbf{a}}+\mathbf{a}_\perp$ and the relationship $d=\frac{1}{2}\|\mathbf{a}\|_2^2$  by $\eqref{1004000}$. By the conditions in $\eqref{104}$,
\begin{align}
&\|\mathbf{a}\|^2_2-\|\widehat{\mathbf{a}}\|_2^2=\|\mathbf{a}\|_2^2O_p(\delta_n),\quad t=1+O_p(\delta_n), \quad d-\widehat{d}
=\|\widehat{\mathbf{a}}\|_2^2O_p(\delta_n),\notag\\
\text{we have}\quad &\|\mathbf{a}_\perp\|_2^2=\|\mathbf{a}\|^2_2-t^2\|\widehat{\mathbf{a}}\|_2^2=\|\widehat{\mathbf{a}}\|_2^2O_p(\delta_n),\quad \widehat{d}
=\|\widehat{\mathbf{a}}\|_2^2(\frac{1}{2}+O_p(\delta_n)).\label{14}
\end{align} 
We first assume that $\mathbf{a}_\perp\neq \mathbf{0}$. Define
\begin{align*}
& \mathbf{W}=\frac{\widehat{\mathbf{a}}\trans}{\|\widehat{\mathbf{a}}\|_2}\mathbf{Z}\sim N(0,1), \quad \mathbf{V}=-\frac{\mathbf{a}_\perp\trans}{\|\mathbf{a}_\perp\|_2}\mathbf{Z}\sim N(0,1),
\end{align*}
where the distributions are conditional on the training sample $\mathbf{X}$. Since $(\mathbf{W},\mathbf{V})$ is jointly normal and $\widehat{\mathbf{a}}$ and $\mathbf{a}_\perp$ are orthogonal, $\mathbf{W}$ and $\mathbf{V}$ are uncorrelated and hence independent. Let $\phi$ and $\Phi$ are the density and cumulative distribution functions of $N(0,1)$, respectively. Define
\begin{align} 
\eta=\frac{|t\widehat{d}-d|}{t}+\frac{\|\mathbf{a}_\perp\|_2}{t}\sqrt{\log{\left[(\|\mathbf{a}\|^2_2\delta_n)^{-1}\right]}}.\label{108}
\end{align}
 Then 
\begin{align}
& P\left(\widehat{\mathbf{a}}\trans\mathbf{Z}>\widehat{d}, \mathbf{a}\trans\mathbf{Z}<d\right)=P\left(\widehat{\mathbf{a}}\trans\mathbf{Z}>\widehat{d}, (t\widehat{\mathbf{a}}+\mathbf{a}_\perp)\trans\mathbf{Z}<d\right)\notag\\
=&P\left(\mathbf{W}>\frac{\widehat{d}}{\|\widehat{\mathbf{a}}\|_2},\quad t\|\widehat{\mathbf{a}}\|_2\mathbf{W}-\|\mathbf{a}_\perp\|_2\mathbf{V}<d\right)\notag\\
=&\int_{\frac{\widehat{d}}{\|\widehat{\mathbf{a}}\|_2}}^\infty\phi(w)P\left(\mathbf{V}>\frac{t\|\widehat{\mathbf{a}}\|_2w-d}{\|\mathbf{a}_\perp\|_2}\right)dw=\int_{\frac{\widehat{d}}{\|\widehat{\mathbf{a}}\|_2}}^\infty\phi(w)\left[1-\Phi\left(\frac{t\|\widehat{\mathbf{a}}\|_2w-d}{\|\mathbf{a}_\perp\|_2}\right)\right]dw\notag\\
=&\int_{\frac{\widehat{d}}{\|\widehat{\mathbf{a}}\|_2}}^{\frac{\widehat{d}+\eta}{\|\widehat{\mathbf{a}}\|_2}}+\int_{\frac{\widehat{d}+\eta}{\|\widehat{\mathbf{a}}\|_2}}^\infty \phi(w)\left[1-\Phi\left(\frac{t\|\widehat{\mathbf{a}}\|_2w-d}{\|\mathbf{a}_\perp\|_2}\right)\right]dw\notag\\
\le&\int_{\frac{\widehat{d}}{\|\widehat{\mathbf{a}}\|_2}}^{\frac{\widehat{d}+\eta}{\|\widehat{\mathbf{a}}\|_2}}\phi(w)dw+\int_{\frac{\widehat{d}+\eta}{\|\widehat{\mathbf{a}}\|_2}}^\infty \phi(w)\left[1-\Phi\left(\frac{t\|\widehat{\mathbf{a}}\|_2\frac{\widehat{d}+\eta}{\|\widehat{\mathbf{a}}\|_2}-d}{\|\mathbf{a}_\perp\|_2}\right)\right]dw\notag\\
\le &\frac{\eta}{\|\widehat{\mathbf{a}}\|_2}\phi(\frac{\widehat{d}}{\|\widehat{\mathbf{a}}\|_2})+\left[1-\Phi\left(\frac{t\widehat{d}-d+t\eta}{\|\mathbf{a}_\perp\|_2}\right)\right]\int_{\frac{\widehat{d}}{\|\widehat{\mathbf{a}}\|_2}}^\infty \phi(w)dw.\label{39}
\end{align}  
Since $\widehat{d}/\|\widehat{\mathbf{a}}\|_2=\|\mathbf{a}\|_2(1/2+O_p(\delta_n))$ and by $\eqref{1304}$, $\|\mathbf{a}\|_2$ is bounded below, it follows from the  inequality:
\begin{align}
(1-\frac{1}{x^2})\phi(x)\le x[1-\Phi(x)]\le \phi(x),\quad \forall x>0,\label{40}
\end{align}  
 that there exists a constant $C_3>0$ independent of $p$ such that with probability converging to 1, 
 \begin{align}
C_3\phi(\frac{\widehat{d}}{\|\widehat{\mathbf{a}}\|_2})\le \frac{\widehat{d}}{\|\widehat{\mathbf{a}}\|_2}[1-\Phi(\frac{\widehat{d}}{\|\widehat{\mathbf{a}}\|_2})].\label{41}
\end{align}  
By $\eqref{1304}$, $\eqref{14}$,$\eqref{40}$, $\eqref{41}$, and the definition $\eqref{108}$ of $\eta$,  the right hand side of $\eqref{39}$,
{\small\begin{align}
 &\frac{\eta}{\|\widehat{\mathbf{a}}\|_2}\phi(\frac{\widehat{d}}{\|\widehat{\mathbf{a}}\|_2})+\left[1-\Phi\left(\frac{t\widehat{d}-d+t\eta}{\|\mathbf{a}_\perp\|_2}\right)\right][1-\Phi(\frac{\widehat{d}}{\|\widehat{\mathbf{a}}\|_2})]\le\frac{1}{C_3}\frac{\eta}{\|\widehat{\mathbf{a}}\|_2}\frac{\widehat{d}}{\|\widehat{\mathbf{a}}\|_2}[1-\Phi(\frac{\widehat{d}}{\|\widehat{\mathbf{a}}\|_2})]\notag\\
&+\left[1-\Phi\left(\frac{t\widehat{d}-d+|t\widehat{d}-d|+\|\mathbf{a}_\perp\|_2\sqrt{\log{\left[(\|\mathbf{a}\|^2_2\delta_n)^{-1}\right]}}}{\|\mathbf{a}_\perp\|_2}\right)\right][1-\Phi(\frac{\widehat{d}}{\|\widehat{\mathbf{a}}\|_2})]\notag\\
&\le\frac{1}{C_3}(\frac{1}{2}+O_p(\delta_n))\eta[1-\Phi(\frac{\widehat{d}}{\|\widehat{\mathbf{a}}\|_2})]+\left[1-\Phi\left(\sqrt{\log{\left[(\|\mathbf{a}\|^2_2\delta_n)^{-1}\right]}}\right)\right][1-\Phi(\frac{\widehat{d}}{\|\widehat{\mathbf{a}}\|_2})]\notag\\
&\le\frac{1}{C_3}(\frac{1}{2}+O_p(\delta_n))\eta[1-\Phi(\frac{\widehat{d}}{\|\widehat{\mathbf{a}}\|_2})]+\left[1-\Phi\left(\sqrt{\log{\left[(2\lambda_{max}(\boldsymbol{\Delta})\delta_n)^{-1}\right]}}\right)\right][1-\Phi(\frac{\widehat{d}}{\|\widehat{\mathbf{a}}\|_2})]\notag\\
&\le \left[\frac{\eta}{C_3}(\frac{1}{2}+O_p(\delta_n))+\frac{\phi\left(\sqrt{\log{\left[(2\lambda_{max}(\boldsymbol{\Delta})\delta_n)^{-1}\right]}}\right)}{\sqrt{\log{\left[(2\lambda_{max}(\boldsymbol{\Delta})\delta_n)^{-1}\right]}}}\right][1-\Phi(\frac{\widehat{d}}{\|\widehat{\mathbf{a}}\|_2})].\label{42}
\end{align} } 
By $\eqref{14}$ and the definition $\eqref{108}$ of $\eta$,
{\small\begin{align}
  &\frac{\eta}{C_3}(\frac{1}{2}+O_p(\delta_n))+\frac{\phi\left(\sqrt{\log{\left[(2\lambda_{max}(\boldsymbol{\Delta})\delta_n)^{-1}\right]}}\right)}{\sqrt{\log{\left[(2\lambda_{max}(\boldsymbol{\Delta})\delta_n)^{-1}\right]}}}\notag \\
	=&\|\mathbf{a}\|_2^2O_p(\delta_n)+\sqrt{\|\mathbf{a}\|_2^2O_p(\delta_n)\log{\left[(\|\mathbf{a}\|^2_2\delta_n)^{-1}\right]}}+O\left(\frac{\exp{\left[-\left(\sqrt{\log{\left[(2\lambda_{max}(\boldsymbol{\Delta})\delta_n)^{-1}\right]}}\right)^2/2\right]}}{\sqrt{\log{\left[(2\lambda_{max}(\boldsymbol{\Delta})\delta_n)^{-1}\right]}}} \right)\notag\\
	\le &2\lambda_{max}(\boldsymbol{\Delta})O_p(\delta_n)+\sqrt{2\lambda_{max}(\boldsymbol{\Delta})O_p(\delta_n)\log{\left[(2\lambda_{max}(\boldsymbol{\Delta})\delta_n)^{-1}\right]}}+O\left(\frac{\sqrt{2\lambda_{max}(\boldsymbol{\Delta})\delta_n}}{\sqrt{\log{\left[(2\lambda_{max}(\boldsymbol{\Delta})\delta_n)^{-1}\right]}}}\right)\notag\\
	=&O_p\left( \sqrt{\lambda_{max}(\boldsymbol{\Delta})\delta_n\log{\left[(\lambda_{max}(\boldsymbol{\Delta})\delta_n)^{-1}\right]}} \right)\label{43}.
\end{align}  }
Next, we estimate 
\begin{align*}
 \left|[1-\Phi(\frac{\widehat{d}}{\|\widehat{\mathbf{a}}\|_2})]-[1-\Phi(\frac{d}{\|\mathbf{a}\|_2})]\right|=\int_{r_1}^{r_2}\phi(x)dx,
\end{align*}  
where $r_1=\min{(\widehat{d}/\|\widehat{\mathbf{a}}\|_2, d/\|\mathbf{a}\|_2)}$, $r_2=\max{(\widehat{d}/\|\widehat{\mathbf{a}}\|_2, d/\|\mathbf{a}\|_2)}$. By $\eqref{40}$ and  $\eqref{14}$,
\begin{align*}
 &\int_{r_1}^{r_2}\phi(x)dx\le (r_2-r_1)\phi(r_1)=(r_2-r_1)O(r_1[1-\Phi(r_1)])=(r_2-r_1)r_1O([1-\Phi(r_1)])\\
=&\left|\frac{\widehat{d}}{\|\widehat{\mathbf{a}}\|_2}-\frac{d}{\|{\mathbf{a}}\|_2}\right|r_1O([1-\Phi(r_1)])=\left|\|\widehat{\mathbf{a}}\|_2(\frac{1}{2}+O_p(\delta_n))-
\frac{1}{2}\|{\mathbf{a}}\|_2\right|\frac{1}{2}\|{\mathbf{a}}\|_2O([1-\Phi(r_1)])\\
\le &2\lambda_{max}(\boldsymbol{\Delta})O_p(\delta_n)O([1-\Phi(r_1)]).
\end{align*}  
Therefore, 
{\small\begin{align}
 &\left|[1-\Phi(\frac{\widehat{d}}{\|\widehat{\mathbf{a}}\|_2})]-[1-\Phi(\frac{d}{\|\mathbf{a}\|_2})]\right|\le 2\lambda_{max}(\boldsymbol{\Delta})O_p(\delta_n)O([1-\Phi(\frac{d}{\|\mathbf{a}\|_2})])=o_p(1)O([1-\Phi(\frac{d}{\|\mathbf{a}\|_2})]),\notag\\
&\text{ and hence } [1-\Phi(\frac{\widehat{d}}{\|\widehat{\mathbf{a}}\|_2})]=[1-\Phi(\frac{d}{\|\mathbf{a}\|_2})](1+o_p(1)),\label{50}
\end{align}  }
By $\eqref{39}$, $\eqref{42}$, $\eqref{43}$ and $\eqref{50}$
\begin{align}
& P\left(\widehat{\mathbf{a}}\trans\mathbf{Z}>\widehat{d}, \mathbf{a}\trans\mathbf{Z}<d\right)\le O_p\left( \sqrt{\lambda_{max}(\boldsymbol{\Delta})\delta_n\log{\left[(\lambda_{max}(\boldsymbol{\Delta})\delta_n)^{-1}\right]}} \right)[1-\Phi(\frac{d}{\|\mathbf{a}\|_2})]\notag\\
&= O_p\left( \sqrt{\lambda_{max}(\boldsymbol{\Delta})\delta_n\log{\left[(\lambda_{max}(\boldsymbol{\Delta})\delta_n)^{-1}\right]}} \right)P\left( \mathbf{a}\trans\mathbf{Z}>d\right).\label{109}
\end{align}  
Now we consider the case of $\mathbf{a}_\perp=\mathbf{0}$. By similar arguments as those for $\eqref{50}$,
\begin{align}
& P\left(\widehat{\mathbf{a}}\trans\mathbf{Z}>\widehat{d}, \mathbf{a}\trans\mathbf{Z}<d\right)=P\left(\widehat{\mathbf{a}}\trans\mathbf{Z}>\widehat{d}, (t\widehat{\mathbf{a}})\trans\mathbf{Z}<d\right)\label{152}\\
=&P\left(\mathbf{W}>\frac{\widehat{d}}{\|\widehat{\mathbf{a}}\|_2},\quad t\|\widehat{\mathbf{a}}\|_2\mathbf{W}<d\right)\le\left|[1-\Phi(\frac{\widehat{d}}{\|\widehat{\mathbf{a}}\|_2})]-[1-\Phi(\frac{d}{t\|\widehat{\mathbf{a}}\|_2})]\right|\notag\\
 \le& \lambda_{max}(\boldsymbol{\Delta})O_p(\delta_n)P\left( \mathbf{a}\trans\mathbf{Z}>d\right)\le O_p\left( \sqrt{\lambda_{max}(\boldsymbol{\Delta})\delta_n\log{\left[(\lambda_{max}(\boldsymbol{\Delta})\delta_n)^{-1}\right]}} \right)P\left( \mathbf{a}\trans\mathbf{Z}>d\right).\notag
\end{align}  
Combining $\eqref{109}$ and $\eqref{152}$, and note the fact that given the new observation $\mathbf{x}$ belonging to the $i$th class, $\boldsymbol{\Sigma}^{-1/2}(\mathbf{x}-\boldsymbol{\mu}_i)$ have the same distribution as $\mathbf{Z}$, we have
\begin{align}
&P_{\mathbf{Z}}\left(\widehat{\mathbf{a}}_{ji}\trans\mathbf{Z}>\widehat{\mathbf{a}}_{ji}\trans\boldsymbol{\Sigma}^{-1/2}(\widehat{\mathbf{b}}_{ji}-\boldsymbol{\mu}_i),  \mathbf{a}_{ji}\trans\mathbf{Z}<\mathbf{a}_{ji}\trans\boldsymbol{\Sigma}^{-1/2}(\mathbf{b}_{ji}-\boldsymbol{\mu}_i)\right)\label{45}\\
\le &O_p\left( \sqrt{\lambda_{max}(\boldsymbol{\Delta})\delta_n\log{\left[(\lambda_{max}(\boldsymbol{\Delta})\delta_n)^{-1}\right]}} \right)P_{\mathbf{Z}}\left(\mathbf{a}_{ji}\trans\mathbf{Z}> d_{ji}\right)\notag\\
\le& O_p\left( \sqrt{\lambda_{max}(\boldsymbol{\Delta})\delta_n\log{\left[(\lambda_{max}(\boldsymbol{\Delta})\delta_n)^{-1}\right]}} \right)\notag\\
&\times P_{\mathbf{Z}}\left(\mathbf{a}_{ji}\trans\boldsymbol{\Sigma}^{-1/2}(\mathbf{x}-\boldsymbol{\mu}_i)> \mathbf{a}_{ji}\trans\boldsymbol{\Sigma}^{-1/2}(\mathbf{b}_{ji}-\boldsymbol{\mu}_i)|\mathbf{x}\in \text{the $i$th class}\right)\notag\\
=&O_p\left( \sqrt{\lambda_{max}(\boldsymbol{\Delta})\delta_n\log{\left[(\lambda_{max}(\boldsymbol{\Delta})\delta_n)^{-1}\right]}} \right)P\left(\mathbf{a}_{ji}\trans\boldsymbol{\Sigma}^{-1/2}(\mathbf{x}-\mathbf{b}_{ji})>0|\mathbf{x}\in \text{the $i$th class}\right)\notag\\
\le &O_p\left( \sqrt{\lambda_{max}(\boldsymbol{\Delta})\delta_n\log{\left[(\lambda_{max}(\boldsymbol{\Delta})\delta_n)^{-1}\right]}} \right)P\left(T_{OPT}(\mathbf{x})\notin \text{the $i$th class}|\mathbf{x}\in \text{the $i$th class}\right)\notag\\
\le &O_p\left( \sqrt{\lambda_{max}(\boldsymbol{\Delta})\delta_n\log{\left[(\lambda_{max}(\boldsymbol{\Delta})\delta_n)^{-1}\right]}} \right)KR_{{OPT}}\notag
\end{align}
 where the $O_p$ term is uniform for all $1\le i\neq j\le K$. Now by  $\eqref{150}$, $\eqref{1335}$ and $\eqref{45}$,
\begin{align*}
&R_{T}(\mathbf{X})-R_{{OPT}} \\
\le& \frac{1}{K}\sum_{i=1}^K\sum_{j\neq i}\left[P_{\mathbf{Z}}\left(\widehat{\mathbf{a}}_{ji}\trans\mathbf{Z}<\widehat{\mathbf{a}}_{ji}\trans\boldsymbol{\Sigma}^{-1/2}(\widehat{\mathbf{b}}_{ji}-\boldsymbol{\mu}_i),  \quad\mathbf{a}_{ji}\trans\mathbf{Z}> \mathbf{a}_{ji}\trans\boldsymbol{\Sigma}^{-1/2}(\mathbf{b}_{ji}-\boldsymbol{\mu}_i)\right) \right],\notag\\
\le &O_p\left(K^2\sqrt{\lambda_{max}(\boldsymbol{\Delta})\delta_n\log{\left[(\lambda_{max}(\boldsymbol{\Delta})\delta_n)^{-1}\right]}} \right)R_{{OPT}}\notag
\end{align*}

By Lemma \ref{lemma_6},   $\lambda_{max}(\boldsymbol{\Delta})=K\lambda_{max}(\boldsymbol{\Xi})=K\lambda_1(\boldsymbol{\Xi})$. Moreover, in this paper, we assume that $K$ is fixed, we have
\begin{align*}
&\frac{R_{T}(\mathbf{X})}{R_{{OPT}}}-1\le O_p\left(\sqrt{\lambda_1(\boldsymbol{\Xi})\delta_n\log{\left[\{\lambda_1(\boldsymbol{\Xi})\delta_n\}^{-1}\right]}} \right).
\end{align*}

\subsection{Proof of Theorem  $\ref{theorem_8}$}
 
To apply Theorem \ref{theorem_7}, we first verify the conditions $\eqref{104}$ for $\delta_n=\Lambda_p^2s_n$. In this proof, let $\boldsymbol{\delta}_{ji}=\boldsymbol{\mu}_j-\boldsymbol{\mu}_i$ and $\widehat{\boldsymbol{\delta}}_{ji}=\bar{\mathbf{x}}_j-\bar{\mathbf{x}}_i$ for any $1\le i,j\le K$.  We only consider the elements in ${\Omega}_n\bigcap \widetilde{\Omega}_n$ (see their definitions $\eqref{1033}$ and $\eqref{1184}$). The complement of ${\Omega}_n\bigcap \widetilde{\Omega}_n$ has a probability less than $3p^{-1}\to 0$ as $n, p\to\infty$. Therefore, by the definition $\eqref{1184}$ of $\widetilde{\Omega}_n$, we have 
\begin{align}
  \|\boldsymbol{\delta}_{ji}-\widehat{\boldsymbol{\delta}}_{ji}\label{10039}\|_\infty\le 2\widetilde{C}\sqrt{\frac{K\log{p}}{n}}=2\widetilde{C}s_n.\label{21111}
\end{align}} 
Let $\mathbf{P}_{K-1}$ and $\widetilde{\mathbf{P}}_{K-1}$ be the orthogonal projection matrices of the two subspaces of  
\begin{align*}
   \mathbf{V}_{K-1}=\text{span}\{\boldsymbol{\gamma}_1,\boldsymbol{\gamma}_2, \cdots, \boldsymbol{\gamma}_{K-1}\},\quad \widetilde{\mathbf{V}}_{K-1}=\text{span}\{\widehat{\boldsymbol{\gamma}}_1,\widehat{\boldsymbol{\gamma}}_2, \cdots, \widehat{\boldsymbol{\gamma}}_{K-1}\}.
\end{align*}
Let $\boldsymbol{\Gamma}=[\boldsymbol{\gamma}_1,\cdots, \boldsymbol{\gamma}_{K-1}]$  and $ \widehat{\boldsymbol{\Gamma}}=[\widehat{\boldsymbol{\gamma}}_1, \cdots, \widehat{\boldsymbol{\gamma}}_{K-1}]$, both of which are $p \times (K-1)$ matrices. Let $\mathbf{K}=\widehat{\boldsymbol{\Gamma}}\trans\widehat{\boldsymbol{\Gamma}}$ which is a symmetric $(K-1)\times (K-1)$ matrix with the $(k,l)$-th entry equal to $\widehat{\boldsymbol{\alpha}}_k\trans \boldsymbol{\Sigma}\widehat{\boldsymbol{\alpha}}_l=\widehat{\boldsymbol{\gamma}}_k\trans\widehat{\boldsymbol{\gamma}}_l$. Then we have
\begin{align}
   \mathbf{P}_{K-1}=\boldsymbol{\Gamma}\boldsymbol{\Gamma}\trans,\qquad \widetilde{\mathbf{P}}_{K-1}=\widehat{\boldsymbol{\Gamma}}\mathbf{K}^{-1}\widehat{\boldsymbol{\Gamma}}\trans,\quad \boldsymbol{\Gamma}\trans\boldsymbol{\Gamma}=\mathbf{I}_{K-1},\label{10039}
\end{align}} 
 where $\mathbf{I}_{K-1}$ is the $K-1$ dimensional identity matrix, because $\boldsymbol{\gamma}_k$, $1\le k\le K-1$, are orthonormal vectors.  By the definition $\eqref{1210}$,
\begin{align}
   \widehat{\mathbf{K}}=\left(\widehat{\boldsymbol{\alpha}}_1,\cdots,\widehat{\boldsymbol{\alpha}}_{K-1}\right)\trans  \widehat{\boldsymbol{\Sigma}}\left(\widehat{\boldsymbol{\alpha}}_1,\cdots,\widehat{\boldsymbol{\alpha}}_{K-1}\right)=\widehat{\boldsymbol{\Gamma}}\trans \boldsymbol{\Sigma}^{-1/2}\widehat{\boldsymbol{\Sigma}}\boldsymbol{\Sigma}^{-1/2}\widehat{\boldsymbol{\Gamma}},\notag\\
	\widehat{\mathbf{D}}=\boldsymbol{\Sigma}^{-1/2}\widehat{\boldsymbol{\Gamma}}\widehat{\mathbf{K}}^{-1}\widehat{\boldsymbol{\Gamma}}\trans\boldsymbol{\Sigma}^{-1/2}.\label{10036}
\end{align}
We consider the first equality in  $\eqref{104}$.  By $\eqref{22}$ and Lemmas \ref{lemma_3} and \ref{lemma_6},
 \begin{align}
& \mathbf{a}_{ji} = \boldsymbol{\Sigma}^{1/2} \mathbf{D}\boldsymbol{\delta}_{ji}= \boldsymbol{\Sigma}^{1/2} \boldsymbol{\Sigma}^{-1}\boldsymbol{\delta}_{ji}=\boldsymbol{\Sigma}^{-1/2}\boldsymbol{\delta}_{ji} ,\notag\\
& 2Kc_1\le (\boldsymbol{\mu}_j-\boldsymbol{\mu}_i)\trans\boldsymbol{\Sigma}^{-1}(\boldsymbol{\mu}_j-\boldsymbol{\mu}_i)=\|\boldsymbol{\Sigma}^{-1}\boldsymbol{\delta}_{ij}\|^2_2= \|\mathbf{a}_{ji}\|^2_2\le 2K\lambda_1(\boldsymbol{\Xi}).\label{1200}
\end{align} 
By $\eqref{10039}$,
 \begin{align}
&\mathbf{D}=\sum_{k=1}^{K-1}\boldsymbol{\alpha}_k\boldsymbol{\alpha}_k\trans=\boldsymbol{\Sigma}^{-1/2} \sum_{k=1}^{K-1}\boldsymbol{\gamma}_k\boldsymbol{\gamma}_k\trans\boldsymbol{\Sigma}^{-1/2}=\boldsymbol{\Sigma}^{-1/2} \boldsymbol{\Gamma}\boldsymbol{\Gamma}\trans\boldsymbol{\Sigma}^{-1/2}\notag\\
&=\boldsymbol{\Sigma}^{-1/2} \mathbf{P}_{K-1}\boldsymbol{\Sigma}^{-1/2}.\label{10038}
\end{align} 
  Hence, by $\eqref{1200}$, $\eqref{10038}$ and Lemma  \ref{lemma_3},
 \begin{align}
& \mathbf{P}_{K-1}\mathbf{a}_{ji} = \mathbf{P}_{K-1}\boldsymbol{\Sigma}^{-1/2}  \boldsymbol{\delta}_{ji}=\boldsymbol{\Sigma}^{1/2} \boldsymbol{\Sigma}^{-1/2} \mathbf{P}_{K-1}\boldsymbol{\Sigma}^{-1/2}\boldsymbol{\delta}_{ij}=\boldsymbol{\Sigma}^{1/2} \mathbf{D}\boldsymbol{\delta}_{ji}\notag\\
=&\boldsymbol{\Sigma}^{1/2} \boldsymbol{\Sigma}^{-1}\boldsymbol{\delta}_{ji}=\mathbf{a}_{ji}.\label{1234}
\end{align} 
For any $(k,l)$, by the definition of $\Omega_n$ and Theorem \ref{theorem_10}, 
\begin{align}
   &|(\widehat{\mathbf{K}})_{kl}-(\mathbf{K})_{kl}|=|\widehat{\boldsymbol{\alpha}}_k\trans \widehat{\boldsymbol{\Sigma}}\widehat{\boldsymbol{\alpha}}_l-\widehat{\boldsymbol{\alpha}}_k\trans \boldsymbol{\Sigma}\widehat{\boldsymbol{\alpha}}_l|\le \|\boldsymbol{\Sigma} -\widehat{\boldsymbol{\Sigma}}\|_{\infty}\|\widehat{\boldsymbol{\alpha}}_k\|_1\|\widehat{\boldsymbol{\alpha}}_l\|_1  \notag\\
	\le & (\tau_n/C_2)D_{k,1}\Lambda_pD_{l,1}\Lambda_p\le h_1 \Lambda_p^2s_n,\label{1230}
\end{align}
where $h_1=(C/C_2)\max_{1\le k,l\le K-1}(D_{k,1}D_{l,1})$. Because $\widehat{\mathbf{K}}-\mathbf{K}$ is symmetric, by $\eqref{1230}$,
\begin{align}
   \|\widehat{\mathbf{K}}-\mathbf{K}\|\le \max_{1\le k\le K-1}\sum_{l=1}^{K-1} |(\widehat{\mathbf{K}})_{kl}-(\mathbf{K})_{kl}|\le (K-1)h_1 \Lambda_p^2s_n=o(1).\label{1213}
\end{align}
 Because $ \boldsymbol{\alpha}_k\trans \boldsymbol{\Sigma}\boldsymbol{\alpha}_l=1$ if $k=l$, and equal to 0 if $k\neq l$, by $\eqref{1053}$ in the proof of Theorem \ref{theorem_10}, as $n$ large enough,
\begin{align*}
   &|(\mathbf{K})_{kl}-(\mathbf{I}_{K-1})_{kl}|=|\widehat{\boldsymbol{\alpha}}_k\trans \boldsymbol{\Sigma}\widehat{\boldsymbol{\alpha}}_l-\boldsymbol{\alpha}_k\trans \boldsymbol{\Sigma}\boldsymbol{\alpha}_l|=|\widehat{\boldsymbol{\gamma}}_k\trans  \widehat{\boldsymbol{\gamma}}_l-\boldsymbol{\gamma}_k\trans  \boldsymbol{\gamma}_l|\\
	\le & |(\widehat{\boldsymbol{\gamma}}_k -\boldsymbol{\gamma}_k)\trans  (\widehat{\boldsymbol{\gamma}}_l-\boldsymbol{\gamma}_l)|+| \boldsymbol{\gamma}_k\trans  (\widehat{\boldsymbol{\gamma}}_l-\boldsymbol{\gamma}_l)|+|(\widehat{\boldsymbol{\gamma}}_k -\boldsymbol{\gamma}_k)\trans  \boldsymbol{\gamma}_l|\notag\\
	\le &\|\widehat{\boldsymbol{\gamma}}_k -\boldsymbol{\gamma}_k\|_2\|\widehat{\boldsymbol{\gamma}}_l-\boldsymbol{\gamma}_l\|_2+\|\widehat{\boldsymbol{\gamma}}_k -\boldsymbol{\gamma}_k\|_2+\|\widehat{\boldsymbol{\gamma}}_l-\boldsymbol{\gamma}_l\|_2\le h_2 \sqrt{\Lambda_p^2s_n},
\end{align*}
  where $h_2$ is a constant independent of $p$ and $n$. Therefore,
\begin{align}
  & \| \mathbf{K}-\mathbf{I}_{K-1}\|\le \max_{1\le k\le K-1}\sum_{l=1}^{K-1} |(\widehat{\mathbf{K}})_{kl}-(\mathbf{I}_{K-1})_{kl}|\le  (K-1)h_2 \sqrt{\Lambda_p^2s_n}=o(1),  \notag\\
	&\text{ and hence } \| \mathbf{K}\|=1+o(1).\label{1236}
\end{align}
By the Taylor's expansion,
\begin{align}
   &\| \mathbf{K}^{-1}\|=  \| [\mathbf{I}_{K-1}-(\mathbf{I}_{K-1}-\mathbf{K})]^{-1}\|=\|\mathbf{I}_{K-1}+(\mathbf{I}_{K-1}-\mathbf{K})+(\mathbf{I}_{K-1}-\mathbf{K})^2+\cdots\|\notag\\
	\le& \|\mathbf{I}_{K-1}\|+\|\mathbf{I}_{K-1}-\mathbf{K}\|+\|\mathbf{I}_{K-1}-\mathbf{K}\|^2+\cdots=\frac{1}{1-\|\mathbf{I}_{K-1}-\mathbf{K}\|}=1+o(1).\label{1215}
\end{align}
$\eqref{1213}$-$\eqref{1215}$ imply that $\| \widehat{\mathbf{K}}-\mathbf{I}_{K-1}\|= o(1)$. By the same argument as in $\eqref{1215}$, $\|\widehat{\mathbf{K}}^{-1} \|\le 1+o(1)$. Then by $\eqref{1213}$,
  \begin{align}
   \|\widehat{\mathbf{K}}^{-1}-\mathbf{K}^{-1}\|\le \|\widehat{\mathbf{K}}^{-1}\|\|\widehat{\mathbf{K}}-\mathbf{K}\|\|\mathbf{K}^{-1}\| \le (K-1)h_1 \Lambda_p^2s_n(1+o(1))^2\le 2(K-1)h_1 \Lambda_p^2s_n,\label{1216}
\end{align} 
as $n$ is large enough. Now by $\eqref{10036}$,
\begin{align}
   &\widehat{\mathbf{a}}_{ji}=\boldsymbol{\Sigma}^{1/2}\widehat{\mathbf{D}}\widehat{\boldsymbol{\delta}}_{ji}=\boldsymbol{\Sigma}^{1/2}\boldsymbol{\Sigma}^{-1/2}\widehat{\boldsymbol{\Gamma}} \widehat{\mathbf{K}}^{-1}\widehat{\boldsymbol{\Gamma}}\trans\boldsymbol{\Sigma}^{-1/2}\widehat{\boldsymbol{\delta}}_{ji}=\widehat{\boldsymbol{\Gamma}} \widehat{\mathbf{K}}^{-1}\widehat{\boldsymbol{\Gamma}}\trans\boldsymbol{\Sigma}^{-1/2}\widehat{\boldsymbol{\delta}}_{ji}\notag\\
=&\widehat{\boldsymbol{\Gamma}} \widehat{\mathbf{K}}^{-1}\widehat{\boldsymbol{\Gamma}}\trans\boldsymbol{\Sigma}^{-1/2}(\widehat{\boldsymbol{\delta}}_{ji}-\boldsymbol{\delta}_{ji})+\widehat{\boldsymbol{\Gamma}} (\widehat{\mathbf{K}}^{-1}-\mathbf{K}^{-1})\widehat{\boldsymbol{\Gamma}}\trans\boldsymbol{\Sigma}^{-1/2} \boldsymbol{\delta}_{ji}+\widehat{\boldsymbol{\Gamma}}\mathbf{K}^{-1}\widehat{\boldsymbol{\Gamma}}\trans\boldsymbol{\Sigma}^{-1/2} \boldsymbol{\delta}_{ji}.\label{1217}
\end{align}
We estimate the first term on the right hand side of $\eqref{1217}$. Let $\mathbf{g}=\widehat{\mathbf{K}}^{-1}\widehat{\boldsymbol{\Gamma}}\trans\boldsymbol{\Sigma}^{-1/2}(\widehat{\boldsymbol{\delta}}_{ji}-\boldsymbol{\delta}_{ji})$ and $g_k$ denote its $k$-th coordinate, $1\le k\le K-1$. By $\eqref{1165}$, $\eqref{21111}$, Theorem \ref{theorem_10} and $\eqref{1184}$, 
 \begin{align}
   & \|\widehat{\boldsymbol{\Gamma}} \widehat{\mathbf{K}}^{-1}\widehat{\boldsymbol{\Gamma}}\trans\boldsymbol{\Sigma}^{-1/2}(\widehat{\boldsymbol{\delta}}_{ji}-\boldsymbol{\delta}_{ji})\|_2=\|\widehat{\boldsymbol{\Gamma}}\mathbf{g}\|_2= \|\sum_{k=1}^{K-1}\widehat{\boldsymbol{\gamma}}_k  g_k\|_2  \le \sum_{k=1}^{K-1}\|\widehat{\boldsymbol{\gamma}}_k\|_2|g_k|\le\|\mathbf{g}\|_1\notag\\
	\le &\sqrt{K-1}\|\mathbf{g}\|_2\le\sqrt{K-1}\|\widehat{\mathbf{K}}^{-1}\|\|\widehat{\boldsymbol{\Gamma}}\trans\boldsymbol{\Sigma}^{-1/2}(\widehat{\boldsymbol{\delta}}_{ji}-\boldsymbol{\delta}_{ji})\|_2\label{1218}\\
	\le& \sqrt{K-1}\|\widehat{\mathbf{K}}^{-1}\|\|\widehat{\boldsymbol{\Gamma}}\trans\boldsymbol{\Sigma}^{-1/2}(\widehat{\boldsymbol{\delta}}_{ji}-\boldsymbol{\delta}_{ji})\|_1\notag\\
	=&\sqrt{K-1}\|\widehat{\mathbf{K}}^{-1}\|\sum_{k=1}^{K-1}|\widehat{\boldsymbol{\alpha}}_k\trans(\widehat{\boldsymbol{\delta}}_{ji}-\boldsymbol{\delta}_{ji})|\le \sqrt{K-1}\|\widehat{\mathbf{K}}^{-1}\|\sum_{k=1}^{K-1}\|\widehat{\boldsymbol{\alpha}}_k\|_1\|\widehat{\boldsymbol{\delta}}_{ji}-\boldsymbol{\delta}_{ji}\|_{\infty}\notag\\
	\le &\sqrt{K-1}(1+o(1))(K-1)D_{k,1}\Lambda_p2\widetilde{C}s_n=O(\Lambda_ps_n)=O(\Lambda_p^2s_n)\le O(\Lambda_p^2s_n)\|\mathbf{a}_{ji}\|_2,\notag
\end{align}
where the second equality in the last line is due to $\Lambda_p\ge \| {\boldsymbol{\alpha}}_k\|_1\ge \| {\boldsymbol{\alpha}}_k\|_2\ge c_0^{-1/2}$ and the last inequality is  due to $\eqref{1200}$. Similarly, by $\eqref{1230}$ and $\eqref{1200}$, for the second term on the right hand side of $\eqref{1217}$, we have
 \begin{align}
   &\|\widehat{\boldsymbol{\Gamma}} (\widehat{\mathbf{K}}^{-1}-\mathbf{K}^{-1})\widehat{\boldsymbol{\Gamma}}\trans\boldsymbol{\Sigma}^{-1/2} \boldsymbol{\delta}_{ji}\|_2\le \|\widehat{\boldsymbol{\Gamma}} (\widehat{\mathbf{K}}^{-1}-\mathbf{K}^{-1})\widehat{\boldsymbol{\Gamma}}\trans\|\|\boldsymbol{\Sigma}^{-1/2} \boldsymbol{\delta}_{ji}\|_2\label{1219}\\
	= &\|\sum_{1\le k,l\le K-1}(\widehat{\mathbf{K}}_{kl}-\mathbf{K}_{kl})\widehat{\boldsymbol{ \gamma}}_k \widehat{\boldsymbol{ \gamma}}_l\trans \|\|\mathbf{a}_{ji}\|_2 \le h_1 \Lambda_p^2s_n\sum_{1\le k,l\le K-1}\|\widehat{\boldsymbol{ \gamma}}_k\|_2\|\widehat{\boldsymbol{ \gamma}}_l\|_2\|\mathbf{a}_{ji}\|_2\notag\\
	=&(K-1)^2h_1\Lambda_p^2s_n\|\mathbf{a}_{ji}\|_2. \notag
\end{align} 
  As for the third term on the right hand side of $\eqref{1217}$, by $\eqref{10039}$, we have
\begin{align}
\|\widehat{\boldsymbol{\Gamma}}\mathbf{K}^{-1}\widehat{\boldsymbol{\Gamma}}\trans\boldsymbol{\Sigma}^{-1/2} \boldsymbol{\delta}_{ji}\|_2=\|\widetilde{\mathbf{P}}_{K-1}\boldsymbol{\Sigma}^{-1/2} \boldsymbol{\delta}_{ji}\|_2= \|\widetilde{\mathbf{P}}_{K-1}\mathbf{a}_{ji}\|_2.\label{1232}
\end{align}
Now by $\eqref{1217}$-$\eqref{1232}$ and $\eqref{1234}$, we have 
\begin{align}
   &\|\widehat{\mathbf{a}}_{ji}\|_2^2=\|\widetilde{\mathbf{P}}_{K-1}\mathbf{a}_{ji}\|^2_2+O(\Lambda_p^2s_n)\|\mathbf{a}_{ji}\|_2^2=\|\mathbf{a}_{ji}\|^2_2-\|\mathbf{a}_{ji}-\widetilde{\mathbf{P}}_{K-1}\mathbf{a}_{ji}\|^2_2+O(\Lambda_p^2s_n)\|\mathbf{a}_{ji}\|_2^2\notag\\
	=&\|\mathbf{a}_{ji}\|^2_2-\|\mathbf{P}_{K-1}\mathbf{a}_{ji}-\widetilde{\mathbf{P}}_{K-1}\mathbf{a}_{ji}\|^2_2+O(\Lambda_p^2s_n)\|\mathbf{a}_{ji}\|_2^2.\label{1233}
\end{align}
By the second bound in $\eqref{1053}$, $\eqref{1236}$ and $\eqref{1215}$,
\begin{align}
   &\|\mathbf{P}_{K-1}\mathbf{a}_{ji}-\widetilde{\mathbf{P}}_{K-1}\mathbf{a}_{ji}\|^2_2\le\|\mathbf{P}_{K-1} -\widetilde{\mathbf{P}}_{K-1} \|^2\|\mathbf{a}_{ji}\|_2^2\le \|\boldsymbol{\Gamma}\boldsymbol{\Gamma}\trans-\widehat{\boldsymbol{\Gamma}}\trans\mathbf{K}^{-1}\widehat{\boldsymbol{\Gamma}}\|^2\|\mathbf{a}_{ji}\|_2^2\notag\\
	\le& \|\boldsymbol{\Gamma}\boldsymbol{\Gamma}\trans-\widehat{\boldsymbol{\Gamma}}\trans \widehat{\boldsymbol{\Gamma}}\|^2\|\mathbf{a}_{ji}\|_2^2+\| \widehat{\boldsymbol{\Gamma}}\trans(\mathbf{I}_{K-1}-\mathbf{K}^{-1})\widehat{\boldsymbol{\Gamma}}\|^2\|\mathbf{a}_{ji}\|_2^2=O(\Lambda_p^2s_n)\|\mathbf{a}_{ji}\|_2^2.\label{1235}
\end{align}
Hence, by $\eqref{1233}$ and $\eqref{1235}$,
\begin{align}
   &\|\widehat{\mathbf{a}}_{ji}\|_2^2=\|\mathbf{a}_{ji}\|^2_2+\|\mathbf{a}_{ji}\|_2^2O(\Lambda_p^2s_n).\label{1340}
\end{align}
Therefore, the first equality in  the condition $\eqref{104}$ is verified. For the second one, by the orthogonal decomposition $\eqref{100400}$, we have $t_{ji}=\mathbf{a}_{ji}\trans\widehat{\mathbf{a}}_{ji}/\|\widehat{\mathbf{a}}_{ji}\|^2_2$. By $\eqref{1200}$ and $\eqref{10036}$,
 \begin{align}
   &\mathbf{a}_{ji}\trans\widehat{\mathbf{a}}_{ji}=\boldsymbol{\delta}_{ij}\trans\boldsymbol{\Sigma}^{-1/2}\boldsymbol{\Sigma}^{1/2}\widehat{\mathbf{D}}\widehat{\boldsymbol{\delta}}_{ji}=\boldsymbol{\delta}_{ij}\trans \boldsymbol{\Sigma}^{-1/2}\widehat{\boldsymbol{\Gamma}} \widehat{\mathbf{K}}^{-1}\widehat{\boldsymbol{\Gamma}}\trans\boldsymbol{\Sigma}^{-1/2}\widehat{\boldsymbol{\delta}}_{ji}\notag\\
	 =& \boldsymbol{\delta}_{ij}\trans \boldsymbol{\Sigma}^{-1/2}\widehat{\boldsymbol{\Gamma}} \widehat{\mathbf{K}}^{-1}\widehat{\boldsymbol{\Gamma}}\trans\boldsymbol{\Sigma}^{-1/2}(\widehat{\boldsymbol{\delta}}_{ji}-\boldsymbol{\delta}_{ij})+\boldsymbol{\delta}_{ij}\trans \boldsymbol{\Sigma}^{-1/2}\widehat{\boldsymbol{\Gamma}} \widehat{\mathbf{K}}^{-1}\widehat{\boldsymbol{\Gamma}}\trans\boldsymbol{\Sigma}^{-1/2} \boldsymbol{\delta}_{ij}\notag\\
	=&\boldsymbol{\delta}_{ij}\trans \boldsymbol{\Sigma}^{-1/2}\widehat{\boldsymbol{\Gamma}} \widehat{\mathbf{K}}^{-1}\widehat{\boldsymbol{\Gamma}}\trans\boldsymbol{\Sigma}^{-1/2}(\widehat{\boldsymbol{\delta}}_{ji}-\boldsymbol{\delta}_{ij})+ \mathbf{a}_{ji}\trans\widetilde{\mathbf{P}}_{K-1}\mathbf{a}_{ji} \notag\\
	=&\boldsymbol{\delta}_{ij}\trans \boldsymbol{\Sigma}^{-1/2}\widehat{\boldsymbol{\Gamma}} \widehat{\mathbf{K}}^{-1}\widehat{\boldsymbol{\Gamma}}\trans\boldsymbol{\Sigma}^{-1/2}(\widehat{\boldsymbol{\delta}}_{ji}-\boldsymbol{\delta}_{ij})+ \mathbf{a}_{ji}\trans \widetilde{\mathbf{P}}_{K-1}\widetilde{\mathbf{P}}_{K-1}\mathbf{a}_{ji} .\label{1237}
\end{align} 
By $\eqref{1218}$, the first term in the last line of $\eqref{1237}$
 \begin{align}
   & |\boldsymbol{\delta}_{ij}\trans \boldsymbol{\Sigma}^{-1/2}\widehat{\boldsymbol{\Gamma}} \widehat{\mathbf{K}}^{-1}\widehat{\boldsymbol{\Gamma}}\trans\boldsymbol{\Sigma}^{-1/2}(\widehat{\boldsymbol{\delta}}_{ji}-\boldsymbol{\delta}_{ij})|\le \|\boldsymbol{\Sigma}^{-1/2}\boldsymbol{\delta}_{ij}\|_2 \|\widehat{\boldsymbol{\Gamma}} \widehat{\mathbf{K}}^{-1}\widehat{\boldsymbol{\Gamma}}\trans\boldsymbol{\Sigma}^{-1/2}(\widehat{\boldsymbol{\delta}}_{ji}-\boldsymbol{\delta}_{ji})\|_2\notag\\
	=&\|\mathbf{a}_{ji}\|_2O(\Lambda_p^2s_n)\|\mathbf{a}_{ji}\|_2=O(\Lambda_p^2s_n)\|\mathbf{a}_{ji}\|_2^2\label{1337}
\end{align} 
By $\eqref{1233}$ and $\eqref{1235}$, the second term in the last line of $\eqref{1237}$ is equal to
 \begin{align*}
   \|\widetilde{\mathbf{P}}_{K-1}\mathbf{a}_{ji}\|^2_2 = \|\mathbf{a}_{ji}\|^2_2-\|(\mathbf{P}_{K-1}-\widetilde{\mathbf{P}}_{K-1})\mathbf{a}_{ji}\|^2_2=\|\mathbf{a}_{ji}\|_2^2-O(\Lambda_p^2s_n)\|\mathbf{a}_{ji}\|_2^2,
\end{align*} 
which together with $\eqref{1237}$ and $\eqref{1337}$ imply that
\begin{align}
   &\mathbf{a}_{ji}\trans\widehat{\mathbf{a}}_{ji}=\|\mathbf{a}_{ji}\|^2_2+\|\mathbf{a}_{ji}\|_2^2O(\Lambda_p^2s_n),\text{ and hence } t_{ji}=\frac{\mathbf{a}_{ji}\trans\widehat{\mathbf{a}}_{ji}}{\|\widehat{\mathbf{a}}_{ji}\|^2_2}=1+O(\Lambda_p^2s_n),\label{1238}
\end{align}
by $\eqref{1340}$. The second condition in  $\eqref{104}$ is verified.  For the last condition in  $\eqref{104}$, by the definitions of $d_{ji}$ and $\widehat{d}_{ji}$, and $\eqref{1200}$,
 \begin{align}
& d_{ji}-\widehat{d}_{ji}=\frac{1}{2}\|\mathbf{a}_{ji}\|_2^2-\widehat{\mathbf{a}}_{ji}\trans\boldsymbol{\Sigma}^{-1/2}(\widehat{\mathbf{b}}_{ji}-\boldsymbol{\mu}_i)=\frac{1}{2}\|\mathbf{a}_{ji}\|_2^2-\widehat{\boldsymbol{\delta}}_{ji}\trans \widehat{\mathbf{D}}\left(\frac{\bar{\mathbf{x}}_j+\bar{\mathbf{x}}_i}{2}-\boldsymbol{\mu}_i\right)\notag\\
=&\frac{1}{2}\|\mathbf{a}_{ji}\|_2^2-\widehat{\boldsymbol{\delta}}_{ji}\trans \widehat{\mathbf{D}}\left(\frac{\bar{\mathbf{x}}_j+\bar{\mathbf{x}}_i}{2}-\frac{\boldsymbol{\mu}_j+\boldsymbol{\mu}_i}{2}+\frac{(\boldsymbol{\mu}_j-\boldsymbol{\mu}_i)}{2}\right)\notag\\
=&\frac{1}{2}\|\mathbf{a}_{ji}\|_2^2-\widehat{\boldsymbol{\delta}}_{ji}\trans \widehat{\mathbf{D}}\left(\frac{\bar{\mathbf{x}}_i-\boldsymbol{\mu}_i}{2}+\frac{\bar{\mathbf{x}}_j-\boldsymbol{\mu}_j}{2}\right)-\frac{1}{2}\widehat{\boldsymbol{\delta}}_{ji}\trans \widehat{\mathbf{D}}\boldsymbol{\delta}_{ji}\notag\\
=&\frac{\|\mathbf{a}_{ji}\|_2^2}{2}-\widehat{\boldsymbol{\delta}}_{ji}\trans \boldsymbol{\Sigma}^{-1/2}\widehat{\boldsymbol{\Gamma}} \widehat{\mathbf{K}}^{-1}\widehat{\boldsymbol{\Gamma}}\trans\boldsymbol{\Sigma}^{-1/2}\left(\frac{\bar{\mathbf{x}}_i-\boldsymbol{\mu}_i}{2}+\frac{\bar{\mathbf{x}}_j-\boldsymbol{\mu}_j}{2}\right)-\frac{\mathbf{a}_{ji}\trans\widehat{\mathbf{a}}_{ji}}{2}\notag\\
=&\frac{\|\mathbf{a}_{ji}\|_2^2}{2}-\frac{\mathbf{a}_{ji}\trans\widehat{\mathbf{a}}_{ji}}{2}-\widehat{\boldsymbol{\delta}}_{ji}\trans \boldsymbol{\Sigma}^{-1/2}\widehat{\boldsymbol{\Gamma}} \widehat{\mathbf{K}}^{-1}\widehat{\boldsymbol{\Gamma}}\trans\widehat{\boldsymbol{\Gamma}} \widehat{\mathbf{K}}^{-1}\widehat{\boldsymbol{\Gamma}}\trans\boldsymbol{\Sigma}^{-1/2}\left(\frac{\bar{\mathbf{x}}_i-\boldsymbol{\mu}_i}{2}+\frac{\bar{\mathbf{x}}_j-\boldsymbol{\mu}_j}{2}\right), \label{1239}
\end{align} 
where in the last line, we use the fact that $\widehat{\boldsymbol{\Gamma}} \widehat{\mathbf{K}}^{-1}\widehat{\boldsymbol{\Gamma}}=\widetilde{\mathbf{P}}_{K-1}$ is a projection matrix and hence $\widetilde{\mathbf{P}}_{K-1}^2=\widetilde{\mathbf{P}}_{K-1}$. We estimate the last term on the right hand side of $\eqref{1239}$,
 \begin{align}
&  \left|\widehat{\boldsymbol{\delta}}_{ji}\trans \boldsymbol{\Sigma}^{-1/2}\widehat{\boldsymbol{\Gamma}} \widehat{\mathbf{K}}^{-1}\widehat{\boldsymbol{\Gamma}}\trans\widehat{\boldsymbol{\Gamma}} \widehat{\mathbf{K}}^{-1}\widehat{\boldsymbol{\Gamma}}\trans\boldsymbol{\Sigma}^{-1/2}\left(\frac{\bar{\mathbf{x}}_i-\boldsymbol{\mu}_i}{2}+\frac{\bar{\mathbf{x}}_j-\boldsymbol{\mu}_j}{2}\right)\right| \notag\\
\le&\|\widehat{\boldsymbol{\Gamma}} \widehat{\mathbf{K}}^{-1}\widehat{\boldsymbol{\Gamma}}\trans\boldsymbol{\Sigma}^{-1/2}\widehat{\boldsymbol{\delta}}_{ji}\|_2\left\|\widehat{\boldsymbol{\Gamma}} \widehat{\mathbf{K}}^{-1}\widehat{\boldsymbol{\Gamma}}\trans\boldsymbol{\Sigma}^{-1/2}\left(\frac{\bar{\mathbf{x}}_i-\boldsymbol{\mu}_i}{2}+\frac{\bar{\mathbf{x}}_j-\boldsymbol{\mu}_j}{2}\right)\right\|_2\notag\\
=&\| \boldsymbol{\Sigma}^{1/2}\widehat{\mathbf{D}}\widehat{\boldsymbol{\delta}}_{ji}\|_2\left\|\widehat{\boldsymbol{\Gamma}} \widehat{\mathbf{K}}^{-1}\widehat{\boldsymbol{\Gamma}}\trans\boldsymbol{\Sigma}^{-1/2}\left(\frac{\bar{\mathbf{x}}_i-\boldsymbol{\mu}_i}{2}+\frac{\bar{\mathbf{x}}_j-\boldsymbol{\mu}_j}{2}\right)\right\|_2 \notag\\
=&\|\widehat{\mathbf{a}}_{ji}\|_2\left\|\widehat{\boldsymbol{\Gamma}} \widehat{\mathbf{K}}^{-1}\widehat{\boldsymbol{\Gamma}}\trans\boldsymbol{\Sigma}^{-1/2}\left(\frac{\bar{\mathbf{x}}_i-\boldsymbol{\mu}_i}{2}+\frac{\bar{\mathbf{x}}_j-\boldsymbol{\mu}_j}{2}\right)\right\|_2 \le\|\widehat{\mathbf{a}}_{ji}\|_2 O(\Lambda_p^2s_n)\|\mathbf{a}_{ji}\|_2,\label{1339}
\end{align}  
where the last inequality can be obtained by a similar argument as in $\eqref{1218}$. it follows from $\eqref{1239}$ and $\eqref{1339}$ that
 \begin{align}
& d_{ji}-\widehat{d}_{ji}=\frac{\|\mathbf{a}_{ji}\|_2^2}{2}-\frac{\mathbf{a}_{ji}\trans\widehat{\mathbf{a}}_{ji}}{2}+\|\widehat{\mathbf{a}}_{ji}\|_2 O(\Lambda_p^2s_n)\|\mathbf{a}_{ji}\|_2= O(\Lambda_p^2s_n)\|\mathbf{a}_{ji}\|_2^2, \label{1239}
\end{align} 
where the last inequality is due to $\eqref{1340}$ and $\eqref{1238}$. Hence, the thir condition in  $\eqref{104}$ is verified. Hence, we can apply Theorem \ref{theorem_7} to obtain the theorem.\\

\section*{Appendix B: Proofs of technical lemmas}

 \renewcommand{\proof}{\noindent\underline{\bf Proof of Lemma}}

 In Appendix B, we provide the proofs for all technical lemmas.\\

\begin{proof} $\ref{lemma_1}$ \\

We will use the Bernstein's inequality for bounded variables (see Lemma 2.2.9 in \citet{van1996weak} or page 855 of \citet{shorack2009empirical}). Given $1\le i\le K$, define  i.i.d. random variables $Z_j$, $1\le j\le n$, where $Z_j=1$ if the $j$th sample observation belongs to the $i$-th class, otherwise $Z_j=0$. Hence, $Z_j$ has the Binomial distribution with parameters $1$ and $1/K$, and it mean and variance are $1/K$ and $(1-1/K)1/K$. Let $Y_j=Z_j-1/K$. Then $EY_j=0$ and $Var(Y_j)=1/K(1-1/K)$. Since $-1\le Y_i\le 1$, by the Bernstein's inequality for bounded variables,
\begin{align*}
&P\left( \left|Y_1+\cdots+Y_n\right|>x\right)\le 2\exp{(-\frac{1}{2}\frac{x^2}{nVar(Y_1)+x/3})},
 \end{align*}
for any $x>0$. For any constant $C>0$, let $x=nC\sqrt{\frac{\log{p}}{Kn}}$. Then we have  
 \begin{align}
&P\left( \left|\frac{n_i}{n}-\frac{1}{K}\right|>C\sqrt{\frac{\log{p}}{Kn}}\right)=P\left( \left|Y_1+\cdots+Y_n\right|>nC\sqrt{\frac{\log{p}}{Kn}}\right)\notag\\
\le& 2\exp{\left(-\frac{1}{2}\frac{n^2C^2\frac{\log{p}}{Kn}}{\frac{n}{K}(1-\frac{1}{K})+n\frac{C}{3}\sqrt{\frac{\log{p}}{Kn}}}\right)}\le 2\exp{\left(-\frac{1}{2}\frac{C^2\log{p}}{1+\frac{C}{3}\sqrt{\frac{K\log{p}}{n}}}\right)}.\label{10006}
 \end{align} 
By the conditions, $p\ge 2$, $\sqrt{K\log{p}/n}\le d_0$ and $K\le p+1$, $\eqref{10006}$ gives
  \begin{align}
&P\left( \max_{1\le i\le K} \left|\frac{n_i}{n}-\frac{1}{K}\right|>C\sqrt{\frac{\log{p}}{Kn}}\right) \le 2K\exp{\left(-\frac{1}{2}\frac{C^2\log{p}}{1+\frac{C}{3}\sqrt{\frac{K\log{p}}{n}}}\right)}\label{100006}\\
\le& 2(p+1)\exp{\left(- \frac{C^2\log{p}}{2+2Cd_0/3}\right)} \le 2(p+1)p^{-C^2/(2+2Cd_0/3)}\le p^{3-C^2/(2+2Cd_0/3)}.\notag
 \end{align} 
For any $M>0$,  when $C\ge (M+3)(d_0+1)$, we have $C>3$ and 
 \begin{align*}
& (M+3)(2+2Cd_0/3)\le (M+3)(C+Cd_0)\le C(M+3)(d_0+1)\le C^2.
 \end{align*} 
Then $C^2/(2+2Cd_0/3)-3\ge M$ and by $\eqref{100006}$, we have
 \begin{align*}
&P\left( \max_{1\le i\le K} \left|\frac{n_i}{n}-\frac{1}{K}\right|>C\sqrt{\frac{\log{p}}{Kn}}\right) \le  p^{3-C^2/(2+2Cd_0/3)}\le p^{-M}.
 \end{align*} \\

 \end{proof}

\begin{proof} $\ref{lemma_8}$ \\

Given $1\le k\le K-1$, let $t_i=\boldsymbol{\mu}_i\trans\boldsymbol{\alpha}_k$ for $i=1,\ldots, K$. Then we have $t_{K}=-\sum_{i=1}^{K-1}t_i$ because $\sum_{i=1}^Kt_i=(\sum_{i=1}^K\boldsymbol{\mu}_i)\trans\boldsymbol{\alpha}_k=0$ by $\eqref{3}$. By $\eqref{10003}$,
 \begin{align}
 &\boldsymbol{\Sigma}\boldsymbol{\alpha}_k= \frac{1}{\lambda_k(\boldsymbol{\Xi})}\mathbf{B}\boldsymbol{\alpha}_k=\frac{1}{\lambda_k(\boldsymbol{\Xi})K}\sum_{i=1}^K\boldsymbol{\mu}_i(\boldsymbol{\mu}_i\trans\boldsymbol{\alpha}_k)=\frac{1}{\lambda_k(\boldsymbol{\Xi})K}\left[\sum_{i=1}^{K-1}t_i\boldsymbol{\mu}_i+ t_{K}\boldsymbol{\mu}_{K}\right]\notag\\
=&\frac{1}{\lambda_k(\boldsymbol{\Xi})K}\left[\sum_{i=1}^{K-1}t_i\boldsymbol{\mu}_i-\sum_{i=1}^{K-1}t_i\boldsymbol{\mu}_{K}\right]=\frac{1}{\lambda_k(\boldsymbol{\Xi})K} \sum_{i=1}^{K-1}t_i(\boldsymbol{\mu}_i-\boldsymbol{\mu}_{K}) \label{1250}
\end{align} 
  It follows from $\eqref{1250}$ that
 \begin{align}
 &\|\boldsymbol{\Sigma}\boldsymbol{\alpha}_k\|_1\le\frac{1}{\lambda_k(\boldsymbol{\Xi})K} \sum_{i=1}^{K-1}|t_i|\|\boldsymbol{\mu}_i-\boldsymbol{\mu}_{K}\|_1 \le\frac{1}{\lambda_k(\boldsymbol{\Xi})K} (\sum_{i=1}^{K}|t_i|)\left(\max_{1\le i\neq j\le K}\|\boldsymbol{\mu}_i-\boldsymbol{\mu}_j\|_1\right) \notag\\
 \le &\frac{1}{\lambda_k(\boldsymbol{\Xi})K} \sqrt{K\sum_{i=1}^{K}t_i^2}\left(\max_{1\le i\neq j\le K}\|\boldsymbol{\mu}_i-\boldsymbol{\mu}_j\|_1\right) =\frac{1}{\lambda_k(\boldsymbol{\Xi})} \sqrt{\frac{\sum_{i=1}^{K}t_i^2}{K}}\left(\max_{1\le i\neq j\le K}\|\boldsymbol{\mu}_i-\boldsymbol{\mu}_j\|_1\right)\notag\\
=&\frac{1}{\sqrt{\lambda_k(\boldsymbol{\Xi})}} \left(\max_{1\le i\neq j\le K}\|\boldsymbol{\mu}_i-\boldsymbol{\mu}_j\|_1\right),\label{1251}
\end{align} 
where the last equality is due to $\sum_{i=1}^{K}t_i^2/K=\sum_{i=1}^{K}(\boldsymbol{\mu}_i\trans\boldsymbol{\alpha}_k)^2/K=\boldsymbol{\alpha}_k\trans\mathbf{B}\boldsymbol{\alpha}_k=\lambda_k(\boldsymbol{\Xi})$. By Condition \ref{condition_2} (c), $\lambda_k(\boldsymbol{\Xi})\ge \lambda_{K-1}(\boldsymbol{\Xi})\ge c_3^{-1}\lambda_1(\boldsymbol{\Xi})$ which together with $\eqref{1251}$ give
 \begin{align*}
 &\max_{1\le k\le K-1}\|\boldsymbol{\Sigma}\boldsymbol{\alpha}_k\|_1\le \frac{
\sqrt{c_3}}{\sqrt{\lambda_1(\boldsymbol{\Xi})}} \left(\max_{1\le i\neq j\le K}\|\boldsymbol{\mu}_i-\boldsymbol{\mu}_j\|_1\right). 
\end{align*} 
On the other hand,  given $1\le i\neq j\le K$, by $\eqref{1173}$ in the proof of Lemma \ref{lemma_3},
\begin{align}
\boldsymbol{\mu}_j-\boldsymbol{\mu}_i= \mathbf{B}\sum_{k=1}^{K-1}s_k\boldsymbol{\alpha}_k=  \sum_{k=1}^{K-1}s_k\lambda_k(\boldsymbol{\Xi})\boldsymbol{\Sigma}\boldsymbol{\alpha}_k,\label{1252}
\end{align}
where $s_k$'s are real numbers. Multiplying $\boldsymbol{\alpha}_k\trans$ on both sides of $\eqref{1252}$, we can obtain $\boldsymbol{\alpha}_k\trans(\boldsymbol{\mu}_j-\boldsymbol{\mu}_i)=s_k\lambda_k(\boldsymbol{\Xi})$. Note that by Lemma \ref{lemma_6} and Condition \ref{condition_1} (b), 
 \begin{align}
\|\boldsymbol{\mu}_j-\boldsymbol{\mu}_i\|_2^2\le c_0 (\boldsymbol{\mu}_i-\boldsymbol{\mu}_j)\trans\boldsymbol{\Sigma} (\boldsymbol{\mu}_i-\boldsymbol{\mu}_j)\le 2c_0\lambda_{max}(\boldsymbol{\Delta})=2c_0\lambda_1(\boldsymbol{\Delta})=2c_0K\lambda_1(\boldsymbol{\Xi}),\label{1253}
\end{align} 
and $\|\boldsymbol{\alpha}_k\|_2^2\le c_0 \boldsymbol{\alpha}_k\trans\boldsymbol{\Sigma} \boldsymbol{\alpha}_k= c_0 $. By $\eqref{1252}$ and $\eqref{1253}$,
\begin{align}
&\|\boldsymbol{\mu}_j-\boldsymbol{\mu}_i\|_1\le\sum_{k=1}^{K-1}|s_k\lambda_k(\boldsymbol{\Xi})|\|\boldsymbol{\Sigma}\boldsymbol{\alpha}_k\|_1=\sum_{k=1}^{K-1}|\boldsymbol{\alpha}_k\trans(\boldsymbol{\mu}_j-\boldsymbol{\mu}_i)|\|\boldsymbol{\Sigma}\boldsymbol{\alpha}_k\|_1 \notag\\
\le& \sum_{k=1}^{K-1}\|\boldsymbol{\alpha}_k\|_2\|\boldsymbol{\mu}_j-\boldsymbol{\mu}_i\|_2\|\boldsymbol{\Sigma}\boldsymbol{\alpha}_k\|_1\le (K-1)\sqrt{c_0}\sqrt{2c_0K\lambda_1(\boldsymbol{\Xi})}\left(\max_{1\le k\le K-1}\|\boldsymbol{\Sigma}\boldsymbol{\alpha}_k\|_1\right).\label{1254}
\end{align}
Therefore,
\begin{align*}
&\frac{1}{(K-1)c_0\sqrt{2K\lambda_1(\boldsymbol{\Xi})}}\left(\max_{1\le i\neq j\le K}\|\boldsymbol{\mu}_i-\boldsymbol{\mu}_j\|_1\right)
\le \max_{1\le k\le K-1}\|\boldsymbol{\Sigma}\boldsymbol{\alpha}_k\|_1 \;.
\end{align*}
\\

\end{proof}

\begin{proof} $\ref{lemma_3}$\\

 By the definition $\eqref{4}$ of $\mathbf{B}$, for any $1\le i\neq j\le K-1$,
\begin{align*}
(\mathbf{U}\trans\boldsymbol{\alpha}_i)\trans(\mathbf{U}\trans\boldsymbol{\alpha}_j)=\boldsymbol{\alpha}_i\trans\mathbf{U}\mathbf{U}\trans\boldsymbol{\alpha}_j=K\boldsymbol{\alpha}_i\trans\mathbf{B}\boldsymbol{\alpha}_j=K\lambda_j(\boldsymbol{\Xi})\boldsymbol{\alpha}_i\trans\boldsymbol{\Sigma}\trans\boldsymbol{\alpha}_j=0,
\end{align*}
and by $\eqref{3}$, $\mathbf{1}_K\trans\mathbf{U}\trans\boldsymbol{\alpha}_i=0$ for any $1\le i\le K-1$. Therefore, $\{\mathbf{1}_K, \mathbf{U}\trans\boldsymbol{\alpha}_1,\cdots,\mathbf{U}\trans\boldsymbol{\alpha}_{K-1}\}$ forms an orthogonal basis of $\mathbb{R}^{K}$. For any  $1\le i\neq j\le K$, let $\mathbf{v}_{ij}$ be the vector in $\mathbb{R}^{K}$ with all coordinates equal to zero except the $i$th and the $j$-th coordinates which are equal to $-1$ and 1, respectively. Let
\begin{align*}
\mathbf{v}_{ij}=a\mathbf{1}_K+\sum_{k=1}^{K-1}b_k\mathbf{U}\trans\boldsymbol{\alpha}_k,
\end{align*}
be the orthogonal expansion of $\mathbf{v}_{ij}$, where  and $a$ and $b_k$ are coefficients. Since $\mathbf{v}_{ij}$ is orthogonal to $\mathbf{1}_K$,   we have $a=0$. Now
\begin{align}
\boldsymbol{\mu}_j-\boldsymbol{\mu}_i=\mathbf{U}\mathbf{v}_{ij}=\mathbf{U}\sum_{k=1}^{K-1}b_k\mathbf{U}\trans\boldsymbol{\alpha}_k=\mathbf{B}\mathbf{z},\label{1173}
\end{align}
where $\mathbf{z}=K\sum_{k=1}^{K-1}b_k\boldsymbol{\alpha}_k$ is a linear combination of $\boldsymbol{\alpha}_k$, $1\le k\le K-1$. By the eigen-decomposition, $\boldsymbol{\Sigma}^{-1/2}\mathbf{B}\boldsymbol{\Sigma}^{-1/2}=\boldsymbol{\Xi}=\sum_{k=1}^{K-1}\lambda_k(\boldsymbol{\Xi})\boldsymbol{\gamma}_k\boldsymbol{\gamma}_k\trans$. Hence, 
\begin{align*}
\mathbf{B}=\sum_{k=1}^{K-1}\lambda_k(\boldsymbol{\Xi})\boldsymbol{\Sigma}^{1/2}\boldsymbol{\gamma}_k\boldsymbol{\gamma}_k\trans\boldsymbol{\Sigma}^{1/2}=\sum_{k=1}^{K-1}\lambda_k(\boldsymbol{\Xi})\boldsymbol{\Sigma} \boldsymbol{\alpha}_k\boldsymbol{\alpha}_k\trans\boldsymbol{\Sigma}. 
\end{align*}
Because $\boldsymbol{\alpha}_k\trans \mathbf{B}=\lambda_k(\boldsymbol{\Xi}) \boldsymbol{\alpha}_k\trans \boldsymbol{\Sigma}$ and 
\begin{align}
&\mathbf{D}(\boldsymbol{\mu}_j-\boldsymbol{\mu}_i)=\sum_{k=1}^{K-1}\boldsymbol{\alpha}_k\boldsymbol{\alpha}_k\trans \mathbf{B}\mathbf{z}=\sum_{k=1}^{K-1}\lambda_k(\boldsymbol{\Xi})\boldsymbol{\alpha}_k\boldsymbol{\alpha}_k\trans \boldsymbol{\Sigma}\mathbf{z}=\boldsymbol{\Sigma}^{-1}\sum_{k=1}^{K-1}\lambda_k(\boldsymbol{\Xi})\boldsymbol{\Sigma}\boldsymbol{\alpha}_k\boldsymbol{\alpha}_k\trans \boldsymbol{\Sigma}\mathbf{z}\notag\\
=&\boldsymbol{\Sigma}^{-1} \mathbf{B}\mathbf{z}=\boldsymbol{\Sigma}^{-1}(\boldsymbol{\mu}_j-\boldsymbol{\mu}_i).\label{1231}
\end{align}
Hence, the lemma is proved.\\

\end{proof}

\begin{proof} $\ref{lemma_6}$\\

Let $\Phi=\boldsymbol{\Sigma}^{-1/2}\mathbf{U}$. Then by the definitions the definition $\eqref{4}$ and  $\eqref{2}$, we have 
\begin{align*}
  \boldsymbol{\Delta}=\Phi\trans\Phi, \quad \boldsymbol{\Xi}=\frac{1}{K}\Phi\Phi\trans.
\end{align*}
By $\eqref{3}$,  $\mathbf{U}\mathbf{1}_K=0$, where $\mathbf{1}_K=(1,1,\cdots,1)\trans$. Hence $\boldsymbol{\Delta}\mathbf{1}_K=0$.  Since $\boldsymbol{\Delta}$ is $K\times K$, the rank of $\boldsymbol{\Delta}$ is at most $K-1$ and it has at most $K-1$ nonzero eigenvalues. For any $1\le i\le K-1$, since $\boldsymbol{\gamma}_i$ is the $i$-th eigenvector of $\boldsymbol{\Xi}$ with the eigenvalue $\lambda_i(\boldsymbol{\Xi})$, we have
\begin{align}
  \boldsymbol{\Delta}\Phi\trans\boldsymbol{\gamma}_i=\Phi\trans\Phi\Phi\trans\boldsymbol{\gamma}_i=K\Phi\trans\boldsymbol{\Xi} \boldsymbol{\gamma}_i=K\Phi\lambda_i(\boldsymbol{\Xi})\boldsymbol{\gamma}_i=K\lambda_i(\boldsymbol{\Xi})\Phi\boldsymbol{\gamma}_i.\label{1300}
\end{align}
Therefore, $\Phi\trans\boldsymbol{\gamma}_i$ is the eigenvector of $\boldsymbol{\Delta}$ with the eigenvalue $K\lambda_i(\boldsymbol{\Xi})$, $1\le i\le K-1$. Hence, all the nonzero eigenvalues of $\boldsymbol{\Delta}$ are 
\begin{align*}
   \lambda_{K-1}(\boldsymbol{\Delta})=K\lambda_{K-1}(\boldsymbol{\Xi})\le \cdots\le \lambda_{2}(\boldsymbol{\Delta})=K\lambda_2(\boldsymbol{\Xi})\le \lambda_1(\boldsymbol{\Delta})=K\lambda_1(\boldsymbol{\Xi}) .
\end{align*}
 To prove the inequalities in the lemma, we define $\mathbf{v}_{ij}$ to be the $K$-vector with all coordinates are equal to zeros except the $i$th and $j$th coordinates which are equal to 1 and -1, respectively, where $1\le i\neq j\le K$. Then
\begin{align*}
 (\boldsymbol{\mu}_i-\boldsymbol{\mu}_j)\trans\boldsymbol{\Sigma}^{-1}(\boldsymbol{\mu}_i-\boldsymbol{\mu}_j)=\mathbf{v}_{ij}\trans\mathbf{U}\trans\boldsymbol{\Sigma}^{-1}\mathbf{U}\mathbf{v}_{ij}=\mathbf{v}_{ij}\trans\boldsymbol{\Delta}\mathbf{v}_{ij}\le \|\mathbf{v}_{ij}\|_2^2\lambda_{max}(\boldsymbol{\Delta})=2\lambda_{max}(\boldsymbol{\Delta}),
\end{align*}
and $\mathbf{v}_{ij}\trans\mathbf{1}_K=0$. Since $\mathbf{1}_K$ is the eigenvector of $\boldsymbol{\Delta}$ with eigenvalue zero, all the eigenvalues of $\boldsymbol{\Delta}+\lambda^+_{min}(\boldsymbol{\Delta})\mathbf{1}_K\mathbf{1}_K\trans/K$ are not less than $\lambda^+_{min}(\boldsymbol{\Delta})$, where $\lambda^+_{min}(\boldsymbol{\Delta})=\lambda_{K-1}(\boldsymbol{\Delta})=K\lambda_{K-1}(\boldsymbol{\Xi})$. Hence,
\begin{align*}
 &(\boldsymbol{\mu}_i-\boldsymbol{\mu}_j)\trans\boldsymbol{\Sigma}^{-1}(\boldsymbol{\mu}_i-\boldsymbol{\mu}_j)=\mathbf{v}_{ij}\trans\boldsymbol{\Delta}\mathbf{v}_{ij}=\mathbf{v}_{ij}\trans\left[\boldsymbol{\Delta}+\lambda^+_{min}(\boldsymbol{\Delta})\mathbf{1}_K\mathbf{1}_K\trans/K\right]\mathbf{v}_{ij}\notag\\
\ge& \|\mathbf{v}_{ij}\|_2^2\lambda^+_{min}(\boldsymbol{\Delta})=2\lambda^+_{min}(\boldsymbol{\Delta}) =2\lambda_{K-1}(\boldsymbol{\Delta})=2K\lambda_{K-1}(\boldsymbol{\Xi})\ge 2Kc_1,
\end{align*}
where the last inequality is due to Condition \ref{condition_2} (a).\\

\end{proof}

\begin{proof} $\ref{lemma_2}$\\

We only consider the case that $(n_1,n_2,\cdots,n_K)$ follows a multinomial distribution. For the nonrandom case, a similar argument can prove the lemma. Let $\bar{\mathbf{x}}^k_j$ denote the $k$-th coordinate of the $j$-th sample class mean $\bar{\mathbf{x}}_j$ and $\sigma_{kk}$ is the $k$-th diagonal element of $\boldsymbol{\Sigma}$. Since $\sqrt{n_i}(\bar{\mathbf{x}}^k_j-\boldsymbol{\mu}^k_j)/\sqrt{\sigma_{kk}}$ has a standard normal distribution,  for any $C_1>0$,
 \begin{align}
&   P\left( |\bar{\mathbf{x}}^k_j-\boldsymbol{\mu}^k_j|\ge C_1\sqrt{\frac{\sigma_{kk}\log{p}}{n_j}}\right)\notag\\
=&1-\Phi(C_1\sqrt{\log{p}})\le \frac{\phi(C_1\sqrt{\log{p}})}{C_1\sqrt{\log{p}}}=\frac{1}{\sqrt{2\pi\log{p}}C_1p^{C_1^2/2}},\label{1261}
\end{align} 
where $\Phi$ and $\phi$ are the cumulative and density functions of the standard normal distribution and we use the inequality $1-\Phi(x)\le \phi(x)/x$ for any $x>0$ (see page 850 in \citet{shorack2009empirical}).   For any $M^\prime>0$, let $C^\prime=2(M^\prime+3)$. Since $K\log{p}/n\to 0$,  $\sqrt{K\log{p}/n}\le \min\{1, 1/(2C^\prime)\}$ for all $n$ large enough.  By Lemma \ref{lemma_1},  
\begin{align}
   P\left( \max_{1\le i\le K} \left|\frac{n_i}{n}-\frac{1}{K}\right|\le C^\prime\sqrt{\frac{\log{p}}{Kn}}\right) \ge 1-p^{-M^\prime}.\label{1260}
\end{align}
Since $C^\prime\sqrt{\log{p}/(Kn)}=(C^\prime\sqrt{K\log{p}/n})/K\le 1/(2K)$, the inequality in the parenthesis in $\eqref{1260}$ implies $\min_{1\le i\le K}n_i\ge n/(2K)$. Hence, we have $P(\min_{1\le i\le K}n_i\ge n/(2K))\ge 1-p^{-M^\prime}$, which together with $\eqref{1261}$ and the inequality $|\sigma_{kk}|\le \lambda_{max}(\boldsymbol{\Sigma})\le c_0$ (see Condition \ref{condition_1} (b)) leads to 
 \begin{align*}
&   P\left( |\bar{\mathbf{x}}^k_j-\boldsymbol{\mu}^k_j|\ge C_1\sqrt{\frac{2c_0K\log{p}}{n}}\right)\le P\left( |\bar{\mathbf{x}}^k_j-\boldsymbol{\mu}^k_j|\ge C_1\sqrt{\frac{\sigma_{kk}\log{p}}{n_j}},\quad \min_{1\le i\le K}n_i\ge n/(2K)\right)\notag\\
&+P(\min_{1\le i\le K}n_i< n/(2K))\le \frac{1}{\sqrt{2\pi\log{p}}C_1p^{C_1^2/2}}+p^{-M^\prime}. 
\end{align*} 
 Hence, it follows from the above inequality that
\begin{align*}
&  P\left(\max_{1\le j\le K}\|\bar{\mathbf{x}}_j-\boldsymbol{\mu}_j\|_{\infty}> C_1\sqrt{\frac{c_0K\log{p}}{n}}\right)\\
\le &\sum_{1\le j\le K}\sum_{1\le k\le p}P\left( |\bar{\mathbf{x}}^k_j-\boldsymbol{\mu}^k_j|\ge C_1\sqrt{\frac{c_0K\log{p}}{n}}\right)\le\frac{Kp}{\sqrt{2\pi\log{p}}C_1p^{C_1^2/2}}+Kpp^{-M^\prime}.
\end{align*} 
Since $K\le p+1\le p^2$, for any $M>0$, we choose $C_1$ large enough such that the first term on the right hand side of the above inequality is less than $p^{-M}/2$. Let $M^\prime=M+4$, then the second  term on the right hand side of the above inequality is less than $p^{-M-1}\le p^M/2$. Hence,  $\eqref{134}$ is true for $C\ge C_1\sqrt{c_0}$ and all $n$ large enough.\\
 
\end{proof}

\begin{proof} $\ref{lemma_10}$\\

In this proof, we only consider the element in $\Omega_n$. Since $\widehat{\boldsymbol{\xi}}_1$ is the solution to $\eqref{1122}$ with $j=1$, we have 
\begin{align}
 \|\widehat{\boldsymbol{\xi}}_1-\widehat{\mathbf{B}}\widehat{\boldsymbol{\alpha}}_1\|^2_2+\kappa_n\|\widehat{\boldsymbol{\xi}}_1\|_1\le \|\mathbf{B}\boldsymbol{\alpha}_1-\widehat{\mathbf{B}}\widehat{\boldsymbol{\alpha}}_1\|^2_2+\kappa_n\|\mathbf{B}\boldsymbol{\alpha}_1\|_1 \label{1121}
\end{align}	
where
\begin{align}
 &\|\widehat{\boldsymbol{\xi}}_1-\widehat{\mathbf{B}}\widehat{\boldsymbol{\alpha}}_1\|^2_2=\|\widehat{\boldsymbol{\xi}}_1- \mathbf{B}\boldsymbol{\alpha}_1\|^2_2 +2(\widehat{\boldsymbol{\xi}}_1- \mathbf{B}\boldsymbol{\alpha}_1)^T(\mathbf{B}\boldsymbol{\alpha}_1-\widehat{\mathbf{B}}\widehat{\boldsymbol{\alpha}}_1)+\|\mathbf{B}\boldsymbol{\alpha}_1-\widehat{\mathbf{B}}\widehat{\boldsymbol{\alpha}}_1\|^2_2.\label{1120}
\end{align}	
It follows from $\eqref{1121}$ and $\eqref{1120}$ that
\begin{align}
 \|\widehat{\boldsymbol{\xi}}_1- \mathbf{B}\boldsymbol{\alpha}_1\|^2_2 +2(\widehat{\boldsymbol{\xi}}_1- \mathbf{B}\boldsymbol{\alpha}_1)^T(\mathbf{B}\boldsymbol{\alpha}_1-\widehat{\mathbf{B}}\widehat{\boldsymbol{\alpha}}_1)+\kappa_n\|\widehat{\boldsymbol{\xi}}_1\|_1\le  \kappa_n\|\mathbf{B}\boldsymbol{\alpha}_1\|_1 \label{1290}
\end{align}	
By Theorem \ref{theorem_12}, the definition $\eqref{1033}$ of $\Omega_n$, the definition $\eqref{1002}$ of $\Lambda_p$ and the fact that $\|\mathbf{B}\|\le \lambda_1(\boldsymbol{\Xi})\|\boldsymbol{\Sigma}\|\le \lambda_1(\boldsymbol{\Xi})c_0$,
{\allowbreak
 \begin{align*}
 & 2(\widehat{\boldsymbol{\xi}}_1- \mathbf{B}\boldsymbol{\alpha}_1)^T(\mathbf{B}\boldsymbol{\alpha}_1-\widehat{\mathbf{B}}\widehat{\boldsymbol{\alpha}}_1)\notag\\
=&2(\widehat{\boldsymbol{\xi}}_1- \mathbf{B}\boldsymbol{\alpha}_1)^T(\mathbf{B}\boldsymbol{\alpha}_1-\mathbf{B}\widehat{\boldsymbol{\alpha}}_1)+2(\widehat{\boldsymbol{\xi}}_1- \mathbf{B}\boldsymbol{\alpha}_1)^T(\mathbf{B}\widehat{\boldsymbol{\alpha}}_1-\widehat{\mathbf{B}}\widehat{\boldsymbol{\alpha}}_1)\notag\\
\ge & -2\|\widehat{\boldsymbol{\xi}}_1- \mathbf{B}\boldsymbol{\alpha}_1\|_2\|\mathbf{B}\boldsymbol{\alpha}_1-\mathbf{B}\widehat{\boldsymbol{\alpha}}_1\|_2-2\|\widehat{\boldsymbol{\xi}}_1-\mathbf{B}\boldsymbol{\alpha}_1\|_1\|\mathbf{B}-\widehat{\mathbf{B}}\|_{\infty}\|\widehat{\boldsymbol{\alpha}}_1\|_1\notag\\
\ge & -2\|\widehat{\boldsymbol{\xi}}_1- \mathbf{B}\boldsymbol{\alpha}_1\|_2\|\mathbf{B}\boldsymbol{\alpha}_1-\mathbf{B}\widehat{\boldsymbol{\alpha}}_1\|_2-2(\|\widehat{\boldsymbol{\xi}}_1\|_1+\|\mathbf{B}\boldsymbol{\alpha}_1\|_1)\|\mathbf{B}-\widehat{\mathbf{B}}\|_{\infty}\|\widehat{\boldsymbol{\alpha}}_1\|_1\notag\\
\ge & -2\|\widehat{\boldsymbol{\xi}}_1- \mathbf{B}\boldsymbol{\alpha}_1\|_2\|\mathbf{B}\boldsymbol{\alpha}_1-\mathbf{B}\widehat{\boldsymbol{\alpha}}_1\|_2-2(\|\widehat{\boldsymbol{\xi}}_1\|_1+\lambda_1(\boldsymbol{\Xi})\|\boldsymbol{\Sigma}\boldsymbol{\alpha}_1\|_1)\|\mathbf{B}-\widehat{\mathbf{B}}\|_{\infty}\sqrt{6\|\boldsymbol{\alpha}_1\|_1^2/\lambda_0}\notag\\
\ge & -2\|\widehat{\boldsymbol{\xi}}_1- \mathbf{B}\boldsymbol{\alpha}_1\|_2\|\mathbf{B}\|\|\boldsymbol{\alpha}_1-\widehat{\boldsymbol{\alpha}}_1\|_2-2(\|\widehat{\boldsymbol{\xi}}_1\|_1+\lambda_1(\boldsymbol{\Xi})\Lambda_p)(\tau_n/C_2)(\sqrt{6\Lambda_p^2/\lambda_0})\notag\\
\ge & -2\|\widehat{\boldsymbol{\xi}}_1- \mathbf{B}\boldsymbol{\alpha}_1\|_2\lambda_1(\boldsymbol{\Xi})c_0\sqrt{C_5c_0}\sqrt{\Lambda_p^2s_n}-2(C/C_2)\sqrt{6/\lambda_0}\lambda_1(\boldsymbol{\Xi})\Lambda_p^2s_n\notag\\
&-2(C/C_2)\sqrt{6/\lambda_0}\Lambda_ps_n\|\widehat{\boldsymbol{\xi}}_1\|_1 
\end{align*} }
 which together with $\eqref{1290}$ lead to
 \begin{align}
 &\|\widehat{\boldsymbol{\xi}}_1- \mathbf{B}\boldsymbol{\alpha}_1\|^2_2-2\|\widehat{\boldsymbol{\xi}}_1- \mathbf{B}\boldsymbol{\alpha}_1\|_2\lambda_1(\boldsymbol{\Xi})c_0\sqrt{C_5c_0}\sqrt{\Lambda_p^2s_n}-2(C/C_2)\sqrt{6/\lambda_0}\lambda_1(\boldsymbol{\Xi})\Lambda_p^2s_n\notag\\
&-2(C/C_2)\sqrt{6/\lambda_0}\Lambda_ps_n\|\widehat{\boldsymbol{\xi}}_1\|_1  +\kappa_n\|\widehat{\boldsymbol{\xi}}_1\|_1\le  \kappa_n\|\mathbf{B}\boldsymbol{\alpha}_1\|_1=\kappa_n\lambda_1(\boldsymbol{\Xi})\|\boldsymbol{\Sigma}\boldsymbol{\alpha}_1\|_1\notag\\
\le &\kappa_n\lambda_1(\boldsymbol{\Xi})\Lambda_n=\widetilde{C} \lambda_1(\boldsymbol{\Xi})^2\Lambda_p^2s_n. \label{1123}
\end{align} 
Then we have
\begin{align}
 &\left(\|\widehat{\boldsymbol{\xi}}_1- \mathbf{B}\boldsymbol{\alpha}_1\|_2-\lambda_1(\boldsymbol{\Xi})c_0\sqrt{C_5c_0}\sqrt{\Lambda_p^2s_n}\right)^2+\left(\kappa_n-2(C/C_2)\sqrt{6/\lambda_0}\Lambda_ps_n\right)\|\widehat{\boldsymbol{\xi}}_1\|_1\notag\\
\le &\left(\widetilde{C}\lambda_1(\boldsymbol{\Xi})^2+2(C/C_2)\sqrt{6/\lambda_0}\lambda_1(\boldsymbol{\Xi})+\lambda_1(\boldsymbol{\Xi})^2(C_5c_0^3)\right) \Lambda_p^2s_n\notag\\
 \le &\left(\widetilde{C} +2(C/C_2)\sqrt{6/\lambda_0}c_1^{-1} +(C_5c_0^3)\right) \lambda_1(\boldsymbol{\Xi})^2\Lambda_p^2s_n\label{1124}
\end{align}
where the last inequality is due to $\lambda_1(\boldsymbol{\Xi})\ge c_1$ by Condition \ref{condition_2} (a). Then it follows from $\eqref{1124}$ that 
\begin{align*}
 &\left(\|\widehat{\boldsymbol{\xi}}_1- \mathbf{B}\boldsymbol{\alpha}_1\|_2-\lambda_1(\boldsymbol{\Xi})c_0\sqrt{C_5c_0}\sqrt{\Lambda_p^2s_n}\right)^2\\
\le &\left(\widetilde{C} +2(C/C_2)\sqrt{6/\lambda_0}c_1^{-1} +(C_5c_0^3)\right) \lambda_1(\boldsymbol{\Xi})^2\Lambda_p^2s_n
\end{align*}
which implies that 
\begin{align}
 & \|\widehat{\boldsymbol{\xi}}_1- \mathbf{B}\boldsymbol{\alpha}_1\|_2\le \left(\sqrt{\widetilde{C} +2(C/C_2)\sqrt{6/\lambda_0}c_1^{-1} +(C_5c_0^3)}+c_0\sqrt{C_5c_0}\right) \lambda_1(\boldsymbol{\Xi})\sqrt{\Lambda_p^2s_n}.\label{10028}
\end{align}
It also follows from $\eqref{1124}$ that 
 \begin{align}
 & \left(\kappa_n-2(C/C_2)\sqrt{6/\lambda_0}\Lambda_ps_n\right)\|\widehat{\boldsymbol{\xi}}_1\|_1\le \left(\widetilde{C} +2(C/C_2)\sqrt{6/\lambda_0}c_1^{-1} +(C_5c_0^3)\right) \lambda_1(\boldsymbol{\Xi})^2\Lambda_p^2s_n.\label{1292}
\end{align} 
 We take $\widetilde{C}> 2(C/C_2)\sqrt{6/\lambda_0}c_1^{-1}$. Then
\begin{align*}
 &  \kappa_n-2(C/C_2)\sqrt{6/\lambda_0}\Lambda_ps_n =\widetilde{C}\lambda_1(\boldsymbol{\Xi})\Lambda_ps_n-2(C/C_2)\sqrt{6/\lambda_0}\Lambda_ps_n\notag\\
&\ge\widetilde{C}\lambda_1(\boldsymbol{\Xi})\Lambda_ps_n-2(C/C_2)\sqrt{6/\lambda_0}c_1^{-1}\lambda_1(\boldsymbol{\Xi})\Lambda_ps_n=\left(\widetilde{C}- 2(C/C_2)\sqrt{6/\lambda_0}c_1^{-1}\right) \lambda_1(\boldsymbol{\Xi})\Lambda_ps_n 
\end{align*}
which together with $\eqref{1292}$ lead to
\begin{align}
 & \|\widehat{\boldsymbol{\xi}}_1\|_1\le\left[\left(\widetilde{C} +2(C/C_2)\sqrt{6/\lambda_0}c_1^{-1} +(C_5c_0^3)\right)/\left(\widetilde{C}- 2(C/C_2)\sqrt{6/\lambda_0}c_1^{-1}\right) \right]\lambda_1(\boldsymbol{\Xi})\Lambda_p. \label{10029}
\end{align}
Therefore, the lemma follows from $\eqref{10028}$ and $\eqref{10029}$ with 
\begin{align*}
 &C_6=\left(\sqrt{\widetilde{C} +2(C/C_2)\sqrt{6/\lambda_0}c_1^{-1} +(C_5c_0^3)}+c_0\sqrt{C_5c_0}\right) ,\notag\\
&C_7=\left(\widetilde{C} +2(C/C_2)\sqrt{6/\lambda_0}c_1^{-1} +(C_5c_0^3)\right)/\left(\widetilde{C}- 2(C/C_2)\sqrt{6/\lambda_0}c_1^{-1}\right).
\end{align*}
\\
\end{proof}

\begin{proof} $\ref{lemma_7}$\\

We only consider  elements in the event $\Omega_n$.  Because $ \boldsymbol{\gamma}_k$ is  the $k$-th eigenvector of $\boldsymbol{\Xi}$, it is the solution to 
 \begin{align*}
\max_{\boldsymbol{\gamma}\in \mathbf{V}_{k-1}^\perp}\frac{\boldsymbol{\gamma}\trans \boldsymbol{\Xi}\boldsymbol{\gamma}}{\|\boldsymbol{\gamma}\|_2^2}. 
\end{align*} 
Since the projection $ (\mathbf{I}-\mathbf{P}_{k-1}) \widehat{\boldsymbol{\gamma}}_k\in   {\mathbf{V}}_{k-1}^\perp$ and $\|(\mathbf{I}-\mathbf{P}_{k-1}) \widehat{\boldsymbol{\gamma}}_k\|_2\le \|\widehat{\boldsymbol{\gamma}}_k\|_2$, we have 
 \begin{align}
&\frac{\widehat{\boldsymbol{\gamma}}_k\trans (\mathbf{I}-\mathbf{P}_{k-1})\boldsymbol{\Xi} (\mathbf{I}-\mathbf{P}_{k-1})\widehat{\boldsymbol{\gamma}}_k}{\|\widehat{\boldsymbol{\gamma}}_k\|_2^2}\le \frac{\widehat{\boldsymbol{\gamma}}_k\trans (\mathbf{I}-\mathbf{P}_{k-1})\boldsymbol{\Xi} (\mathbf{I}-\mathbf{P}_{k-1})\widehat{\boldsymbol{\gamma}}_k}{\|(\mathbf{I}-\mathbf{P}_{k-1}) \widehat{\boldsymbol{\gamma}}_k\|_2^2}\le \frac{\boldsymbol{\gamma}_k\trans \boldsymbol{\Xi}\boldsymbol{\gamma}_k}{\boldsymbol{\gamma}_k\trans  \boldsymbol{\gamma}_k}=\lambda_k(\boldsymbol{\Xi}).\label{1071}
\end{align} 
It can be seen that $\mathbf{P}_{k-1}=\sum_{i=1}^{k-1}\boldsymbol{\gamma}_i\boldsymbol{\gamma}_i\trans$, hence 
\begin{align}
&\boldsymbol{\Xi}\mathbf{P}_{k-1}=\sum_{i=1}^{k-1}\boldsymbol{\Xi}\boldsymbol{\gamma}_i\boldsymbol{\gamma}_i\trans=\sum_{i=1}^{k-1}\lambda_i(\boldsymbol{\Xi})\boldsymbol{\gamma}_i\boldsymbol{\gamma}_i\trans,\qquad  \mathbf{P}_{k-1}\boldsymbol{\Xi}\mathbf{P}_{k-1}=\boldsymbol{\Xi}\mathbf{P}_{k-1}.\label{10030}
\end{align}
  By $\eqref{10030}$,
\begin{align*}
& \widehat{\boldsymbol{\gamma}}_k\trans (\mathbf{I}-\mathbf{P}_{k-1})\boldsymbol{\Xi} (\mathbf{I}-\mathbf{P}_{k-1})\widehat{\boldsymbol{\gamma}}_k=\widehat{\boldsymbol{\gamma}}_k\trans \boldsymbol{\Xi} \widehat{\boldsymbol{\gamma}}_k-2\widehat{\boldsymbol{\gamma}}_k\trans  \boldsymbol{\Xi}\mathbf{P}_{k-1} \widehat{\boldsymbol{\gamma}}_k+\widehat{\boldsymbol{\gamma}}_k\trans  \mathbf{P}_{k-1}\boldsymbol{\Xi} \mathbf{P}_{k-1} \widehat{\boldsymbol{\gamma}}_k\notag\\
=&\widehat{\boldsymbol{\gamma}}_k\trans \boldsymbol{\Xi} \widehat{\boldsymbol{\gamma}}_k-2\widehat{\boldsymbol{\gamma}}_k\trans  \boldsymbol{\Xi}\mathbf{P}_{k-1} \widehat{\boldsymbol{\gamma}}_k+\widehat{\boldsymbol{\gamma}}_k\trans   \boldsymbol{\Xi} \mathbf{P}_{k-1} \widehat{\boldsymbol{\gamma}}_k=\widehat{\boldsymbol{\gamma}}_k\trans \boldsymbol{\Xi} \widehat{\boldsymbol{\gamma}}_k-\widehat{\boldsymbol{\gamma}}_k\trans  \boldsymbol{\Xi}\mathbf{P}_{k-1} \widehat{\boldsymbol{\gamma}}_k\notag\\
=&\widehat{\boldsymbol{\gamma}}_k\trans \boldsymbol{\Xi} \widehat{\boldsymbol{\gamma}}_k-\sum_{i=1}^{k-1}\lambda_i(\boldsymbol{\Xi}) (\boldsymbol{\gamma}_i\trans\widehat{\boldsymbol{\gamma}}_k)^2,
\end{align*}
which combined with $\eqref{1071}$ lead to
\begin{align*}
&  \widehat{\boldsymbol{\gamma}}_k\trans \boldsymbol{\Xi} \widehat{\boldsymbol{\gamma}}_k-\sum_{i=1}^{k-1}\lambda_i(\boldsymbol{\Xi}) (\boldsymbol{\gamma}_i\trans\widehat{\boldsymbol{\gamma}}_k)^2\le \lambda_k(\boldsymbol{\Xi})\|\widehat{\boldsymbol{\gamma}}_k\|^2_2=\lambda_k(\boldsymbol{\Xi})\widehat{\boldsymbol{\gamma}}_k\trans   \widehat{\boldsymbol{\gamma}}_k.
\end{align*}
Then we have
\begin{align}
&  \widehat{\boldsymbol{\gamma}}_k\trans \boldsymbol{\Xi} \widehat{\boldsymbol{\gamma}}_k \le \lambda_k(\boldsymbol{\Xi})\widehat{\boldsymbol{\gamma}}_k\trans   \widehat{\boldsymbol{\gamma}}_k+\sum_{i=1}^{k-1} \lambda_i(\boldsymbol{\Xi}) (\boldsymbol{\gamma}_i\trans\widehat{\boldsymbol{\gamma}}_k)^2\le \lambda_k(\boldsymbol{\Xi})\widehat{\boldsymbol{\gamma}}_k\trans   \widehat{\boldsymbol{\gamma}}_k+\lambda_1(\boldsymbol{\Xi}) \sum_{i=1}^{k-1} (\boldsymbol{\gamma}_i\trans\widehat{\boldsymbol{\gamma}}_k)^2\notag\\
\le & \lambda_k(\boldsymbol{\Xi})\left[\widehat{\boldsymbol{\gamma}}_k\trans   \widehat{\boldsymbol{\gamma}}_k+c_3\sum_{i=1}^{k-1} (\boldsymbol{\gamma}_i\trans\widehat{\boldsymbol{\gamma}}_k)^2\right],\label{1093}
\end{align}
where the last inequality is due to  Condition \ref{condition_2}(b). Because $\widehat{\boldsymbol{\alpha}}_k$ is the solution to 
\begin{align}
 \max_{\boldsymbol{\alpha}\in \widehat{\mathbf{W}}_{k-1}^\perp}\frac{\boldsymbol{\alpha}\trans \widehat{\mathbf{B}}\boldsymbol{\alpha}}{\boldsymbol{\alpha}\trans \widehat{\boldsymbol{\Sigma}}\boldsymbol{\alpha}+\tau_n\|\boldsymbol{\alpha}\|^2_{\lambda_n}},\label{1060}
\end{align}
 and noting that $ (\mathbf{I}-\widehat{\mathbf{Q}}_{k-1})  {\boldsymbol{\alpha}}_k\in  \widehat{\mathbf{W}}_{k-1}^\perp$, $\widehat{\boldsymbol{\alpha}}_k=\boldsymbol{\Sigma}^{-1/2}\widehat{\boldsymbol{\gamma}}_k$ and $\widehat{\boldsymbol{\alpha}}_k\trans \widehat{\boldsymbol{\Sigma}}\widehat{\boldsymbol{\alpha}}_k+\tau_n\|\widehat{\boldsymbol{\alpha}}_k\|^2_{\lambda_n}=1$, we have
\begin{align}
  &\widehat{\boldsymbol{\gamma}}_k\trans \boldsymbol{\Sigma}^{-1/2}\widehat{\mathbf{B}}\boldsymbol{\Sigma}^{-1/2}\widehat{\boldsymbol{\gamma}}_k=\widehat{\boldsymbol{\alpha}}_k\trans \widehat{\mathbf{B}}\widehat{\boldsymbol{\alpha}}_k= \frac{\widehat{\boldsymbol{\alpha}}_k\trans \widehat{\mathbf{B}}\widehat{\boldsymbol{\alpha}}_k}{\widehat{\boldsymbol{\alpha}}_k\trans \widehat{\boldsymbol{\Sigma}}\widehat{\boldsymbol{\alpha}}_k+\tau_n\|\widehat{\boldsymbol{\alpha}}_k\|^2_{\lambda_n}}\notag\\
	\ge&\frac{ {\boldsymbol{\alpha}}_k\trans(\mathbf{I}-\widehat{\mathbf{Q}}_{k-1}) \widehat{\mathbf{B}}(\mathbf{I}-\widehat{\mathbf{Q}}_{k-1})  {\boldsymbol{\alpha}}_k}{{\boldsymbol{\alpha}}_k\trans(\mathbf{I}-\widehat{\mathbf{Q}}_{k-1}) \widehat{\boldsymbol{\Sigma}}(\mathbf{I}-\widehat{\mathbf{Q}}_{k-1})  {\boldsymbol{\alpha}}_k+\tau_n\|(\mathbf{I}-\widehat{\mathbf{Q}}_{k-1})  {\boldsymbol{\alpha}}_k\|^2_{\lambda_n}},\label{1072}
\end{align}
By $\eqref{1093}$  and the definition of $\Omega_n$, the left hand side of $\eqref{1072}$
 \begin{align}
  &\widehat{\boldsymbol{\gamma}}_k\trans \boldsymbol{\Sigma}^{-1/2}\widehat{\mathbf{B}}\boldsymbol{\Sigma}^{-1/2}\widehat{\boldsymbol{\gamma}}_k\le \widehat{\boldsymbol{\gamma}}_k\trans \boldsymbol{\Sigma}^{-1/2} {\mathbf{B}}\boldsymbol{\Sigma}^{-1/2}\widehat{\boldsymbol{\gamma}}_k+\|\widehat{\mathbf{B}}-\mathbf{B}\|_{\infty}\|\boldsymbol{\Sigma}^{-1/2}\widehat{\boldsymbol{\gamma}}_k\|^2_1\notag\\
	\le& \widehat{\boldsymbol{\gamma}}_k\trans  \boldsymbol{\Xi} \widehat{\boldsymbol{\gamma}}_k+ \frac{1}{C_2}\tau_n\|\boldsymbol{\Sigma}^{-1/2}\widehat{\boldsymbol{\gamma}}_k\|^2_1\le \widehat{\boldsymbol{\gamma}}_k\trans  \boldsymbol{\Xi} \widehat{\boldsymbol{\gamma}}_k+ \frac{\lambda_k(\boldsymbol{\Xi})}{c_1}\frac{1}{C_2}\tau_n\|\boldsymbol{\Sigma}^{-1/2}\widehat{\boldsymbol{\gamma}}_k\|^2_1\notag\\
	\le &\lambda_k(\boldsymbol{\Xi})\left[\widehat{\boldsymbol{\gamma}}_k\trans  \widehat{\boldsymbol{\gamma}}_k+c_3\sum_{i=1}^{k-1} (\boldsymbol{\gamma}_i\trans\widehat{\boldsymbol{\gamma}}_k)^2+\frac{c_1^{-1}}{C_2}\tau_n\|\widehat{\boldsymbol{\alpha}}_k\|^2_1\right]\label{1073}
\end{align}
where we use $\lambda_k(\boldsymbol{\Xi}) \ge c_1$. Now
\begin{align}
  & \widehat{\boldsymbol{\gamma}}_k\trans  \widehat{\boldsymbol{\gamma}}_k=\widehat{\boldsymbol{\gamma}}_k\trans  \widehat{\boldsymbol{\gamma}}_k-\widehat{\boldsymbol{\gamma}}_k\trans \boldsymbol{\Sigma}^{-1/2}\widehat{\boldsymbol{\Sigma}}\boldsymbol{\Sigma}^{-1/2}\widehat{\boldsymbol{\gamma}}_k+\widehat{\boldsymbol{\gamma}}_k\trans \boldsymbol{\Sigma}^{-1/2}\widehat{\boldsymbol{\Sigma}}\boldsymbol{\Sigma}^{-1/2}\widehat{\boldsymbol{\gamma}}_k  \label{1293}\\
	=&\widehat{\boldsymbol{\gamma}}_k\trans \boldsymbol{\Sigma}^{-1/2}(\boldsymbol{\Sigma}-\widehat{\boldsymbol{\Sigma}})\boldsymbol{\Sigma}^{-1/2}\widehat{\boldsymbol{\gamma}}_k+\widehat{\boldsymbol{\alpha}}_k\trans \widehat{\boldsymbol{\Sigma}}\widehat{\boldsymbol{\alpha}}_k \le \|\boldsymbol{\Sigma}-\widehat{\boldsymbol{\Sigma}}\|_{\infty}\|\boldsymbol{\Sigma}^{-1/2}\widehat{\boldsymbol{\gamma}}_k\|^2_1+\widehat{\boldsymbol{\alpha}}_k\trans \widehat{\boldsymbol{\Sigma}}\widehat{\boldsymbol{\alpha}}_k\notag\\
	\le& \frac{1}{C_2}\tau_n\|\widehat{\boldsymbol{\alpha}}_k\|^2_1+\widehat{\boldsymbol{\alpha}}_k\trans \widehat{\boldsymbol{\Sigma}}\widehat{\boldsymbol{\alpha}}_k=\widehat{\boldsymbol{\alpha}}_k\trans \widehat{\boldsymbol{\Sigma}}\widehat{\boldsymbol{\alpha}}_k+ \frac{1}{C_2}\tau_n\|\widehat{\boldsymbol{\alpha}}_k\|^2_1=1-\tau_n\|\widehat{\boldsymbol{\alpha}}_k \|_{\lambda_n}+ \frac{1}{C_2}\tau_n\|\widehat{\boldsymbol{\alpha}}_k\|^2_1\notag\\
	\le &1-\lambda_n\tau_n\|\widehat{\boldsymbol{\alpha}}_k \|_1+ \frac{1}{C_2}\tau_n\|\widehat{\boldsymbol{\alpha}}_k\|^2_1=1-(\lambda_n-1/C_2)\tau_n\|\widehat{\boldsymbol{\alpha}}_k\|^2_1.\notag
\end{align}
  Because as $n$ is large enough, $\lambda_n-1/C_2\ge \lambda_n-(1+c_1^{-1})/C_2=\lambda_n-\lambda_0/2>\lambda_0/2$,  $\eqref{1293}$ gives $\|\widehat{\boldsymbol{\gamma}}_k\|_2^2\le 1 $. On the other hand,
\begin{align*}
  & \widehat{\boldsymbol{\gamma}}_k\trans  \widehat{\boldsymbol{\gamma}}_k=\widehat{\boldsymbol{\gamma}}_k\trans  \widehat{\boldsymbol{\gamma}}_k-\widehat{\boldsymbol{\gamma}}_k\trans \boldsymbol{\Sigma}^{-1/2}\widehat{\boldsymbol{\Sigma}}\boldsymbol{\Sigma}^{-1/2}\widehat{\boldsymbol{\gamma}}_k+\widehat{\boldsymbol{\gamma}}_k\trans \boldsymbol{\Sigma}^{-1/2}\widehat{\boldsymbol{\Sigma}}\boldsymbol{\Sigma}^{-1/2}\widehat{\boldsymbol{\gamma}}_k\\
	=&\widehat{\boldsymbol{\gamma}}_k\trans \boldsymbol{\Sigma}^{-1/2}(\boldsymbol{\Sigma}-\widehat{\boldsymbol{\Sigma}})\boldsymbol{\Sigma}^{-1/2}\widehat{\boldsymbol{\gamma}}_k+\widehat{\boldsymbol{\alpha}}_k\trans \widehat{\boldsymbol{\Sigma}}\widehat{\boldsymbol{\alpha}}_k \ge \widehat{\boldsymbol{\alpha}}_k\trans \widehat{\boldsymbol{\Sigma}}\widehat{\boldsymbol{\alpha}}_k-\|\boldsymbol{\Sigma}-\widehat{\boldsymbol{\Sigma}}\|_{\infty}\|\boldsymbol{\Sigma}^{-1/2}\widehat{\boldsymbol{\gamma}}_k\|^2_1\notag\\
	\ge& \widehat{\boldsymbol{\alpha}}_k\trans \widehat{\boldsymbol{\Sigma}}\widehat{\boldsymbol{\alpha}}_k- \frac{1}{C_2}\tau_n\|\widehat{\boldsymbol{\alpha}}_k\|^2_1=1-\tau_n\|\widehat{\boldsymbol{\alpha}}_k \|_{\lambda_n}- \frac{1}{C_2}\tau_n\|\widehat{\boldsymbol{\alpha}}_k\|^2_1
	\ge 1-(1+1/C_2)\tau_n\|\widehat{\boldsymbol{\alpha}}_k\|^2_1,\notag
\end{align*}
 which together with $\eqref{1293}$ lead to
\begin{align}
  &  1-(1+1/C_2)\tau_n\|\widehat{\boldsymbol{\alpha}}_k\|^2_1\le \|\widehat{\boldsymbol{\gamma}}_k\|_2^2\le 1,\label{1165}
\end{align}
as $n$ is large enough. It follows from $\eqref{1073}$ and $\eqref{1293}$ that
\begin{align}
  &\widehat{\boldsymbol{\gamma}}_k\trans \boldsymbol{\Sigma}^{-1/2}\widehat{\mathbf{B}}\boldsymbol{\Sigma}^{-1/2}\widehat{\boldsymbol{\gamma}}_k\le \lambda_k(\boldsymbol{\Xi})\left[1-(\lambda_n-1/C_2)\tau_n\|\widehat{\boldsymbol{\alpha}}_k\|^2_1+c_3\sum_{i=1}^{k-1} (\boldsymbol{\gamma}_i\trans\widehat{\boldsymbol{\gamma}}_k)^2+\frac{c_1^{-1}}{C_2}\tau_n\|\widehat{\boldsymbol{\alpha}}_k\|^2_1\right]\notag\\
	=&\lambda_k(\boldsymbol{\Xi})\left[1-(\lambda_n-(1+c_1^{-1})/C_2)\tau_n\|\widehat{\boldsymbol{\alpha}}_k\|^2_1+c_3\sum_{i=1}^{k-1} (\boldsymbol{\gamma}_i\trans\widehat{\boldsymbol{\gamma}}_k)^2 \right]\notag\\
	=&\lambda_k(\boldsymbol{\Xi})\left[1-(\lambda_n-\lambda_0/2)\tau_n\|\widehat{\boldsymbol{\alpha}}_k\|^2_1+c_3\sum_{i=1}^{k-1} (\boldsymbol{\gamma}_i\trans\widehat{\boldsymbol{\gamma}}_k)^2 \right]\label{1074}
\end{align}
where we use $(1+c_1^{-1})/C_2=\lambda_0/2$ by the definition $\eqref{1271}$ of $C_2$. Next, we calculate the right hand side of $\eqref{1072}$. Let $\boldsymbol{\beta}_k=(\mathbf{I}-\widehat{\mathbf{Q}}_{k-1})\boldsymbol{\alpha}_k$. 
\begin{align}
  & \boldsymbol{\beta}_k\trans  {\mathbf{B}}\boldsymbol{\beta}_k= {\boldsymbol{\alpha}}_k\trans(\mathbf{I}-\widehat{\mathbf{Q}}_{k-1})  \mathbf{B}(\mathbf{I}-\widehat{\mathbf{Q}}_{k-1})  {\boldsymbol{\alpha}}_k={\boldsymbol{\alpha}}_k\trans \mathbf{B} {\boldsymbol{\alpha}}_k-2{\boldsymbol{\alpha}}_k\trans\widehat{\mathbf{Q}}_{k-1} \mathbf{B} {\boldsymbol{\alpha}}_k+{\boldsymbol{\alpha}}_k\trans\widehat{\mathbf{Q}}_{k-1} \mathbf{B} \widehat{\mathbf{Q}}_{k-1}{\boldsymbol{\alpha}}_k\notag\\
	\ge &{\boldsymbol{\alpha}}_k \trans\mathbf{B} {\boldsymbol{\alpha}}_k-2{\boldsymbol{\alpha}}_k\trans\widehat{\mathbf{Q}}_{k-1} \mathbf{B} {\boldsymbol{\alpha}}_k=\lambda_k(\boldsymbol{\Xi})\left[{\boldsymbol{\alpha}}_k \trans\boldsymbol{\Sigma} {\boldsymbol{\alpha}}_k-2{\boldsymbol{\alpha}}_k\trans\widehat{\mathbf{Q}}_{k-1} \boldsymbol{\Sigma} {\boldsymbol{\alpha}}_k\right].\label{1075}
\end{align}
 On the other hand, 
 \begin{align}
  & \boldsymbol{\beta}_k\trans  {\boldsymbol{\Sigma}}\boldsymbol{\beta}_k= {\boldsymbol{\alpha}}_k\trans(\mathbf{I}-\widehat{\mathbf{Q}}_{k-1})  \boldsymbol{\Sigma}(\mathbf{I}-\widehat{\mathbf{Q}}_{k-1})  {\boldsymbol{\alpha}}_k
	={\boldsymbol{\alpha}}_k\trans \boldsymbol{\Sigma} {\boldsymbol{\alpha}}_k-2{\boldsymbol{\alpha}}_k\trans\widehat{\mathbf{Q}}_{k-1} \boldsymbol{\Sigma} {\boldsymbol{\alpha}}_k+{\boldsymbol{\alpha}}_k\trans\widehat{\mathbf{Q}}_{k-1} \boldsymbol{\Sigma} \widehat{\mathbf{Q}}_{k-1}{\boldsymbol{\alpha}}_k\notag\\
	\le &{\boldsymbol{\alpha}}_k\trans \boldsymbol{\Sigma} {\boldsymbol{\alpha}}_k-2{\boldsymbol{\alpha}}_k\trans\widehat{\mathbf{Q}}_{k-1} \boldsymbol{\Sigma} {\boldsymbol{\alpha}}_k+\|\boldsymbol{\Sigma}\|\|\widehat{\mathbf{Q}}_{k-1}{\boldsymbol{\alpha}}_k\|_2^2
	\le  {\boldsymbol{\alpha}}_k\trans \boldsymbol{\Sigma} {\boldsymbol{\alpha}}_k-2{\boldsymbol{\alpha}}_k\trans\widehat{\mathbf{Q}}_{k-1} \boldsymbol{\Sigma} {\boldsymbol{\alpha}}_k+c_0\|\widehat{\mathbf{Q}}_{k-1}{\boldsymbol{\alpha}}_k\|_2^2.\label{1076}
\end{align} 
Then by the definition $\eqref{1033}$ of $\Omega_n$, $\eqref{1075}$, $\eqref{1076}$ and Condition \ref{condition_2} (a), the right hand side of $\eqref{1072}$ is equal to
 \begin{align}
  & \frac{\boldsymbol{\beta}_k\trans \widehat{\mathbf{B}}\boldsymbol{\beta}_k}{\boldsymbol{\beta}_k\trans \widehat{\boldsymbol{\Sigma}}\boldsymbol{\beta}_k+\tau_n\|\boldsymbol{\beta}_k\|^2_{\lambda_n}}\ge \frac{\boldsymbol{\beta}_k\trans  {\mathbf{B}}\boldsymbol{\beta}_k-\|\widehat{\mathbf{B}}-\mathbf{B}\|_{\infty}\|\boldsymbol{\beta}_k\|^2_1}{\boldsymbol{\beta}_k\trans  {\boldsymbol{\Sigma}}\boldsymbol{\beta}_k+\|\widehat{\boldsymbol{\Sigma}}- \boldsymbol{\Sigma}\|_{\infty}\|\boldsymbol{\beta}_k\|^2_1+\tau_n\|\boldsymbol{\beta}_k\|^2_1}\notag\\
	\ge& \frac{\boldsymbol{\beta}_k\trans  {\mathbf{B}}\boldsymbol{\beta}_k-\frac{1}{C_2}\tau_n \|\boldsymbol{\beta}_k\|^2_1}{\boldsymbol{\beta}_k\trans  {\boldsymbol{\Sigma}}\boldsymbol{\beta}_k+\frac{1}{C_2}\tau_n \|\boldsymbol{\beta}_k\|^2_1+\tau_n\|\boldsymbol{\beta}_k\|^2_1}\notag\\
	\ge& \frac{\lambda_k(\boldsymbol{\Xi})\left[{\boldsymbol{\alpha}}_k\trans \boldsymbol{\Sigma} {\boldsymbol{\alpha}}_k-2{\boldsymbol{\alpha}}_k\trans\widehat{\mathbf{Q}}_{k-1} \boldsymbol{\Sigma} {\boldsymbol{\alpha}}_k\right]-\lambda_k(\boldsymbol{\Xi})\frac{c_1^{-1}}{C_2}\tau_n \|\boldsymbol{\beta}_k\|^2_1}{{\boldsymbol{\alpha}}_k\trans \boldsymbol{\Sigma} {\boldsymbol{\alpha}}_k-2{\boldsymbol{\alpha}}_k\trans\widehat{\mathbf{Q}}_{k-1} \boldsymbol{\Sigma} {\boldsymbol{\alpha}}_k+c_0\|\widehat{\mathbf{Q}}_{k-1}{\boldsymbol{\alpha}}_k\|_2^2+(1+1/C_2)\tau_n\|\boldsymbol{\beta}_k\|^2_1}.\label{1077}
\end{align} 
Now by $\eqref{1072}$, $\eqref{1074}$ and $\eqref{1077}$,
\begin{align*}
  &\frac{\lambda_k(\boldsymbol{\Xi})\left[{\boldsymbol{\alpha}}_k\trans \boldsymbol{\Sigma} {\boldsymbol{\alpha}}_k-2{\boldsymbol{\alpha}}_k\trans\widehat{\mathbf{Q}}_{k-1} \boldsymbol{\Sigma} {\boldsymbol{\alpha}}_k\right]-\lambda_k(\boldsymbol{\Xi})\frac{c_1^{-1}}{C_2}\tau_n \|\boldsymbol{\beta}_k\|^2_1}{{\boldsymbol{\alpha}}_k\trans \boldsymbol{\Sigma} {\boldsymbol{\alpha}}_k-2{\boldsymbol{\alpha}}_k\trans\widehat{\mathbf{Q}}_{k-1} \boldsymbol{\Sigma} {\boldsymbol{\alpha}}_k+c_0\|\widehat{\mathbf{Q}}_{k-1}{\boldsymbol{\alpha}}_k\|_2^2+(1+1/C_2)\tau_n\|\boldsymbol{\beta}_k\|^2_1}\notag\\
	\le &\lambda_k(\boldsymbol{\Xi})\left[1-(\lambda_n-\lambda_0/2)\tau_n\| \widehat{\boldsymbol{\alpha}}_k\|^2_1+c_3\sum_{i=1}^{k-1} (\boldsymbol{\gamma}_i\trans\widehat{\boldsymbol{\gamma}}_k)^2\right], 
\end{align*}
 which, by a simple calculation, leads to 
 \begin{align}
  &(\lambda_n-\lambda_0/2)\tau_n\| \widehat{\boldsymbol{\alpha}}_k\|^2_1\notag\\
	\le& \frac{c_0\|\widehat{\mathbf{Q}}_{k-1}{\boldsymbol{\alpha}}_k\|_2^2+(1+1/C_2)\tau_n\|\boldsymbol{\beta}_k\|^2_1+\frac{c_1^{-1}}{C_2}\tau_n \|\boldsymbol{\beta}_k\|^2_1}{{\boldsymbol{\alpha}}_k\trans \boldsymbol{\Sigma} {\boldsymbol{\alpha}}_k-2{\boldsymbol{\alpha}}_k\trans\widehat{\mathbf{Q}}_{k-1} \boldsymbol{\Sigma} {\boldsymbol{\alpha}}_k+c_0\|\widehat{\mathbf{Q}}_{k-1}{\boldsymbol{\alpha}}_k\|_2^2+(1+1/C_2)\tau_n\|\boldsymbol{\beta}_k\|^2_1}+c_3\sum_{i=1}^{k-1} (\boldsymbol{\gamma}_i\trans\widehat{\boldsymbol{\gamma}}_k)^2 \notag\\
=&\frac{c_0\|\widehat{\mathbf{Q}}_{k-1}{\boldsymbol{\alpha}}_k\|_2^2+(1+\lambda_0/2)\tau_n\|\boldsymbol{\beta}_k\|^2_1 }{1-2{\boldsymbol{\alpha}}_k\trans\widehat{\mathbf{Q}}_{k-1} \boldsymbol{\Sigma} {\boldsymbol{\alpha}}_k+c_0\|\widehat{\mathbf{Q}}_{k-1}{\boldsymbol{\alpha}}_k\|_2^2+(1+1/C_2)\tau_n\|\boldsymbol{\beta}_k\|^2_1}+c_3\sum_{i=1}^{k-1} (\boldsymbol{\gamma}_i\trans\widehat{\boldsymbol{\gamma}}_k)^2 \notag\\
\le &\frac{c_0\|\widehat{\mathbf{Q}}_{k-1}{\boldsymbol{\alpha}}_k\|_2^2+(1+\lambda_0/2)\tau_n\|\boldsymbol{\beta}_k\|^2_1 }{1-2c_0^{3/2}\|\widehat{\mathbf{Q}}_{k-1}{\boldsymbol{\alpha}}_k\|_2}+c_3\sum_{i=1}^{k-1} (\boldsymbol{\gamma}_i\trans\widehat{\boldsymbol{\gamma}}_k)^2,  \label{1163}
\end{align} 
where we use  
\begin{align}
 \|{\boldsymbol{\alpha}}_i\|_2^2\le \|\boldsymbol{\Sigma}^{-1}\|\|\boldsymbol{\Sigma}^{1/2}{\boldsymbol{\alpha}}_i\|_2^2=\|\boldsymbol{\Sigma}^{-1}\|\|\boldsymbol{\gamma}_k\|_2^2=\|\boldsymbol{\Sigma}^{-1}\|\le c_0.\label{10032}
\end{align}
We will estimate terms on the right hand side of $\eqref{1163}$. By $\eqref{1154}$, we have
\begin{align}
 \widehat{\mathbf{P}}_{k-1}\widehat{\boldsymbol{\gamma}}_k=\mathbf{0},\quad \mathbf{Q}_{k-1}\boldsymbol{\alpha}_k=\mathbf{0}.\label{10033}
\end{align}
  Since we assume that $\eqref{1053}$ is true for all $1\le i\le k-1$ and all $n$ large enough, by $\eqref{1165}$,  $\eqref{10032}$ and $\eqref{10033}$, we have 
\begin{align}
  &\sum_{i=1}^{k-1} (\boldsymbol{\gamma}_i\trans\widehat{\boldsymbol{\gamma}}_k)^2=\|\mathbf{P}_{k-1}\widehat{\boldsymbol{\gamma}}_k\|^2_2=\|\mathbf{P}_{k-1}\widehat{\boldsymbol{\gamma}}_k-\widehat{\mathbf{P}}_{k-1}\widehat{\boldsymbol{\gamma}}_k\|^2_2\le \|\widehat{\mathbf{P}}_{k-1} -\mathbf{P}_{k-1} \|^2\|\widehat{\boldsymbol{\gamma}}_k\|^2_2\notag\\
	\le& \|\widehat{\mathbf{P}}_{k-1} -\mathbf{P}_{k-1} \|^2\le C_{k-1,3}\Lambda_p^2s_n\notag\\
	\text{and }& \|\widehat{\mathbf{Q}}_{k-1}{\boldsymbol{\alpha}}_k\|_2^2=\|\widehat{\mathbf{Q}}_{k-1}{\boldsymbol{\alpha}}_k-\mathbf{Q}_{k-1}\boldsymbol{\alpha}_k\|_2^2\le \|\widehat{\mathbf{Q}}_{k-1}-\mathbf{Q}_{k-1}\|^2\|{\boldsymbol{\alpha}}_k\|_2^2,\notag\\
	&\le c_0\|\widehat{\mathbf{Q}}_{k-1}-\mathbf{Q}_{k-1}\|^2\le  c_0C_{k-1,4}\Lambda_p^2s_n.\label{1131}
\end{align}
  Next, we estimate $\|\boldsymbol{\beta}_k\|_1$. Let $\widehat{\mathbf{Q}}_{k-1}{\boldsymbol{\alpha}}_k=\sum_{i=1}^{k-1}t_i\widehat{\boldsymbol{\xi}}_i$, where $\mathbf{t}=(t_1,\cdots,t_{k-1})$ is the coefficient vector. By $\eqref{1053}$,
\begin{align}
  &\|\widehat{\mathbf{Q}}_{k-1}{\boldsymbol{\alpha}}_k\|_1\le \sum_{i=1}^{k-1}|t_i|\max_{1\le i\le k-1}\|\widehat{\boldsymbol{\xi}}_i\|_1\le \|\mathbf{t}\|_1\left(\max_{1\le i\le k-1}C_{i,5}\right)\lambda_1(\boldsymbol{\Xi})\Lambda_p. \label{1158}
\end{align}
  To find an upper bound for $\|\mathbf{t}\|_1$, we multiply ${\boldsymbol{\alpha}}_j\trans$, $1\le j\le k-1$, on both sides of $\widehat{\mathbf{Q}}_{k-1}{\boldsymbol{\alpha}}_k=\sum_{i=1}^{k-1}t_i\widehat{\boldsymbol{\xi}}_i=\sum_{i=1}^{k-1}t_i\boldsymbol{\xi}_i-\sum_{i=1}^{k-1}t_i(\boldsymbol{\xi}_i-\widehat{\boldsymbol{\xi}}_i)$, then by $\eqref{1053}$ and $\eqref{10032}$, we have 
 \begin{align}
  &|{\boldsymbol{\alpha}}_j\trans\widehat{\mathbf{Q}}_{k-1}{\boldsymbol{\alpha}}_k|=\left|\sum_{i=1}^{k-1}t_i{\boldsymbol{\alpha}}_j\trans\boldsymbol{\xi}_i-\sum_{i=1}^{k-1}t_i{\boldsymbol{\alpha}}_j\trans(\boldsymbol{\xi}_i-\widehat{\boldsymbol{\xi}}_i)\right|=\left|\lambda_j(\boldsymbol{\Xi}) t_j - \sum_{i=1}^{k-1}t_i{\boldsymbol{\alpha}}_j\trans(\boldsymbol{\xi}_i-\widehat{\boldsymbol{\xi}}_i)\right|\notag\\
\ge &\lambda_j(\boldsymbol{\Xi})|t_j|- \sum_{i=1}^{k-1}|t_i|\|\boldsymbol{\xi}_i-\widehat{\boldsymbol{\xi}}_i\|_2\|\boldsymbol{\alpha}_j\|_2\ge \lambda_j(\boldsymbol{\Xi})|t_j|-c_0^{1/2}\sum_{i=1}^{k-1}|t_i|\|\boldsymbol{\xi}_i-\widehat{\boldsymbol{\xi}}_i\|_2\notag\\
\ge &\lambda_j(\boldsymbol{\Xi})|t_j| -c_0^{1/2}\sum_{i=1}^{k-1}|t_i|\left(\max_{1\le i\le k-1}\|\boldsymbol{\xi}_i-\widehat{\boldsymbol{\xi}}_i\|_2\right)\notag\\
\ge& \lambda_j(\boldsymbol{\Xi})|t_j|-c_0^{1/2}\|\mathbf{t}\|_1	\left(\max_{1\le i\le k-1}\sqrt{C_{i,6}}\right)\lambda_1(\boldsymbol{\Xi})\sqrt{\Lambda_p^2s_n}\notag\\
\ge& \lambda_1(\boldsymbol{\Xi})\left[c_3^{-1}|t_j|-c_0^{1/2}\|\mathbf{t}\|_1	\left(\max_{1\le i\le k-1}\sqrt{C_{i,6}}\right)\sqrt{\Lambda_p^2s_n}\right],\label{1159}
\end{align} 
where the last inequality is due to Condition \ref{condition_2} (c). On the other hand, by $\eqref{10033}$ and $\eqref{10032}$, $|{\boldsymbol{\alpha}}_j\trans\widehat{\mathbf{Q}}_{k-1}{\boldsymbol{\alpha}}_k|=|{\boldsymbol{\alpha}}_j\trans(\widehat{\mathbf{Q}}_{k-1}-\mathbf{Q}_{k-1}){\boldsymbol{\alpha}}_k|\le \|\widehat{\mathbf{Q}}_{k-1}-\mathbf{Q}_{k-1}\|\|{\boldsymbol{\alpha}}_j\|_2\|{\boldsymbol{\alpha}}_k\|_2\le c_0\|\widehat{\mathbf{Q}}_{k-1}-\mathbf{Q}_{k-1}\|$ which together with $\eqref{1159}$ leads to
\begin{align*}
  &\lambda_1(\boldsymbol{\Xi}) c_3^{-1}\|\mathbf{t}\|_1=\lambda_1(\boldsymbol{\Xi}) c_3^{-1}\sum_{j=1}^{k-1}|t_j|\\
	\le& \sum_{j=1}^{k-1}|{\boldsymbol{\alpha}}_j\trans\widehat{\mathbf{Q}}_{k-1}{\boldsymbol{\alpha}}_k|+(k-1)c_0^{1/2}\|\mathbf{t}\|_1	\left(\max_{1\le i\le k-1}\sqrt{C_{i,6}}\right)\lambda_1(\boldsymbol{\Xi})\sqrt{\Lambda_p^2s_n}\\
	\le & (k-1)c_0\|\widehat{\mathbf{Q}}_{k-1}-\mathbf{Q}_{k-1}\|+(k-1)c_0^{1/2}\|\mathbf{t}\|_1	\left(\max_{1\le i\le k-1}\sqrt{C_{i,6}}\right)\lambda_1(\boldsymbol{\Xi})\sqrt{\Lambda_p^2s_n}
\end{align*}
By solving the above inequality, we obtain
 \begin{align*}
  &\|\mathbf{t}\|_1\le \left[c_3^{-1}-(k-1)c_0^{1/2} \left(\max_{1\le i\le k-1}\sqrt{C_{i,6}}\right)\sqrt{\Lambda_p^2s_n}\right]^{-1}\lambda_1(\boldsymbol{\Xi})^{-1}(k-1)c_0\|\widehat{\mathbf{Q}}_{k-1}-\mathbf{Q}_{k-1}\|,
\end{align*}
which together  $\eqref{1158}$ imply $\|\widehat{\mathbf{Q}}_{k-1}{\boldsymbol{\alpha}}_k\|_1\le O(1)\|\widehat{\mathbf{Q}}_{k-1}-\mathbf{Q}_{k-1}\|\Lambda_p=o(1)\Lambda_p$ by $\eqref{1053}$. Therefore, 
\begin{align}
  &\|\boldsymbol{\beta}_k\|_1=\|\boldsymbol{\alpha}_k-\widehat{\mathbf{Q}}_{k-1}{\boldsymbol{\alpha}}_k\|_1\le \|\boldsymbol{\alpha}_k\|_1+\|\widehat{\mathbf{Q}}_{k-1}{\boldsymbol{\alpha}}_k\|_1=(1+o(1))\Lambda_p\le 2\Lambda_p,\label{1162}
\end{align}
as $n$ is large enough. By  $\eqref{1163}$, $\eqref{1131}$ and $\eqref{1162}$,  and noting that $\lambda_n> \lambda_0$ as $n$ is large enough, we have
\begin{align}
  &  \| \widehat{\boldsymbol{\alpha}}_k\|_1\le C_{k,1}\Lambda_p, \label{1094}
\end{align}
for all $n$ large enough, where $C_{k,1}$ is a constant only depending $C_{i,j}$, $1\le i\le k-1$ and $1\le j\le 6$, $\lambda_0$, $C$, $C_2$, $\widetilde{C}$ and the constants in Conditions \ref{condition_1} and \ref{condition_2}. By $\eqref{1093}$, $\eqref{1165}$ and $\eqref{1131}$,
\begin{align}
&  \widehat{\boldsymbol{\gamma}}_k\trans \boldsymbol{\Xi} \widehat{\boldsymbol{\gamma}}_k \le \lambda_k(\boldsymbol{\Xi})\left[\widehat{\boldsymbol{\gamma}}_k\trans   \widehat{\boldsymbol{\gamma}}_k+c_3\sum_{i=1}^{k-1} (\boldsymbol{\gamma}_i\trans\widehat{\boldsymbol{\gamma}}_k)^2\right]\le \lambda_k(\boldsymbol{\Xi})\left[1+c_3C_{k-1,3}\Lambda_p^2s_n\right].\label{1135}
\end{align}
By $\eqref{1072}$ and $\eqref{1077}$,
\begin{align}
  & \widehat{\boldsymbol{\gamma}}_k\trans \boldsymbol{\Xi} \widehat{\boldsymbol{\gamma}}_k=\widehat{\boldsymbol{\gamma}}_k\trans \boldsymbol{\Sigma}^{-1/2} {\mathbf{B}}\boldsymbol{\Sigma}^{-1/2}\widehat{\boldsymbol{\gamma}}_k\ge \widehat{\boldsymbol{\gamma}}_k\trans \boldsymbol{\Sigma}^{-1/2}\widehat{\mathbf{B}}\boldsymbol{\Sigma}^{-1/2}\widehat{\boldsymbol{\gamma}}_k-\|\widehat{\mathbf{B}}-\mathbf{B}\|_{\infty}\|\boldsymbol{\Sigma}^{-1/2}\widehat{\boldsymbol{\gamma}}_k\|^2_1	\notag\\
	\ge& \frac{\lambda_k(\boldsymbol{\Xi})\left[{\boldsymbol{\alpha}}_k \boldsymbol{\Sigma} {\boldsymbol{\alpha}}_k-2{\boldsymbol{\alpha}}_k\widehat{\mathbf{Q}}_{k-1} \boldsymbol{\Sigma} {\boldsymbol{\alpha}}_k\right]-\lambda_k(\boldsymbol{\Xi})\frac{c_1^{-1}}{C_2}\tau_n \|\boldsymbol{\beta}_k\|^2_1}{{\boldsymbol{\alpha}}_k \boldsymbol{\Sigma} {\boldsymbol{\alpha}}_k-2{\boldsymbol{\alpha}}_k\widehat{\mathbf{Q}}_{k-1} \boldsymbol{\Sigma} {\boldsymbol{\alpha}}_k+c_0\|\widehat{\mathbf{Q}}_{k-1}{\boldsymbol{\alpha}}_k\|_2^2+(1+1/C_2)\tau_n\|\boldsymbol{\beta}_k\|^2_1}-\tau_n\| \widehat{\boldsymbol{\alpha}}_k\|^2_1/C_2.\label{1140}
\end{align}
By the similar arguments as in $\eqref{1163}$ and $\eqref{1094}$, we have 
\begin{align*}
  & \widehat{\boldsymbol{\gamma}}_k\trans \boldsymbol{\Xi} \widehat{\boldsymbol{\gamma}}_k/\lambda_k(\boldsymbol{\Xi})-1 \\
	\ge& \frac{c_0\|\widehat{\mathbf{Q}}_{k-1}{\boldsymbol{\alpha}}_k\|_2^2+(1+1/C_2)\tau_n\|\boldsymbol{\beta}_k\|^2_1+\frac{c_1^{-1}}{C_2}\tau_n \|\boldsymbol{\beta}_k\|^2_1}{{\boldsymbol{\alpha}}_k\trans \boldsymbol{\Sigma} {\boldsymbol{\alpha}}_k-2{\boldsymbol{\alpha}}_k\trans\widehat{\mathbf{Q}}_{k-1} \boldsymbol{\Sigma} {\boldsymbol{\alpha}}_k+c_0\|\widehat{\mathbf{Q}}_{k-1}{\boldsymbol{\alpha}}_k\|_2^2+(1+1/C_2)\tau_n\|\boldsymbol{\beta}_k\|^2_1}-\tau_n\| \widehat{\boldsymbol{\alpha}}_k\|^2_1/(C_2\lambda_k(\boldsymbol{\Xi}))\notag\\
=&\frac{c_0\|\widehat{\mathbf{Q}}_{k-1}{\boldsymbol{\alpha}}_k\|_2^2+(1+\lambda_0/2)\tau_n\|\boldsymbol{\beta}_k\|^2_1 }{1-2{\boldsymbol{\alpha}}_k\trans\widehat{\mathbf{Q}}_{k-1} \boldsymbol{\Sigma} {\boldsymbol{\alpha}}_k+c_0\|\widehat{\mathbf{Q}}_{k-1}{\boldsymbol{\alpha}}_k\|_2^2+(1+1/C_2)\tau_n\|\boldsymbol{\beta}_k\|^2_1}-\tau_n\| \widehat{\boldsymbol{\alpha}}_k\|^2_1/(C_2\lambda_k(\boldsymbol{\Xi}))\notag\\
\ge &\frac{c_0\|\widehat{\mathbf{Q}}_{k-1}{\boldsymbol{\alpha}}_k\|_2^2+(1+\lambda_0/2)\tau_n\|\boldsymbol{\beta}_k\|^2_1 }{1-2c_0^{3/2}\|\widehat{\mathbf{Q}}_{k-1}{\boldsymbol{\alpha}}_k\|_2}-CC_2^{-1}C_{k,1}^2\lambda_k(\boldsymbol{\Xi})^{-1}\Lambda_p^2s_n
\end{align*}
 which together with $\eqref{1131}$ and $\eqref{1162}$  imply that as $n$ is large enough,
\begin{align}
  & \widehat{\boldsymbol{\gamma}}_k\trans \boldsymbol{\Xi} \widehat{\boldsymbol{\gamma}}_k/\lambda_k(\boldsymbol{\Xi})-1\ge - C_8\Lambda_p^2s_n,\label{1141}
\end{align}
where $C_8$ is a constant independent of $n$ and $p$. Combining $\eqref{1135}$ and $\eqref{1141}$, we obtain
\begin{align}
  & |\widehat{\boldsymbol{\gamma}}_k\trans \boldsymbol{\Xi} \widehat{\boldsymbol{\gamma}}_k-\lambda_k(\boldsymbol{\Xi})|\le C_9\lambda_k(\boldsymbol{\Xi})\Lambda_p^2s_n,\label{1142}
\end{align}
where $C_9=\max{(C_8, c_3C_{k-1,3})}$.  Let
\begin{align}
  \widehat{\boldsymbol{\gamma}}_k=d_1\boldsymbol{\gamma}_1+d_2\boldsymbol{\gamma}_2+\cdots+d_{K-1}\boldsymbol{\gamma}_{K-1}+\widehat{c}\widehat{\boldsymbol{\beta}}  \label{1149}
\end{align}
be the orthogonal expansion of $\widehat{\boldsymbol{\gamma}}_k$, where $\widehat{\boldsymbol{\beta}}$ is an vector orthogonal to each of $\boldsymbol{\gamma}_{K-1}$, $\cdots$, $\boldsymbol{\gamma}_1$, with $\|\widehat{\boldsymbol{\beta}}\|_2=1$.    
\begin{align}
    &  |\widehat{\boldsymbol{\gamma}}_k\trans \boldsymbol{\Xi} \widehat{\boldsymbol{\gamma}}_k-\lambda_k(\boldsymbol{\Xi})|\notag\\
		=&\biggr\rvert d_1^2\lambda_1(\boldsymbol{\Xi})+d_2^2\lambda_2(\boldsymbol{\Xi})+\cdots+d_{K-1}^2\lambda_{K-1}(\boldsymbol{\Xi})-\lambda_k(\boldsymbol{\Xi})\biggr\rvert\notag\\
		\ge&\biggr\rvert d_k^2-1\biggr\rvert\lambda_k(\boldsymbol{\Xi})-\lambda_{k+1}(\boldsymbol{\Xi})\sum_{i=k+1}^{K-1}d_i^2-\lambda_1(\boldsymbol{\Xi})\sum_{i=1}^{k-1}d_i^2 \notag\\
		\ge&\biggr\rvert d_k^2-1\biggr\rvert\lambda_k(\boldsymbol{\Xi})-\biggr\rvert d_k^2-1\biggr\rvert \lambda_{k+1}(\boldsymbol{\Xi})- (d_k^2-1)\lambda_{k+1}(\boldsymbol{\Xi})-\lambda_{k+1}(\boldsymbol{\Xi})\sum_{i=k+1}^{K-1}d_i^2-\lambda_{k+1}(\boldsymbol{\Xi})\sum_{i=1}^{k-1}d_i^2\notag\\
		&-(\lambda_1(\boldsymbol{\Xi})-\lambda_{k+1}(\boldsymbol{\Xi}))\sum_{i=1}^{k-1}d_i^2 \notag\\
	=& \biggr\rvert d_k^2-1\biggr\rvert[\lambda_k(\boldsymbol{\Xi})-\lambda_{k+1}(\boldsymbol{\Xi})]-\lambda_{k+1}(\boldsymbol{\Xi})\left[\sum_{i=1}^{K-1}d_i^2-1\right]-(\lambda_1(\boldsymbol{\Xi})-\lambda_{k+1}(\boldsymbol{\Xi}))\sum_{i=1}^{k-1}d_i^2 \notag\\
	\ge& \biggr\rvert d_k^2-1\biggr\rvert[\lambda_k(\boldsymbol{\Xi})-\lambda_{k+1}(\boldsymbol{\Xi})]-\lambda_{k+1}(\boldsymbol{\Xi})\left[\|\widehat{\boldsymbol{\gamma}}_k\|_2^2-1\right]-\lambda_1(\boldsymbol{\Xi})\sum_{i=1}^{k-1}d_i^2\notag\\
	=& \biggr\rvert d_k^2-1\biggr\rvert[\lambda_k(\boldsymbol{\Xi})-\lambda_{k+1}(\boldsymbol{\Xi})]-\lambda_{k+1}(\boldsymbol{\Xi})\left[\|\widehat{\boldsymbol{\gamma}}_k\|_2^2-1\right]-\lambda_1(\boldsymbol{\Xi})\|\mathbf{P}_{k-1}\widehat{\boldsymbol{\gamma}}_k\|_2^2\notag\\
	=& \biggr\rvert d_k^2-1\biggr\rvert[\lambda_k(\boldsymbol{\Xi})-\lambda_{k+1}(\boldsymbol{\Xi})]-\lambda_{k+1}(\boldsymbol{\Xi})\left[\|\widehat{\boldsymbol{\gamma}}_k\|_2^2-1\right]-\lambda_1(\boldsymbol{\Xi})\|\mathbf{P}_{k-1}\widehat{\boldsymbol{\gamma}}_k- \mathbf{P}_{k-1}\widehat{\boldsymbol{\gamma}}_k \|_2^2\notag\\
	\ge& \biggr\rvert d_k^2-1\biggr\rvert[\lambda_k(\boldsymbol{\Xi})-\lambda_{k+1}(\boldsymbol{\Xi})]-\lambda_{k+1}(\boldsymbol{\Xi})\left[\|\widehat{\boldsymbol{\gamma}}_k\|_2^2-1\right]-\lambda_1(\boldsymbol{\Xi})\|\widehat{\mathbf{P}}_{k-1} -\mathbf{P}_{k-1} \|^2,\label{1150}
\end{align}
 where  by $\eqref{1165}$, $\eqref{1094}$ and $\eqref{1053}$, as $n$ is large enough,
\begin{align}
  &\biggr\rvert\|\widehat{\boldsymbol{\gamma}}_k\|_2^2-1\biggr\rvert=\biggr\rvert \widehat{\boldsymbol{\gamma}}_k\trans\widehat{\boldsymbol{\gamma}}_k-1\biggr\rvert  \le (1+1/C_2)\tau_n\|\widehat{\boldsymbol{\alpha}}_k\|^2_1 \le (1+1/C_2) C C_{k,1}\Lambda_p^2s_n.\label{1185}
\end{align}
Hence by $\eqref{1131}$,$\eqref{1142}$,$\eqref{1150}$,   $\eqref{1185}$, $\eqref{1053}$ and Condition \ref{condition_2},
 \begin{align}
   \biggr\rvert d_k^2-1\biggr\rvert\le  C_{10} \Lambda_p^2s_n,\label{1148}
\end{align}
where $C_{10}$ is a constant independent of $n$ and $p$. Since $\widehat{\boldsymbol{\gamma}}_k\trans\boldsymbol{\gamma}_k=d_k>0$, by the orthogonal decomposition $\eqref{1149}$,  $\eqref{1185}$ and $\eqref{1148}$
 \begin{align}
   \biggr\rvert \widehat{\boldsymbol{\gamma}}_k\trans\boldsymbol{\gamma}_k-\| \boldsymbol{\gamma}_k\|_2^2\biggr\rvert
	=\biggr\rvert d_k-1\biggr\rvert\le  \biggr\rvert d_k-1\biggr\rvert(d_k+1)=\biggr\rvert d_k^2-1\biggr\rvert\le  C_{10} \Lambda_p^2s_n.\label{1152}
\end{align}
and 
 \begin{align*}
   &\|\widehat{\boldsymbol{\gamma}}_k-\boldsymbol{\gamma}_k\|_2^2=\biggr\rvert \| \widehat{\boldsymbol{\gamma}}_k\|_2^2-2\widehat{\boldsymbol{\gamma}}_k\trans\boldsymbol{\gamma}_k+\| \boldsymbol{\gamma}_k\|_2^2\biggr\rvert\le \biggr\rvert \| \widehat{\boldsymbol{\gamma}}_k\|_2^2- \| \boldsymbol{\gamma}_k\|_2^2\biggr\rvert+\biggr\rvert    -2\widehat{\boldsymbol{\gamma}}_k\trans\boldsymbol{\gamma}_k+2\| \boldsymbol{\gamma}_k\|_2^2\biggr\rvert\notag\\
	=&\biggr\rvert\|\widehat{\boldsymbol{\gamma}}_k\|_2^2-1\biggr\rvert+2\biggr\rvert \widehat{\boldsymbol{\gamma}}_k\trans\boldsymbol{\gamma}_k-\| \boldsymbol{\gamma}_k\|_2^2\biggr\rvert\le C_{k,2}\Lambda_p^2s_n,\label{1153}
\end{align*}
where $C_{k,2}=(1+1/C_2) C C_{k,1}+2C_{10}$. By $\eqref{1094}$, a similar argument as in the proof of Lemma \ref{lemma_10} leads to
\begin{align}
  &  \| \widehat{\boldsymbol{\xi}}_k\|_1\le C_{k,5}\lambda_1(\boldsymbol{\Xi})\Lambda_p, \quad \| \widehat{\boldsymbol{\xi}}_k-\boldsymbol{\xi}_k\|^2_2\le C_{k,6}\lambda_1(\boldsymbol{\Xi})\Lambda^2_ps_n\label{1294}
\end{align}
where $C_{k,5}$ and $C_{k,6}$ are constants independent of $n$  and $p$. Now we estimate $\|\widehat{\mathbf{P}}_{k}-\mathbf{P}_{k}\|$ and $\|\widehat{\mathbf{Q}}_{k}-\mathbf{Q}_{k}\|$. Let $\widehat{\mathbf{w}}_k=(\mathbf{I}-\widehat{\mathbf{P}}_{k-1})\widehat{\boldsymbol{\zeta}}_k/\|(\mathbf{I}-\widehat{\mathbf{P}}_{k-1})\widehat{\boldsymbol{\zeta}}_k\|_2$. Then it is easy to show that $\widehat{\mathbf{P}}_{k}=\widehat{\mathbf{w}}_k\widehat{\mathbf{w}}_k\trans+\widehat{\mathbf{P}}_{k-1}$ and $\mathbf{P}_{k}=\boldsymbol{\gamma}_k\boldsymbol{\gamma}_k\trans+\mathbf{P}_{k-1}$. Hence,
\begin{align}
  &  \|\widehat{\mathbf{P}}_{k}-\mathbf{P}_{k}\|=\|\widehat{\mathbf{w}}_k\widehat{\mathbf{w}}_k\trans+\widehat{\mathbf{P}}_{k-1}-\boldsymbol{\gamma}_k\boldsymbol{\gamma}_k\trans-\mathbf{P}_{k-1}\|\le\|\widehat{\mathbf{P}}_{k-1}- \mathbf{P}_{k-1}\|\label{1167}\\
	&+2\|\widehat{\mathbf{w}}_k -\boldsymbol{\gamma}_k\|_2\|\boldsymbol{\gamma}_k\|_2+\|\widehat{\mathbf{w}}_k -\boldsymbol{\gamma}_k\|_2^2 = \|\widehat{\mathbf{P}}_{k-1}- \mathbf{P}_{k-1}\|+2\|\widehat{\mathbf{w}}_k -\boldsymbol{\gamma}_k\|_2+\|\widehat{\mathbf{w}}_k -\boldsymbol{\gamma}_k\|_2^2,\notag
\end{align}
where
\begin{align}
  &  \|\widehat{\mathbf{w}}_k -\boldsymbol{\gamma}_k\|_2=\left\|\frac{(\mathbf{I}-\widehat{\mathbf{P}}_{k-1})\widehat{\boldsymbol{\zeta}}_k}{\|(\mathbf{I}-\widehat{\mathbf{P}}_{k-1})\widehat{\boldsymbol{\zeta}}_k\|_2}-\boldsymbol{\gamma}_k\right\|_2 \notag\\
	\le &\left\|\frac{(\mathbf{I}-\widehat{\mathbf{P}}_{k-1})\widehat{\boldsymbol{\zeta}}_k}{\|(\mathbf{I}-\widehat{\mathbf{P}}_{k-1})\widehat{\boldsymbol{\zeta}}_k\|_2}
-(\mathbf{I}-\widehat{\mathbf{P}}_{k-1})\right\|_2+\|(\mathbf{I}-\widehat{\mathbf{P}}_{k-1})-\boldsymbol{\gamma}_k\|_2 \notag\\
	\le &\left|1-\|(\mathbf{I}-\widehat{\mathbf{P}}_{k-1})\widehat{\boldsymbol{\zeta}}_k\|_2 \right|+\|(\mathbf{I}-\widehat{\mathbf{P}}_{k-1})\widehat{\boldsymbol{\zeta}}_k -\boldsymbol{\gamma}_k\|_2.\label{1168}
\end{align}
By $\eqref{1131}$ , $\eqref{1294}$ and $\eqref{1053}$
\begin{align}
  &   \|(\mathbf{I}-\widehat{\mathbf{P}}_{k-1})\widehat{\boldsymbol{\zeta}}_k -\boldsymbol{\gamma}_k\|_2=\|(\mathbf{I}-\widehat{\mathbf{P}}_{k-1})(\widehat{\boldsymbol{\zeta}}_k -\boldsymbol{\gamma}_k)-\widehat{\mathbf{P}}_{k-1}\boldsymbol{\gamma}_k \|_2\notag\\
	=&\|(\mathbf{I}-\widehat{\mathbf{P}}_{k-1})(\widehat{\boldsymbol{\zeta}}_k -\boldsymbol{\gamma}_k)-(\widehat{\mathbf{P}}_{k-1}-\mathbf{P}_{k-1})\boldsymbol{\gamma}_k \|_2\le \|\widehat{\boldsymbol{\zeta}}_k -\boldsymbol{\gamma}_k\|_2+\|\widehat{\mathbf{P}}_{k-1}-\mathbf{P}_{k-1}\|\notag\\
	=&\|\lambda_k(\boldsymbol{\Xi})^{-1}\boldsymbol{\Sigma}^{-1/2}\widehat{\boldsymbol{\xi}}_k-\boldsymbol{\Sigma}^{1/2}\boldsymbol{\alpha}_k\|_2+\|\widehat{\mathbf{P}}_{k-1}-\mathbf{P}_{k-1}\|\notag\\
	\le& \lambda_k(\boldsymbol{\Xi})^{-1}\|\boldsymbol{\Sigma}^{-1/2}\|\|\widehat{\boldsymbol{\xi}}_k-\mathbf{B}\boldsymbol{\alpha}_k\|_2+\|\widehat{\mathbf{P}}_{k-1}-\mathbf{P}_{k-1}\|\le \sqrt{C_{11} \Lambda_p^2s_n},\label{1169}
\end{align}
where $C_{11}$ is a constant independent of $n$ and $p$, and we use Condition \ref{condition_2} (c). Hence
\begin{align}
  &   \left|\|(\mathbf{I}-\widehat{\mathbf{P}}_{k-1})\widehat{\boldsymbol{\zeta}}_k\|_2-1\right|=\left|\|(\mathbf{I}-\widehat{\mathbf{P}}_{k-1})\widehat{\boldsymbol{\zeta}}_k\|_2-\|\boldsymbol{\gamma}_k\|_2\right|\le \|(\mathbf{I}-\widehat{\mathbf{P}}_{k-1})\widehat{\boldsymbol{\zeta}}_k -\boldsymbol{\gamma}_k\|_2 \le \sqrt{C_{11} \Lambda_p^2s_n} .	\label{1169}
\end{align}
Therefore, by $\eqref{1168}$-$\eqref{1169}$, $\|\widehat{\mathbf{w}}_k -\boldsymbol{\gamma}_k\|_2=2\sqrt{C_{11} \Lambda_p^2s_n}$ which together with $\eqref{1167}$ imply that $\|\widehat{\mathbf{P}}_{k}-\mathbf{P}_{k}\|\le \sqrt{C_{k,3}\Lambda_p^2s_n}$, where $C_{k,3}$ is a constant independent of $n$ and $p$.  Let $\widehat{\mathbf{v}}_k=(\mathbf{I}-\widehat{\mathbf{Q}}_{k-1})\widehat{\boldsymbol{\xi}}_k/\|(\mathbf{I}-\widehat{\mathbf{Q}}_{k-1})\widehat{\boldsymbol{\xi}}_k\|_2$. Then it is easy to show that $\widehat{\mathbf{Q}}_{k}=\widehat{\mathbf{v}}_k\widehat{\mathbf{v}}_k\trans+\widehat{\mathbf{Q}}_{k-1}$ and $\mathbf{Q}_{k}=\boldsymbol{\xi}_k\boldsymbol{\xi}_k\trans/\|\boldsymbol{\xi}_k\|^2_2+\mathbf{Q}_{k-1}$. A similar argument leads to   $\|\widehat{\mathbf{Q}}_{k}-\mathbf{Q}_{k}\|\le \sqrt{C_{k,4}\Lambda_p^2s_n}$, where $C_{k,4}$ is a constant independent of $n$ and $p$. We have proved the lemma.
\end{proof}
